\documentclass[preprint,12pt]{elsarticle}
\usepackage{mathrsfs}
\usepackage{amsbsy}
\usepackage{amsmath}
\usepackage{stmaryrd}
\usepackage{bbding}
\usepackage{dcolumn}
\usepackage{graphicx}
\usepackage{amsfonts}
\usepackage{amssymb}
\usepackage{psfrag}
\usepackage{wrapfig}
\usepackage{subfigure}
\usepackage{makeidx}
\usepackage{bm}
\usepackage{epsf}
\usepackage{epsfig}
\usepackage{setspace}
\usepackage{graphicx}
\usepackage{epstopdf}
\usepackage{psfrag}
\usepackage{subfigure}
\usepackage{fancyhdr}
\usepackage[parfill]{parskip}
\usepackage{setspace}
\usepackage{mathrsfs}
\usepackage{amscd,extarrows}
\usepackage{pict2e}
\usepackage{fullpage}
\usepackage{mathtools}
\newcommand*\Bell{\ensuremath{\boldsymbol\ell}}

\DeclarePairedDelimiter\floor{\lfloor}{\rfloor}

\allowdisplaybreaks
\DeclareMathAlphabet{\mathscrbf}{OMS}{mdugm}{b}{n}
\newdefinition{rmk}{Remark}
\begin{document}

\title{A Unified Gas Kinetic Scheme for Multi-scale and Multi-component Plasma Transport}

\author[HKUST]{Chang Liu}
\ead{cliuaa@connect.ust.hk}

\author[HKUST,HKUST2]{Kun Xu}
\ead{makxu@ust.hk}

\address[HKUST]{Department of Mathematics, Hong Kong University of Science and Technology, Clear Water Bay, Kowloon, Hong Kong}

\address[HKUST2]{Department of Mechanical and Aerospace Engineering, Hong Kong University of Science and Technology, Clear Water Bay, Kowloon, Hong Kong}

\begin{abstract}

A unified gas kinetic scheme (UGKS) for multi-scale and multi-component plasma transport is constructed.
The current scheme is a direct modeling method,
where the time evolution solutions from the Vlasov-BGK equations of electron and ion
and the Maxwell equations are used to construct a scale-dependent plasma simulation model.
As a result, with the changing of modeling scales of mesh size and time step and with a variation of Knudsen number and Larmor radius,
the discretized governing equations for a wide range of plasma evolution regimes can be obtained.
The physics recovered in UGKS ranges from the Vlasov equation in the kinetic scale to different-type magnetohydrodynamic (MHD) equations in the
hydrodynamic scale.
The unified treatment covers
all scales from the collisionless particle transport to the hydrodynamic wave interactions.
The UGKS provides a more general evolution model than the Vlasov equation in the kinetic scale and
many types of MHD equations in the hydrodynamic scale, such as the two fluids model, the Hall, the resistive,
and the ideal MHD equations.
All above specific governing equations become the subsets of the UGKS.
The key dynamics in UGKS is the un-splitting treatment of particle collision, acceleration, and transport
in the construction of numerical flux across a cell interface,
and this flux is a scale-dependent evolving solution of the Vlasov-BGK model.
At the same time, the plasma evolution is coupled with the Maxwell equations in an implicit way,
which automatically provides a smooth transition between
the Amp$\grave{\text{e}}$re's law and the Ohm's law for the calculation of electric field.
The time step of UGKS is not limited by the relaxation time, the cyclotron period, and the speed of light in the MHD regime.
Our scheme is able to give a physically accurate solution for plasma simulation with a large variation of Knudsen number and normalized Larmor radius.
It can be used to simulate the phenomena from the Vlasov limit to the scale of plasma skin depth
for the capturing of two-fluid effect,
and the phenomena in the plasma transition regime with
a modest Knudsen number and Larmor radius in a unified smoothly transitional way.
The UGKS is validated by numerical test cases, such as the Landau damping and two stream instability in the kinetic regime, and the Brio-Wu shock tube problem and the Orszag-Tang MHD turbulence problem
in the hydrodynamic regime.
The scheme is also used to study the geospace environment modeling (GEM),
such as the challenging magnetic reconnection problem in the transition regime.
At the same time, the magnetic reconnection mechanism of the Sweet-Parker model and the Hall effect model
can be connected smoothly through the variation of Larmor radius in the UGKS simulations.
Overall, the UGKS is a physically reliable multi-scale plasma simulation method.
It provides a powerful and unified approach for the study of plasma physics.

\end{abstract}

\maketitle

\section{Introduction}
\begin{figure}[t!]
  \centering
  \includegraphics[width=0.65\textwidth]{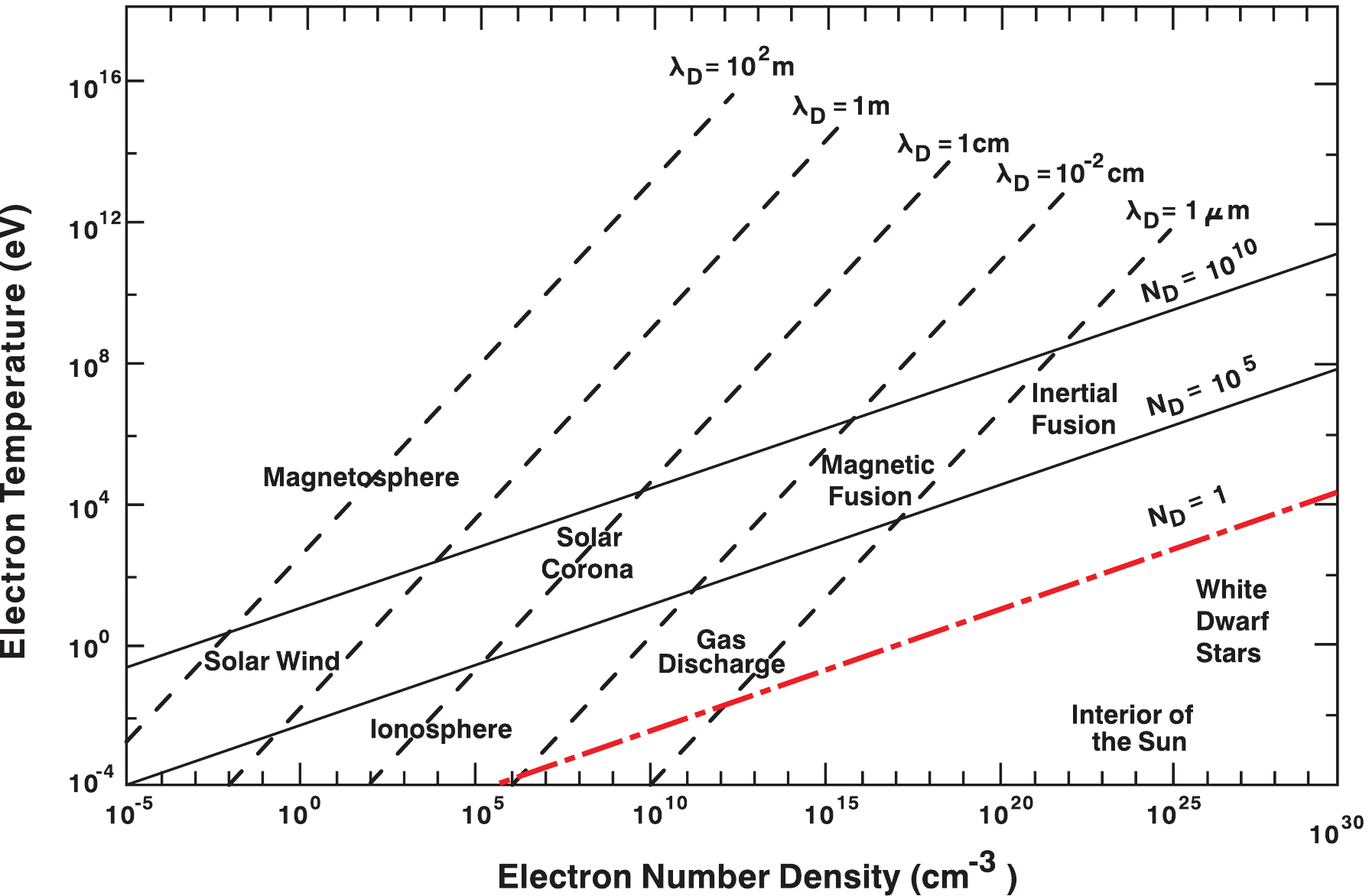}
  \caption{The range of plasma phenomena.
  The temperature $k_BT$ is plotted in units of electron volts.
  The UGKS is mainly used for the region above the red dot-dashed
  line $N_D=1$.}
  \label{weakly}
\end{figure}

Generally, plasma is a medium with positive, negative, and neutral particles.
It is quasi-neutral on the length scale larger than the Debye length $r_D$.
The plasma we concerned is weakly coupled plasma,
such as the solar corona, the magnetosphere around the Earth, plasma inside a Tokamak, etc. as shown in Fig. \ref{weakly}.
The dynamics of a weakly coupled plasma can be described by the kinetic equations.
The number of electron inside a Debye cubic, namely the plasma parameter $N_D$ is much larger than one (or the coupling parameter $\Gamma\ll1$).
In a weakly coupled plasma, the ratio of the plasma frequency $\omega_p$ to the collision frequency $\nu$ is large than one,
so that the collective behavior is observed on the time scale longer than the plasma period $\omega_p^{-1}$,
and on the length scale larger than the Debye length ($\lambda_D=U_t\omega_p^{-1}$).
In this work, we propose a unified gas kinetic scheme (UGKS) that can be applied to the fully ionized weakly coupled plasma composed of electrons and ions.

\begin{figure}[t!]
  \centering
  \includegraphics[width=0.65\textwidth]{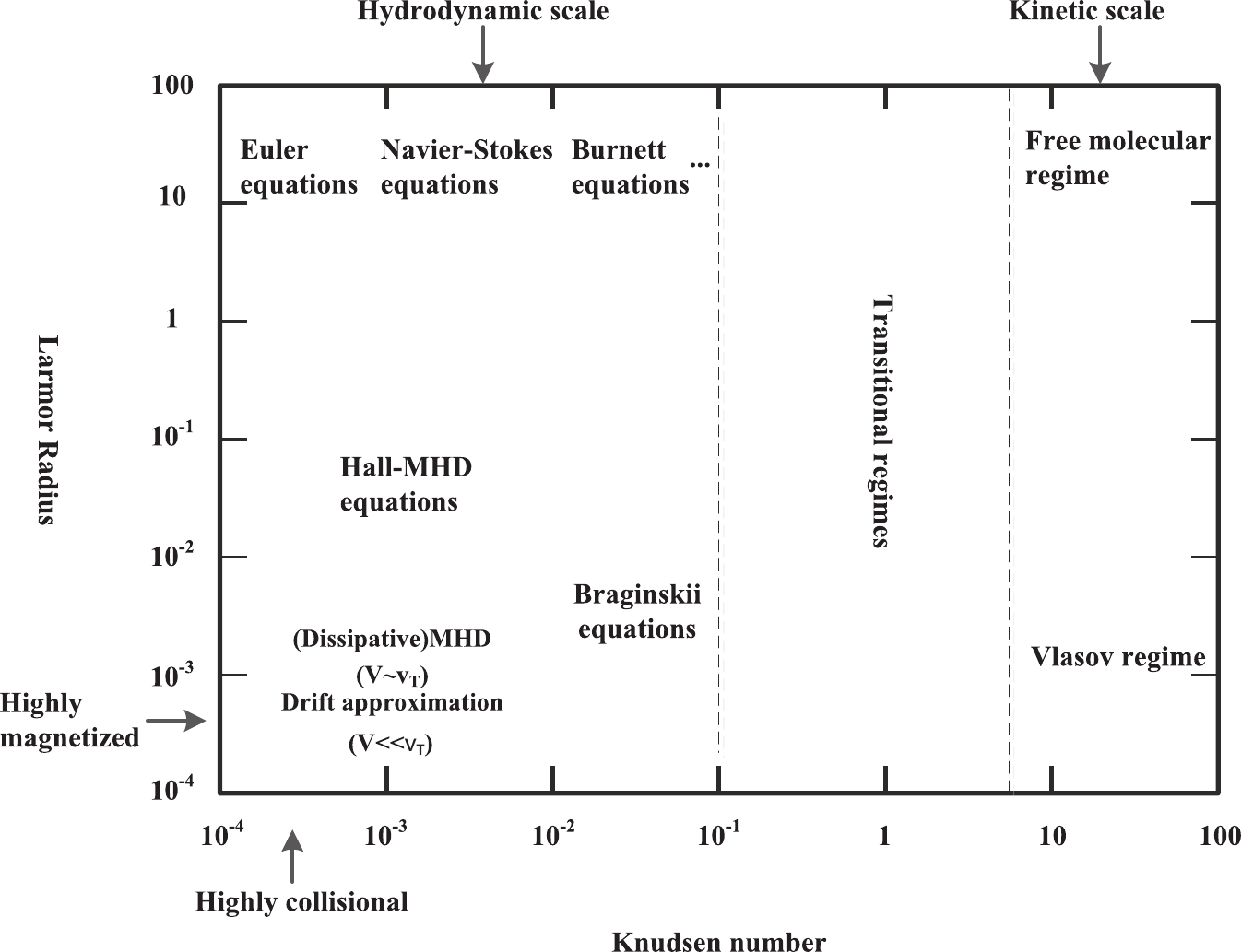}
  \caption{Flow Regimes of Plasma.}
  \label{regime}
\end{figure}

The motion of charged particles in a plasma is coupled with the evolution of electromagnetic field.
The flow regime of a plasma is more complex than that of a neutral one.
Many plasma parameters are important in characterizing plasma flow property, such as the Debye length, the plasma frequency, the ion inertial length, and the plasma beta, etc..
Among those parameters, two parameters are important in characterizing the flow regimes of plasma,
namely the Knudsen number $\text{Kn}$ and the normalized Larmor radius $\hat{r}_{L_i}$.
The Knudsen number is the ratio between the particle mean free path to a characteristic length,
and the normalized Larmor radius is the ratio between the Larmor radius and the characteristic length.
The Knudsen number indicates the collision intensity and the normalized Larmor radius is about how strong the plasma is magnetized.
On the kinetic scale,
the dynamics of plasma is described by
the kinetic equation, such as Fokker-Planck-Landau equation \cite{chen1984plasma}.
In kinetic equation,
the averaged electromagnetic field effect is modeled
to the order of the reciprocal of the normalized Larmor radius
and the collision term effect is to the order of the reciprocal of the Knudsen number.
As shown in Fig. \ref{regime}, in the rarefied regime with large Knudsen number,
the plasma follows the collisionless Vlasov equation. In the highly collisional regime, the plasma is described by the hydrodynamic-type equations.
When $\hat{r}_{L_i}\ll1$, the plasma can be described by single fluid MHD equations.
The two-fluid effect or the Hall effect becomes important in large $\hat{r}_{L_i}$ regimes.

Numerical methods have been developed
for plasma simulation since 1950s,
such as the particle-in-cell (PIC) method,
the kinetic Vlasov solvers and
the hydrodynamic MHD solvers.
The traditional PIC method suffers from
statistical noise and restricted time step \cite{vahedi1995monte,filbet2003comparison}.
In order to overcome those shortcomings,
a series of asymptotic preserving PIC methods are developed by Degond \emph{et al.} \cite{degon2010,degon2017},
which remain stable and are consistent with
a quasi-neutral model in the quasi-neutral limit.
Reformulated Vlasov-Poisson/Maxwell equations are used
in the AP-PIC methods to unify models
in different regimes.



Based on the high accurate phase space reconstruction techniques,
many high order Vlasov solvers are proposed
such as the conservative method \cite{filbet2001conservative},
the semi-Lagrange method \cite{qiu2011conservative},
the finite element method \cite{guo2013hybrid}, and finite difference method \cite{xiong2014high}.
The Vlasov solvers can capture solutions of the collisionless Vlasov equation accurately.
However only few work has been done to study the full kinetic equation including both the electromagnetic acceleration and the collision integral term for multi-species.

In the highly collisional and highly magnetized regimes,
the plasma flow is governed by the hydrodynamic-type equations.
Riemann solution based MHD solvers \cite{powell1999solution,brio1988upwind}
and the kinetic based MHD solvers \cite{xu1999gas,araya2015magneto} have been proposed,
which can recover the ideal MHD equations,
and the extended Hall-MHD or dissipative MHD equations.
Those methods are based on the single-fluid hydrodynamic equations,
and hence can only be applied on the scale much larger than the Larmor radius and in the regime close to equilibrium.
Another type of schemes are proposed for the two-fluid system \cite{shumlak2003approximate,hakim2006high,loverich2005discontinuous,loverich2011discontinuous,srinivasan2011analytical}.
The two-fluid model takes into account the electron mass and the separation of electrons and ions,
which can recover the flow regime from MHD to Euler regime.

In the transitional regime, hybrid methods are usually proposed
to connect the kinetic solver with the hydrodynamic one \cite{crestetto2012kinetic,dimarco2014asymptotic}.
In the hybrid models, the PIC method is used in the collisionless regime
and the hydrodynamic models are used in the collisional regime.
The main difficulty for the hybrid method is to find a proper criteria to couple different numerical models.
And the solutions are not physically reliable in the transitional regime.
Another asymptotic preserving (AP) scheme of the FPL equation in \cite{jin2011class,dimarco2015numerical}
can preserve the collisionless and Euler regimes,
but the scheme is currently built on one species, where the two species effect and the MHD limit cannot be resolved.
A good review about AP methods and multiscale models for plasma physics was given recently by Degond and Deluzet \cite{degond}.
Even with asymptotic preserving property, the cell size is still limited by the mean free path scale for accuracy consideration for the capturing of dissipative solution.

In the past years, based on the methodology of direct modeling on the mesh size and time step scales, the UGKS has been
developed to simulate multiple scale transport problems.
For the rarefied gas dynamics,
radiative transfer, and phonon transport,
the UGKS becomes a successful multi-scale method and provides accurate solutions in all regimes \cite{xu2010,Huangxuyu2,Huangxuyu3,sun-wj,sun-wj2,guo-xu,xu-book}.
In this paper, the UGKS is developed for the plasma simulation.
The corresponding scheme is based on the space and time evolution solution of the Vlasov-BGK equations for electron and ion,
and the Maxwell equations for electromagnetic field.
During this modeling process, the evolution solution on the numerical cell size and time step is used
 for the construction of the scheme.
The coupling of the particle transport, collision, acceleration in the flux calculation, and the implicit treatment
of the source terms inside each control volume, endow the scheme with multi-scale nature.
With the variation of Knudsen number, inter-species collision frequency, and dimensionless Larmor radius,
the scheme unifies the solutions in the kinetic Vlasov regime, the two-fluid regime, and the MHD regime with a smooth transition
among them.

The outline of this paper is the following.
Section 2 reviews the kinetic equations, the Maxwell equations, and their asymptotic behavior in limiting flow regimes.
The detailed formulation of UGKS is proposed in Section 3.
Section 4 studies the numerical property of UGKS as well as the stability constraint.
The numerical test cases are given in Section 5 to validate the UGKS in different flow regimes.
The last section is the conclusion.

\section{A review of kinetic and hydrodynamic model equations of plasma}
\subsection{Kinetic equations and Maxwell equations}
In this paper, we consider the kinetic equations
\begin{equation}\label{kinetic}
  \frac{\partial f_\alpha}{\partial t}
  +\mathbf{u}\cdot \nabla_\mathbf{x} f_{\alpha}
  +\frac{F_\alpha}{m_\alpha}
  \cdot\nabla_\mathbf{u}f_{\alpha}=
  \frac{f_\alpha^+-f_\alpha}{\tau_\alpha}
\end{equation}
for electron ($\alpha=e$) and ion ($\alpha=i$),
where $F_\alpha$ is the averaged electromagnetic force.
The collisional term is modeled by a single BGK-type collision operator proposed by Andries, Aoki, and Perthame (AAP model) \cite{AAP}.
Other collision terms such as the Landau collision operator can be built into our scheme using the method we proposed for Boltzmann collision operator \cite{liu2016}.
In AAP model, one global collision operator is used
for each component to take account of both
self-collision and cross-collision.
The post collision distribution $f^+$ is
\begin{equation}\label{f+}
  f^+_\alpha=\rho_\alpha\left(\frac{m_\alpha}{2\pi k_B \bar{T}_\alpha}\right)^{3/2}
  \exp\left(-\frac{m_\alpha}{2\pi k_B \bar{T}_\alpha}(u-\bar{\mathbf{U}}_\alpha)^2\right).
\end{equation}
The parameters $\bar{T}_\alpha$ and $\bar{\mathbf{U}}_\alpha$ are connected to the
macroscopic properties of individual components by \cite{AAP}
\begin{equation}\label{W-bar}
\begin{aligned}
  \bar{\mathbf{U}}_\alpha&=\mathbf{U}_\alpha+\tau_\alpha\sum_r 2\frac{m_r}{m_\alpha+m_r}\nu_{\alpha r}(\mathbf{U}_r-\mathbf{U}_\alpha),\\
  \frac32 k_B \bar{T}_\alpha&=\frac32 k_B T_\alpha-\frac{m_\alpha}{2}(\bar{\mathbf{U}}_\alpha-\mathbf{U}_\alpha)^2\\
  &+\tau_\alpha\sum_r 4m_\alpha\frac{m_r}{(m_\alpha+m_r)^2}\nu_{\alpha r}\left(\frac32k_BT_r-\frac32 k_BT_\alpha+\frac{m_r}{2}(\mathbf{U}_r-\mathbf{U}_{\alpha})^2\right),
\end{aligned}
\end{equation}
where $\nu_{\alpha r}$ are the interaction coefficients between particles that measure the strength of intermolecular collision.
The relaxation time is determined by
$\tau_\alpha=1/\sum_r\nu_{\alpha r}$.

The averaged electromagnetic force $\mathbf{F} = \mathbf{E}+\mathbf{u}_\alpha\times \mathbf{B}$,
where the averaged electric field $\mathbf{E}$ and magnetic field $\mathbf{B}$ follow the Maxwell equations,
\begin{equation}\label{maxwell1}
  \begin{cases}
  &\frac{\partial \mathbf{B}}{\partial t}=-\nabla_\mathbf{x} \times \mathbf{E},\\
  &\frac{\partial \mathbf{E}}{\partial t}=c^2\nabla_\mathbf{x} \times \mathbf{B}-\frac{1}{\epsilon_0} \mathbf{j},
\end{cases}
\end{equation}
with divergence constraints of
\begin{equation}\label{maxwell2}
\nabla_\mathbf{x}\cdot \mathbf{E}=\frac{1}{\epsilon_0}
\sum_\alpha q_\alpha n_\alpha,
\qquad \nabla_\mathbf{x} \cdot \mathbf{B}=0.
\end{equation}
The perfectly hyperbolic Maxwell equations (PHM) \cite{munz2000divergence} are used in current work
to evolve the electromagnetic field in order to preserve the divergence constraints,
\begin{equation}\label{Maxwell3}
\begin{aligned}
  &\frac{\partial \mathbf{E}}{\partial t}-c^2\nabla_\mathbf{x}\times \mathbf{B}+\chi c^2\nabla_\mathbf{x}\phi=-\frac{1}{\epsilon_0}\mathbf{j},\\
  &\frac{\partial \mathbf{B}}{\partial t}+\nabla_\mathbf{x} \times \mathbf{E}+\gamma \nabla_\mathbf{x} \psi=0,\\
  &\frac{1}{\chi}\frac{\partial \phi}{\partial t}+\nabla_\mathbf{x} \cdot \mathbf{E}=\frac{\rho}{\epsilon_0},\\
  &\frac{\epsilon_0\mu_0}{\gamma}\frac{\partial \psi}{\partial t}+\nabla_\mathbf{x} \cdot \mathbf{B}=0,
\end{aligned}
\end{equation}
where $\mathbf{j}$ is the electric current density, $\phi$, $\psi$ are correction potentials, and $\gamma$, $\chi$ are error propagation speeds.
Our scheme is built on the BGK-Maxwell system
Eqs.\eqref{kinetic},\eqref{maxwell1},\eqref{maxwell2}, which are able to cover the flow regimes of plasma from the collisionless Vlasov regime to the continuum MHD regime.

\subsection{Asymptotic limits of BGK-Maxwell system}

In order to study the asymptotic limits, we introduce the following scaling of the BGK-Maxwell system, the scaled variables are given by
\begin{equation}\label{scale}
\begin{aligned}
  &\bar{x}=\frac{x}{l_0},\quad
  \bar{\mathbf{u}}=\frac{\mathbf{u}}{u_0},\quad
  \bar{t}=\frac{u_0}{l_0}t,\quad
  \bar{m}=\frac{m}{m_i},\quad
  \bar{n}_\alpha=\frac{n_\alpha}{n_0},\\
  &\bar{q}=\frac{q}{q_i},\quad
  \bar{f}\alpha=\frac{m_in_0}{u_0^3}f_\alpha,\quad
  \bar{\mathbf{B}}=\frac{\mathbf{B}}{B_0},\quad
  \bar{\mathbf{E}}=\frac{\mathbf{E}}{B_0 u_0}.
\end{aligned}
\end{equation}
where $u_0$ is the ion thermal velocity scale given by $u_0=\sqrt{k_BT_0/m_i}$.
Inserting this scaling into the BGK-Maxwell system and omitting the bars, we get the following scaled BGK-Maxwell system
\begin{equation}\label{scaled-Vlasov-BGK-Maxwell}
\begin{aligned}
  &\frac{\partial f_\alpha}{\partial t}
  +\mathbf{u}\cdot \nabla_\mathbf{x} f_{\alpha}
  +\frac{q_\alpha}{r_{L_i}m_\alpha}(\mathbf{E}+\mathbf{u}\times \mathbf{B})
  \cdot\nabla_\mathbf{u}f_{\alpha}=\frac{f_\alpha^+-f_\alpha}{\tau_\alpha},\\
  &\frac{\partial \mathbf{B}}{\partial t}+\nabla_\mathbf{x} \times \mathbf{E}=0,\\
  &\frac{\partial \mathbf{E}}{\partial t}-c^2\nabla_\mathbf{x} \times \mathbf{B}=
  -\frac{1}{\hat{\lambda}_D^2r_{L_i}} \mathbf{j},
\end{aligned}
\end{equation}
where the physically significant similarity parameters are: normalized relaxation time $\tau_\alpha$,
the scaled Debye length $\hat{\lambda}_D=\lambda_D/r_{L_i}$,
the normalized ion Larmor radius $r_{L_i}$,
the normalized speed of light $c$,
and the normalized electron mass $m_e$
in the non-dimensional equations.
In the following, we study the asymptotic behavior of the BGK-Maxwell system with respect to the similarity parameters.

\begin{figure}
  \centering
  \includegraphics[width=0.9\textwidth]{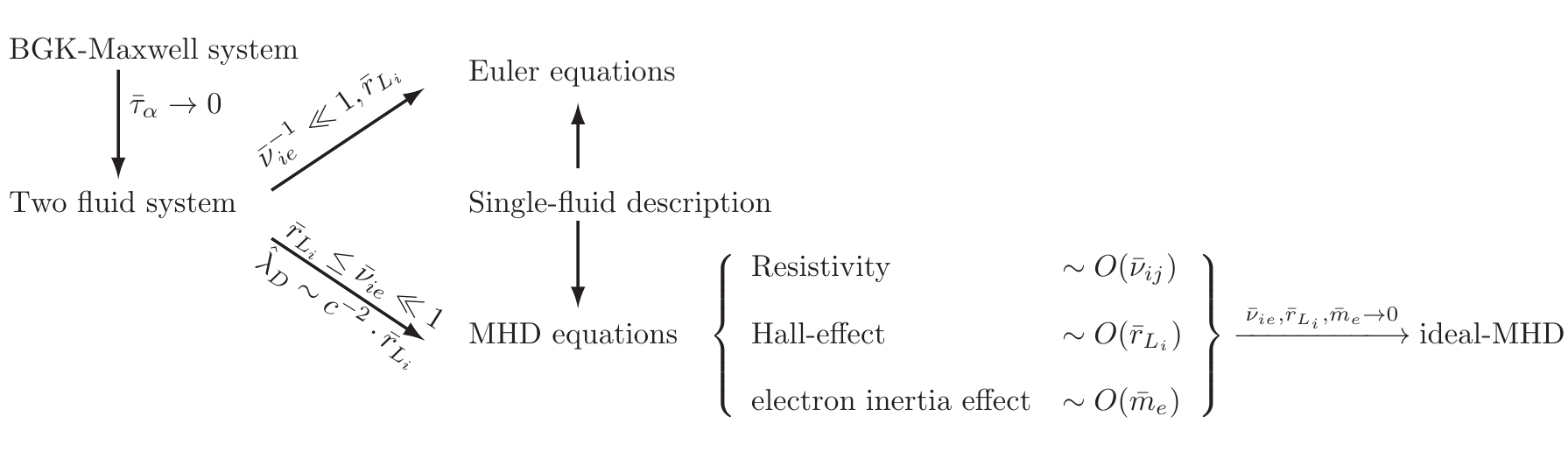}\\
  \caption{Asymptotic limits of BGK-Maxwell system.}\label{asymptotic}
\end{figure}

The normalized relaxation time $\tau_\alpha$ is proportional to the Knudsen number $\text{Kn}_\alpha$,
which is defined as the ratio between the mean free path $\ell_\alpha$ and the length scale $l_0$.
When $\tau_\alpha\ll 1$, the zero-th order of the BGK-Maxwell system with respect to $\tau_\alpha$ gives the following
hydrodynamic two-fluid equations
\begin{equation}\label{two-fluid}
\begin{aligned}
  \partial_t \rho_\alpha+
  \nabla_\mathbf{x}\cdot(\rho_\alpha\mathbf{U}_\alpha)&=0,\\
  \partial_t(\rho_\alpha \mathbf{U}_\alpha)
  +\nabla_\mathbf{x}\cdot(\rho_\alpha \mathbf{U}_\alpha\mathbf{U}_\alpha+p_\alpha I)&
  =\frac{n_\alpha q_\alpha}{r_{L_i}}
  (\mathbf{E}+\mathbf{U}_\alpha\times \mathbf{B})+S_\alpha,\\
  \partial_t \mathscr{E}_\alpha
  +\nabla_\mathbf{x}\cdot((\mathscr{E}_\alpha
  +p_\alpha)\mathbf{U}_\alpha)&
  =\frac{n_\alpha q_\alpha}{r_{L_i}}
  \mathbf{U}_\alpha\cdot \mathbf{E}+Q_\alpha,
\end{aligned}
\end{equation}
where $S_i=-S_e$ and $Q_i=-Q_e$ are the corresponding momentum and energy exchange between electron and ion,
\begin{equation}
\begin{aligned}
  S_\alpha=&\int \mathbf{u}\frac{f^+_\alpha-f_\alpha}{\tau_\alpha} d\mathbf{u}
  =\sum_r\frac{2m_\alpha m_r}{m_\alpha+m_r}
   n_\alpha\nu_{\alpha r}
  (\mathbf{U}_r-\mathbf{U}_\alpha),\\
  Q_\alpha=&\int\frac12|\mathbf{u}-\mathbf{U}|^2
  \frac{f^+_\alpha-f_\alpha}{\tau_\alpha}d\mathbf{u}\\
  =&\sum_r\frac{4m_\alpha m_r}{(m_\alpha+m_r)^2}
  n_r \nu_{\alpha r}
  \left(\frac32k_BT_r-\frac32k_BT_\alpha +\frac{m_r}{2}(\mathbf{U}_r-\mathbf{U}_\alpha)^2\right).
\end{aligned}
\end{equation}
When the interspecies molecular interaction is intensive, i.e. $\nu_{i,e}^{-1}\ll (1,r_{L_i})$,
the zero-th order of the two-fluid system with respect to $(r_{L_i}\nu_{i,e})^{-1}$
gives the Euler equations for the total densities, common velocity, and common temperature,
\begin{equation}\label{Euler}
\begin{aligned}
  \partial_t \rho+\nabla_\mathbf{x}\cdot(\rho \mathbf{U})&=0,\\
  \partial_t(\rho \mathbf{U})+\nabla_\mathbf{x}\cdot(nk_BT I+\rho \mathbf{U} \mathbf{U})&=0,\\
  \partial_t \mathscr{E}+\nabla_\mathbf{x}\cdot((\mathscr{E}+nk_BT)\mathbf{U})&=0,
\end{aligned}
\end{equation}
where
\begin{equation}
\begin{aligned}
  &\rho=\sum_\alpha \rho_\alpha, \quad \rho \mathbf{U}=\sum_\alpha \rho_\alpha \mathbf{U}_\alpha,\\
  &\frac32nk_BT+\frac{\rho}{2}\mathbf{U}^2=\mathscr{E}=\sum_\alpha \mathscr{E}_\alpha.
\end{aligned}
\end{equation}
When $r_{L_i}\nu_{i,e}\leq 1$, $r_{L_i}\ll 1$, $m_e\ll m_i$, and $\hat{\lambda}_D\sim c^{-1}\ll 1$,
the first order with respect to $r_{L_i}$,
the zero-th order with respect of $m_e/m_i$ and $\hat{\lambda}_D$
of the two-fluid system give the MHD equations with Hall effect and magnetic diffusion,
\begin{equation}\label{Hall-MHD}
  \begin{aligned}
  &\partial_t\rho_i+\nabla_\mathbf{x}\cdot(\rho_i \mathbf{U}_i)=0,\\[1.3mm]
  &\partial_t(\rho_i \mathbf{U}_i)+\nabla_\mathbf{x}\cdot(\rho_i \mathbf{U}_i \mathbf{U}_i+p_i \mathbf{I})
  =\frac{n_i q_i}{r_{L_i}} \mathbf{E}+\mathbf{j}_i\times \mathbf{B},\\[1.3mm]
  &\mathbf{E}+\mathbf{U}\times \mathbf{B}=
  \frac{r_{L_i}}{\sigma} \mathbf{j}+\frac{1}{n_e q_e}\mathbf{j}\times \mathbf{B}
  +\frac{r_{L_i}}{n_e q_e}\nabla_\mathbf{x} p_e,\\[1.3mm]
  &\partial_t \mathscr{E}_\alpha+\nabla_\mathbf{x}\cdot((\mathscr{E}_\alpha+p_\alpha)\mathbf{U}_\alpha)
  =\frac{1}{r_{L_i}}\mathbf{j}_\alpha\cdot \mathbf{E},\\[1.3mm]
  &\partial_t \mathbf{B}+\nabla_\mathbf{x} \times \mathbf{E}=0,
   \quad \\[1.3mm]
 &\mathbf{j}=r_{L_i}\hat{\lambda}_D^2c^2\nabla_\mathbf{x}\times \mathbf{B},
  \end{aligned}
\end{equation}
where
$\mathbf{j}=n_e q_e \mathbf{U}_e+n_i q_i \mathbf{U}_i$ is the plasma electric current density and
$$\sigma=\frac{n_i e^2(m_i+m_e)}{2m_im_e\nu_{ie}}
\approx\frac{n_ie^2}{2m_e\nu_{ie}}$$ is the electrical conductivity, which is related to the magnetic Reynolds number $R_m$ by
$\sigma=R_m\hat{\lambda}_D^2 r_{L_i}^2 c^2$.

The ideal-MHD equations is the zero-th order approximation of the MHD equations \eqref{Hall-MHD} with respect to the Larmor radius, which read
\begin{equation}
  \begin{aligned}
  &\partial_t\rho+\nabla_\mathbf{x}\cdot(\rho \mathbf{U})=0,\\[1.2mm]
  &\partial_t(\rho \mathbf{U})+\nabla_\mathbf{x}\cdot(\rho \mathbf{U}\mathbf{U}+pI)=\frac{(\mathbf{B}\cdot \nabla_\mathbf{x})\mathbf{B}}{\mu_0}-\nabla_\mathbf{x}\left(\frac{\mathbf{B}^2}{2\mu_0}\right),\\[1.2mm]
  &\partial_t \mathscr{E}+\nabla_\mathbf{x}\cdot((\mathscr{E}+p)\mathbf{U})=\frac{1}{\mu_0}\rho \mathbf{U}\cdot(\nabla_\mathbf{x}\times \mathbf{B}\times \mathbf{B}),\\[1.2mm]
  &\partial_t \mathbf{B}+\nabla_\mathbf{x} \times (\mathbf{U}\times \mathbf{B})=0,
  \end{aligned}
\end{equation}
where $\rho=\rho_i+\rho_e$ is the total mass,
$p=p_i+p_e$ is the total pressure,
$(\mathbf{B}\cdot \nabla_\mathbf{x})\mathbf{B}/\mu_0$ is the magnetic tension force, and
$\nabla_\mathbf{x}(\mathbf{B}^2/2\mu_0)$ is magnetic pressure.

The above analysis shows that in the continuum regime,
when the interspecies collisions are strong, the gas mixture behaves like dielectric material, the BGK-Maxwell equations goes to Euler equations Eq.\eqref{Euler}.
For a conductive plasma, the BGK-Maxwell equations can span the complete range from the neutral two-fluid system to the resistive-MHD, Hall-MHD, and ideal MHD equations as shown in Fig.\ref{asymptotic}.
The BGK-Maxwell equations can be applied in the transition regime as well with modest Knudsen number, Debye length, and Larmor radius.

\section{Unified gas kinetic scheme}
\subsection{General framework}
The UGKS is a finite volume scheme built on the phase space $\mathbf{X}=\sum_i \Omega_{\mathbf{x} i}\otimes\sum_j \Omega_{\mathbf{u} j}=
\sum_{i,j} \Omega_{\mathbf{x}i\times\mathbf{u}j}$.
The averaged conservative variables in a physical cell $\Omega_{\mathbf{x}i}$ is
$$ \left(\mathbf{W}_{\alpha}\right)_{ i}=\frac{1}{|\Omega_{\mathbf{x}i}|}
\int_{\Omega_{\mathbf{x}i}} \mathbf{W}_\alpha d\mathbf{x},$$
similar for the cell averaged electromagnetic field
and divergence correction terms $\mathbf{Q}_i=(E_{1i},E_{2i},E_{3i},
B_{1i},B_{2i},B_{3i},\phi_{i},\psi_{i})^T$.
The averaged distribution function in a phase cell
$\Omega_{\mathbf{x}i\times\mathbf{u}j}$ is
$$\left(f_{\alpha}\right)_{ij}=
\frac{1}{|\Omega_{\mathbf{x}i\times\mathbf{u}j}|}
\int_{\Omega_{\mathbf{x}i\times\mathbf{v}j}} f_\alpha d\mathbf{x}d\mathbf{u}.$$
The time evolution of the velocity distribution function
\begin{equation}\label{update-f}
  \left(f_{\alpha}\right)_{ij}^{n+1}=
  \left(f_{\alpha}\right)_{ij}^{n}
  -\frac{1}{|\Omega_{\mathbf{x}}|}
  \sum_{s_i\in\partial_{\Omega_{\mathbf{x}}}}
  |s_i|\mathscr{F}^x_{f_\alpha s_i}
  -\frac{1}{|\Omega_{\mathbf{u}}|}
  \sum_{s_j\in\partial_{\Omega_{\mathbf{u}}}}
  |s_j|\mathscr{F}^v_{f_\alpha s_j}
  +\int_{t^n}^{t^{n+1}}
  Q_c(f_{\alpha ij},f_{\alpha ij}) dt,
\end{equation}
is coupled with the time evolution of the conservative variables
\begin{equation}\label{update-w}
  \left(\mathbf{W}_{\alpha}\right)_{i}^{n+1}=
  \left(\mathbf{W}_{\alpha}\right)_{i}^n
  -\frac{1}{|\Omega_{\mathbf{x}}|}
  \sum_{s_i\in\partial{\Omega_{\mathbf{x}}}}
  |s_i|\mathscr{F}_{\mathbf{W}_\alpha s_i}
  +\frac{\Delta t}{\tau_\alpha}
  \left(\left(\bar{\mathbf{W}}\right)_{i}^{n}-
  \left(\mathbf{W}_{\alpha}\right)_{i}^{n}\right)
  +\Delta t \mathbf{S}_{\mathbf{W}_{\alpha}i}^{n+1},
\end{equation}
and the time evolution of the electromagnetic fields
\begin{equation}\label{update-em}
  \mathbf{Q}_i^{n+1}=\mathbf{Q}_i^{n}
  +\frac{\Delta t}{|\Omega_\mathbf{x}|}\sum_{s_i\in\partial \Omega}|s_i|\mathscr{F}_{\mathbf{Q} s_i}+\Delta t \mathbf{S}_{Q_i}^{n+1},
\end{equation}
where $|\Omega|$ is the volume of the cell, $s_i\in\partial \Omega$ is the cell interface. The numerical flux in UGKS
for the distribution function and conservative variables are calculated from the time-dependent distribution function at a cell interface, for example
\begin{align}
  \mathscr{F}^x_{f_\alpha s_i}&=
  \int_{t^n}^{t^{n+1}} \mathbf{u}\cdot \mathbf{n}_{s_i} f_\alpha(\mathbf{x}_{s_i},t,\mathbf{u}) dt,\label{Ffx}\\
  \mathscr{F}_{\mathbf{W}_\alpha s_i}&=
  \int_{t^n}^{t^{n+1}}\int \mathbf{\psi}
  \mathbf{u}\cdot \mathbf{n}_{s_i} f_\alpha(\mathbf{x}_{s_i},t,\mathbf{u}) d\mathbf{u} dt,\label{Fwx}\\
  \mathscr{F}^v_{f_\alpha s_j}&=
  \int_{t^n}^{t^{n+1}} \Bell \cdot \mathbf{n}_{s_j} f(\mathbf{x},t,\mathbf{u}_{s_j}) dt \label{Ffv},
\end{align}
where $\mathbf{n}$ is the outer normal direction at interface, and $\Bell =(l_u,l_v,l_w)^T$ is the acceleration due to Lorenz force.

\subsection{Numerical flux}
In this subsection, the detailed formulation of numerical flux is derived
(subscript $\alpha$ is dropped for simplicity).
The time dependent distribution function at a cell interface plays an important role in the UGKS flux calculation, which is modeled based on the integral solution of the kinetic equation Eq.\eqref{kinetic},
\begin{equation}\label{integral-solution}
  f(\mathbf{x},t,\mathbf{u})=\frac{1}{\tau}\int_{t^n}^{t}
  f^+(\mathbf{x}',t',\mathbf{u}')\text{e}^{-\frac{t-t'}{\tau}}dt'
  +\text{e}^{-\frac{t-t^n}{\tau}}f_0\left(\mathbf{x}-\mathbf{u}(t-t^n),
  \mathbf{u}-\Bell (t-t^{n})\right),
\end{equation}
where $\mathbf{x}'=\mathbf{x}-\mathbf{u}(t-t^n-t')$, $\mathbf{u}'=\mathbf{u}-\Bell (t-t^n-t')$, and $f_0(\mathbf{x},\mathbf{u})$ is the initial distribution function at $t=t^{n}$.
Assume that
the cell interface is located at $\mathbf{x}_0$,
the velocity cell center is located at $\mathbf{u}_k$,
with the normal direction $\mathbf{e}_1$,
and the local basis
$(\mathbf{e}_1,\mathbf{e}_2,\mathbf{e}_3)$.
The initial distribution function is reconstructed as
\begin{equation}\label{f0}
\begin{aligned}
  f_0(\mathbf{x},\mathbf{u})=&\left(f_0^L(x_0)+\Delta \mathbf{x}\cdot\frac{\partial f_0^L}{\partial \mathbf{x}}+\Delta \mathbf{u}\frac{\partial f_0^L}{\partial \mathbf{u}}\right)
  \left(1-H[\Delta \mathbf{x}\cdot \mathbf{e}_1]\right)\\
  +&\left(f_0^R(\mathbf{x}_0)+\Delta \mathbf{x}\frac{\partial f_0^R}{\partial \mathbf{x}}+\Delta \mathbf{u}\frac{\partial f_0^R}{\partial \mathbf{u}}\right)
  H[\Delta \mathbf{x}\cdot \mathbf{e}_1],
\end{aligned}
\end{equation}
where $\Delta \mathbf{x}=\mathbf{x}-\mathbf{x}_0$,
$\Delta \mathbf{u}=\mathbf{u}-\mathbf{u}_k$.
Slope limit such as the van-Leer limiter is used to reconstruct the slope $\partial _\mathbf{x} f_0$ and $\partial_\mathbf{u} f_0$ in each phase space cells.

The post collision distribution function is expanded around the cell interface as
\begin{equation}\label{f+0}
\begin{aligned}
  &f^+(\mathbf{x},t,\mathbf{u})=f_0^+(\mathbf{x}_0,t,\mathbf{u})
  \left[1+(1-H[\bar{x}])a^L \bar{x}+H[\bar{x}]a^R\bar{x}+b\bar{y}+c\bar{z}+A(t-t^n)\right],\\[1.3mm]
  =&f_0^+(\mathbf{x}_0,t,\mathbf{u}_k)\left[1-2\lambda(\mathbf{u}-\bar{\mathbf{U}})\cdot\Delta\mathbf{u}\right]
  \left[1+(1-H[\bar{x}])a^L \bar{x}+H[\bar{x}]a^R\bar{x}+b\bar{y}+c\bar{z}+A(t-t^n)\right],\\[1.3mm]
  =&f_0^+(\mathbf{x}_0,t,\mathbf{u}_k)
  \left[1-2\lambda(\mathbf{u}-\bar{\mathbf{U}})\cdot\Delta\mathbf{u}
  +(1-H[\bar{x}])a^L \bar{x}+H[\bar{x}]a^R\bar{x}+b\bar{y}+c\bar{z}+A(t-t^n)\right],
\end{aligned}
\end{equation}
where $\bar{x}=\Delta \mathbf{x}\cdot \mathbf{e}_1$,
$\bar{y}=\Delta \mathbf{y}\cdot\mathbf{e}_2$, $\bar{z}=\Delta \mathbf{x}\cdot \mathbf{e}_3$.
The coefficients $a^{L,R},b,c,A$  are related to the spatial and time derivatives of $f^+$, for example
\begin{equation}
a^{L,R}=\left.\frac{1}{f^+_0}\frac{\partial f^+_0}{\partial \mathbf{W}_0}\frac{\partial \mathbf{W}_0^{L,R}}{\partial x}\right|_{x=x_0},\quad
A=\left.\frac{1}{f^+_0}\frac{\partial f^+_0}{\partial \mathbf{W}_0}\frac{\partial \mathbf{W}_0^{L,R}}{\partial t}\right|_{t=t^n},
\end{equation}
and an analogous expression can be derived for $b$, $c$.
From the reconstructed distribution $f_0(\mathbf{x}_0,\mathbf{u})$, the macroscopic conservative variables at a cell interface can be calculated
\begin{equation}\label{ftow}
  \mathbf{W}_0(\mathbf{x}_0)=\int \mathbf{\psi}\left(f_0^L(\mathbf{x}_0)H[\mathbf{u}\cdot\mathbf{e}_1]+
  f_0^R(\mathbf{x}_0)(1-H[\mathbf{u}\cdot\mathbf{e}_1])\right)d\mathbf{u}.
\end{equation}
Then, the averaged macroscopic variables $\bar{\mathbf{W}}_0$ in Eq.\eqref{W-bar} can be evaluated, and its spacial derivative is reconstructed to be
\begin{equation}\label{wx}
  \left.\frac{\partial \bar{\mathbf{W}}_0^{L,R}}
  {\partial \bar{x}}\right|_{\mathbf{x}=\mathbf{x}_0}
  =\frac{\mathbf{\bar{W}}_0^{L,R}(\mathbf{x}_0)-\mathbf{\bar{W}}_0(\mathbf{x}_{L,R})}
  {(\mathbf{x}_0-\mathbf{x}_{L,R})\cdot\mathbf{n}}.
\end{equation}
The time derivative can be derived from the compatible condition
 \begin{equation}
   \left.\frac{d}{dt}\int (f^+-f) \mathbf{\psi} d\mathbf{u}\right|_{t=t^n} =0,
 \end{equation}
which gives
 \begin{equation}\footnotesize
\frac{\partial \mathbf{\bar{W}}_0}{\partial t} =
 \int\left((a_x^lH[u]+a_x^r(1-H[u])uf^+_0+a_yvf^+_0+
 (\mathbf{E}+\mathbf{u}\times\mathbf{B})\cdot\frac{\partial f^+_0}{\partial \mathbf{u}}\right)
 \mathbf{\psi}  d\mathbf{u}.
 \end{equation}

 Substituting Eq.\eqref{f0} and Eq.\eqref{f+0} into the integral solution Eq.\eqref{integral-solution}, the time dependent distribution at the cell interface is
 \begin{equation}\label{integral-solution2}
   \begin{aligned}
   f(\mathbf{x}_0,t,\mathbf{u}_0)=&\left(1-\text{e}^{(t-t^n)/\tau}\right)f_0^+(\mathbf{x}_0)\\[1.3mm]
   +&\left((t-t^n+\tau)\text{e}^{-(t-t^n)/\tau}-\tau\right)
   \left(a^LH[\bar{u}]+a^{R}(1-H[\bar{u}])\right)\bar{u}f^+_0(\mathbf{x}_0,\mathbf{u}_0)\\[1.3mm]
   +&\left((t-t^n+\tau)\text{e}^{-(t-t^n)/\tau}-\tau\right)
   \left(b\bar{v}+c\bar{w}\right)\bar{u}f^+_0(\mathbf{x}_0,\mathbf{u}_0)\\[1.3mm]
   +&\left(t-t^n+\tau\left(\text{e}^{-(t-t^n)/\tau}-1\right)\right)Af^+_0(\mathbf{x}_0,\mathbf{u}_0)\\[1mm]
   +&\text{e}^{-(t-t^n)/\tau}\left(f_0^L(\mathbf{x}_0,\mathbf{u}_0)-(t-t^n)\mathbf{u}\cdot\frac{\partial f_0^L}{\partial \mathbf{x}}
   -(t-t^n)\Bell \cdot\frac{\partial f_0^L}{\partial \mathbf{x}}\right)H[\bar{u}]\\[1mm]
   +&\text{e}^{-(t-t^n)/\tau}\left(f_0^R(\mathbf{x}_0,\mathbf{u}_0)-(t-t^n)\mathbf{u}\cdot\frac{\partial f_0^R}{\partial \mathbf{x}}
   -(t-t^n)\Bell \cdot\frac{\partial f_0^R}{\partial \mathbf{x}}\right)(1-H[\bar{u}]),
   \end{aligned}
 \end{equation}
 where $\bar{u}=\mathbf{u}\cdot \mathbf{e}_1$,
 $\bar{v}=\mathbf{u}\cdot \mathbf{e}_2$,
 $\bar{w}=\mathbf{u}\cdot \mathbf{e}_3$.
Based on the time dependent distribution function we constructed at the cell interface, the UGKS flux can be calculated by Eq.\eqref{Ffx}, \eqref{Fwx}, and \eqref{Ffv}.

Denote the Jacobian matrix of the PHM system Eq.\eqref{Maxwell3} as $A_x$ for x direction and
$A_y$ for y direction.
The eigen-systems of the Jacobian matrixes are given in Appendix.
The wave propagation method proposed by LeVeque \cite{hakim2006high, LeVeque} is used to construct the numerical flux in Eq.\eqref{update-em}, for example
\begin{equation}\label{Fqx}
\begin{aligned}
  \left(\mathscr{F}^x_Q\right)_{i-1/2,j}
  =&\frac12(A_1 Q_{i,j}+A_1 Q_{i-1,j})-
  \frac12(A_1^+\Delta Q_{i-1/2}-A_1^-\Delta Q_{i-1/2})\\[1.3mm]
  +&\frac12 \sum_p \text{sign} (s_{i-1/2,j}^p)
    \left(1-\frac{\Delta t}{\Delta x}|s^p_{i-1/2,j}|\right)
    \mathscrbf{L}^p_{1,i-1/2,j}\Phi(\theta)\\[1.3mm]
  -&\frac{\Delta t}{2\Delta x}\mathbf{A_1^+}\mathscrbf{A}^-_2\Delta \mathbf{Q}_{i,j+1/2}
    -\frac{\Delta t}{2\Delta x}\mathbf{A_1^+}\mathscrbf{A}^+_2\Delta \mathbf{Q}_{i,j-1/2}\\[1.3mm]
  -&\frac{\Delta t}{2\Delta x}\mathbf{A_1^-}\mathscrbf{A}^-_2\Delta \mathbf{Q}_{i+1,j+1/2}
    -\frac{\Delta t}{2\Delta x}\mathbf{A_1^-}\mathscrbf{A}^+_2\Delta \mathbf{Q}_{i+1,j-1/2},
\end{aligned}
\end{equation}
where
\begin{equation}
  \mathscrbf{L}^p_{1,i-1/2,j}=\Bell ^p_{i,i-1/2,j}\cdot(\mathbf{f}_{1,i,j}-\mathbf{f}_{1,i-1,j})\mathbf{r}^p_{1,i-1/2,j}.
\end{equation}
$\Phi$ is a limiter function with  $$\theta^p_{i-1/2}\equiv\frac{\mathscrbf{L}^p_{I-1/2}\cdot\mathscrbf{L}^p_{i-1/2}}{\mathscrbf{L}^p_{i-1/2}\cdot\mathscrbf{L}^p_{i-1/2}},$$
with $I=i-1$ if $s^p_{i-1/2}>0$ and $I=i+1$ if $s^p_{i-1/2}<0$.
The left and right going fluctuations are
\begin{equation}
  \mathscrbf{A}_2^\pm\Delta  \mathbf{Q}_{i,j-1/2}=
  \mathbf{A_1^\pm}\Delta \mathbf{Q}_{i,j-1/2}\mp\sum_p
  \text{sign}(s^p_{i,j-1/2})\left(1-\frac{\Delta t}{\Delta x}|s^p_{i,j-1/2}|\right)
  \mathbf{\mathscrbf{L}}^p_{1,i,j-1/2}\Phi(\theta).
\end{equation}
An analogous expression can be derived for the Y directional flux.

\subsection{Numerical treatment of particle acceleration and collision}
\begin{figure}
  \centering
  \includegraphics[width=0.6\textwidth]{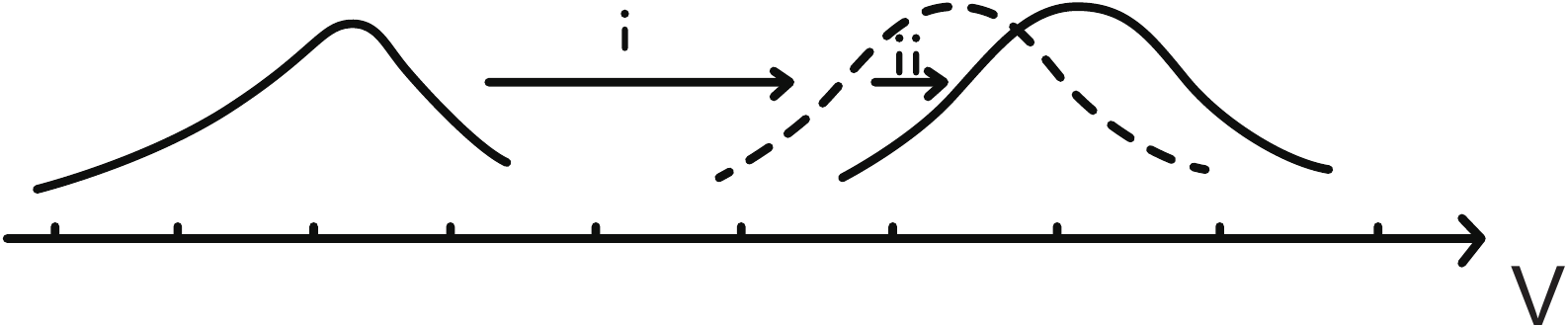}\\
  \caption{Two steps to update distribution function.}\label{Update-distribution-function}
\end{figure}

In many cases, the electromagnetic acceleration is so large
that the time step is restricted to be very small.
In order to remove the constraint, we split the particle acceleration and collision process into two steps.
First we shift the velocity distribution between cell centers as shown in
Fig. \ref{Update-distribution-function}-i.

Kinetic equation
\begin{equation}
  \frac{\partial f}{\partial t}+\Bell \cdot \nabla_\mathbf{v} f=\frac{f^+-f}{\tau}
\end{equation}
has exact solution
\begin{equation}\label{sln}
  f(\mathbf{x},t,\mathbf{u})=f_0(\mathbf{x},\mathbf{v}-\Bell t)\text{e}^{-t/\tau}
  +f^{+}(\mathbf{x},\mathbf{v}-\Bell t)(1-\text{e}^{-t/\tau}).
\end{equation}
Based the exact solution, the distribution functions is shifted as
\begin{equation}
\begin{aligned}
f^{**}(u_{k},v_{l},w_{m})=
f^{*}(u_{k-su},v_{l-sv},w_{m-sw})
\text{e}^{-\Delta t_1/\tau}
+f^{+*}(u_{k-su},v_{l-sv},w_{m-sw})
(1-\text{e}^{-\Delta t_1/\tau}),
\end{aligned}
\end{equation}
where $su=\Delta t_1 l_u/\Delta u$,
$sv=\Delta t_1 l_v/\Delta v$,
$sw=\Delta t_1 l_w/\Delta w$,
and $$\Delta t_1=\min\left(
\frac{\Delta u}{|l_u|}
\floor*{\frac{|l_u|\Delta t}{\Delta u}},
\frac{\Delta v}{|l_v|}
\floor*{\frac{|l_v|\Delta t}{\Delta v}},
\frac{\Delta w}{|l_w|}
\floor*{\frac{|l_w|\Delta t}{\Delta w}}\right).$$

\begin{figure}
  \centering
  \includegraphics[width=0.9\textwidth]{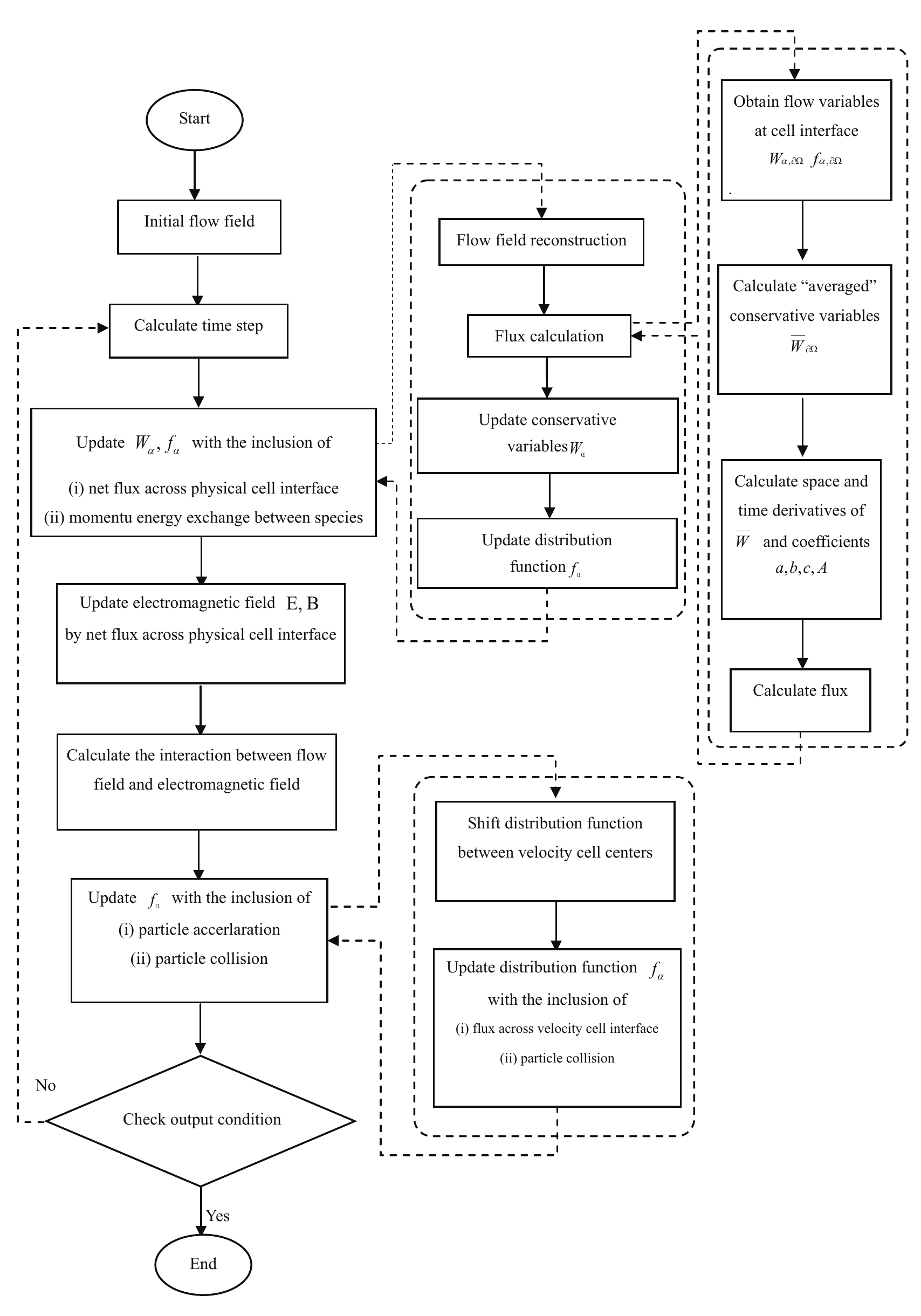}\\
  \caption{Flowchart of UGKS algorithm.}\label{flowchart}
\end{figure}

Then the distribution function is updated to the next time step as shown in Fig. \ref{Update-distribution-function}-ii by
\begin{equation}\label{update-f2}
  f^{n+1}=\left(f^{**}
  -\sum_{s_j\in\partial_{\Omega_{\mathbf{v}}}}
  |s_j|\int_{t^n+\Delta t_1}^{t^{n+1}} \Bell \cdot \mathbf{n} f(\mathbf{x}_i,t,\mathbf{u}_{k+1/2}) dt
  +\frac{\Delta t}{\tau^{n+1}}(f^{+})^{n+1}\right)/
  \left(1+\frac{\Delta t}{\tau^{n+1}}\right),
\end{equation}
where $f^{+,n+1}$ and $\tau^{n+1}$ are obtained from the updated conservative variables $\bar{\mathbf{W}}$.
As $t^{n+1}-t^{n}-\Delta t_1$ is small,
the simplified upwind flux can be used
in Eq.\eqref{update-f2},
\begin{equation}
f(\mathbf{x}_i,t,\mathbf{u}_{k+1/2})=
f(\mathbf{x}_i,t^n+\Delta t_1,\mathbf{u}_{k+1/2}).
\end{equation}
In summary, the UGKS algorithm is shown as the flowchart in Fig. \ref{flowchart}.

\section{Limiting solutions of UGKS}
\subsection{Limits of UGKS flux}
Define the time averaged flux of UGKS as
\begin{equation}\label{t-averaged}
\tilde{F}_{f}=\frac{1}{\Delta t}\int_0^{\Delta t} \mathbf{u}f_{j+1/2}(t) dt.
\end{equation}
In the limit of $\Delta t \ll \tau$, $\tilde{F}_{f}$ follows the particle transport and acceleration, which gives the Vlasov flux
\begin{equation}
\lim_{\Delta t/\tau\rightarrow0} \tilde{F}_{f}=\mathbf{u}\left[f_{i+1/2}-\frac{1}{2}\Delta t\mathbf{u}\cdot\partial_\mathbf{x} f_{i+1/2}
-\frac12 \Delta t \Bell \cdot\partial_\mathbf{u} f\right],
\end{equation}
which is consistent with the collisionless Vlasov equation.
In the limit of $\Delta t\gg \tau$, $\tilde{F}_{f}$  converge to the hydrodynamic flux, which gives
\begin{equation}
\lim_{\Delta t/\tau\rightarrow\infty} \tilde{F}_{f}=\mathbf{u}\left[g_{i+1/2}-\tau(\partial_t g_{1+1/2}+\mathbf{u}\cdot\partial_\mathbf{x}g_{i+1/2}+\Bell \cdot\partial_\mathbf{u}g_{i+1/2})+\frac12\Delta t \partial_t g)\right],
\end{equation}
from which the hydrodynamic two-fluid system can be recovered as well as the MHD equations.
The flow dynamics depends on the ratio of local time step $\Delta t$ to the relaxation parameter $\tau$.
In other words, the UGKS presents the plasma evolution model on the scales of the cell size and time step.

\subsection{Limits of the source terms}
\begin{figure}
  \centering
  \includegraphics[width=0.45\textwidth]{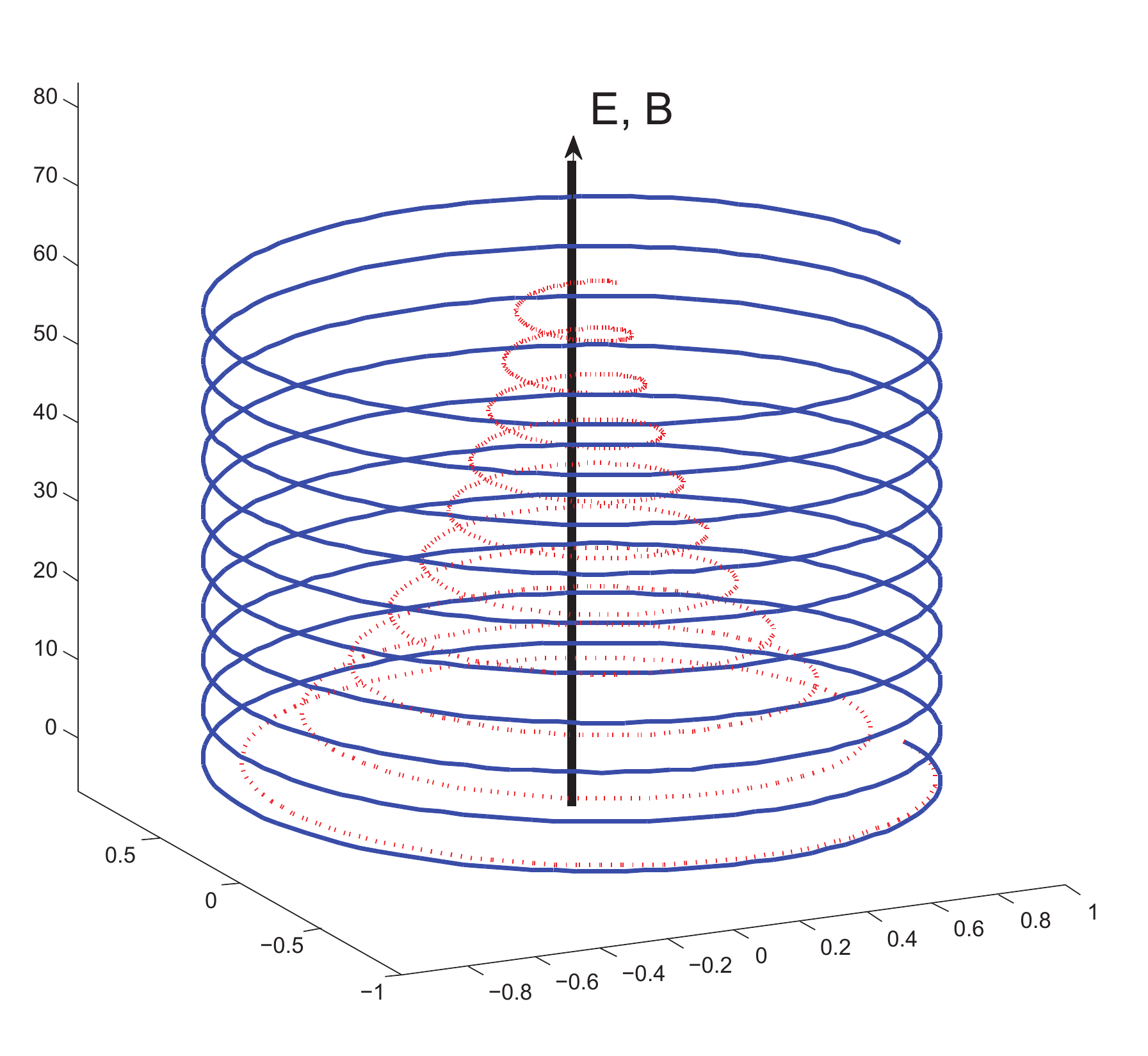}{a}
  \includegraphics[width=0.4\textwidth]{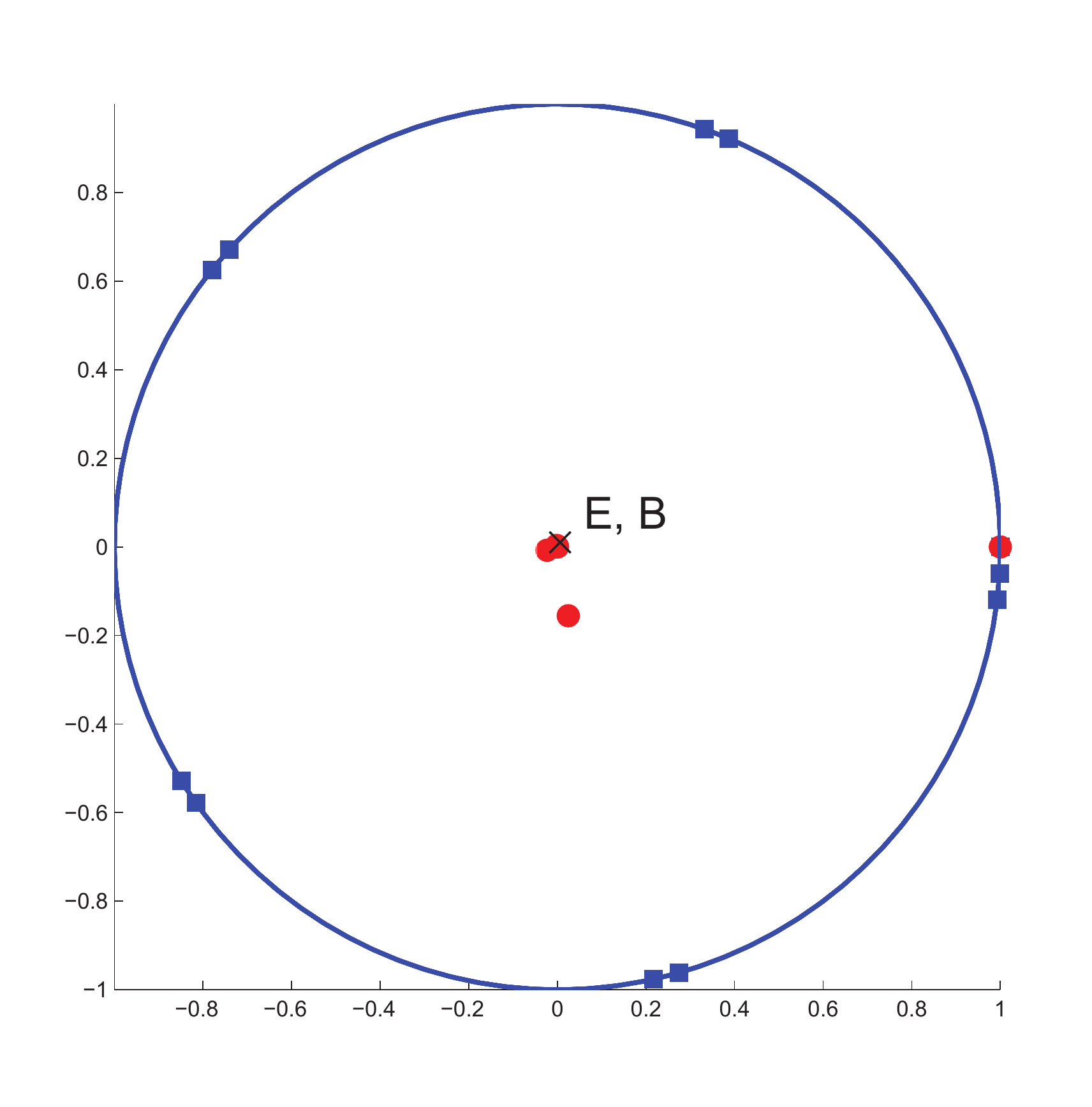}{b}
  \caption{Helical trajectory calculated by semi-implicit scheme (solid line and square symbols) and by fully implicit scheme (dotted line and circular symbols). Time step is set to be (a) $\Delta t=0.2\omega_c^{-1}$, and (b) $\Delta t=2\omega_c^{-1}$.}\label{helix}
\end{figure}

In Eq.\eqref{update-w} and \eqref{update-em},
the interactions between the electromagnetic field and the flow field, i.e. the electromagnetic source term and the electric current density term, are treated implicitly.
The interaction can be written down into the following linear system
\begin{align}
\frac{m_i \mathbf{j}_i^{n+1}-m_i \mathbf{j}_i^*}{\Delta t}&=\left(\frac{n_i\mathbf{E}}{r_L}+\frac{\mathbf{j}_i}{r_L}\times \mathbf{B}\right)^{n+1},\label{implicit1}\\
\frac{m_e \mathbf{j}_e^{n+1}-m_e \mathbf{j}_e^*}{\Delta t}&=\left(\frac{n_e\mathbf{E}}{r_L}-\frac{\mathbf{j}_e}{r_L}\times \mathbf{B}\right)^{n+1},\label{implicit2}\\
-\lambda^2_D\frac{\mathbf{E}^{n+1}-\mathbf{E}^*}{\Delta t}&=\left(\frac{\mathbf{j}_i}{r_L}+\frac{\mathbf{j}_e}{r_L}\right)^{n+1},\label{implicit3}\\
\frac{\phi^{n+1}-\phi^{*}}{\Delta t}&=\frac{\chi}{\hat{\lambda}_D^2r_{L_i}}\rho^{n+1},\label{implicit4}
\end{align}
where the numerical fluxes are included in terms $\mathbf{j}_i^*$, $\mathbf{j}_e^*$, and $\mathbf{E}^*$.
The implicit source terms treatment endows the UGKS with the following two properties.

First, the time step is not be restricted by the cyclotron period. As shown in Fig. \ref{helix} that the linear system Eq.\eqref{implicit1}-\eqref{implicit4} are contract projection with respect to the helical motion of the charged particles.
The motion of particles will be confined to the magnetic field lines as time step getting large,
which ensures the stability of the scheme.

Second, the asymptotic limits of the BGK-Maxwell system are preserved.
As $r_{L_i},\hat{\lambda_D}\to 0$, Eq.\eqref{implicit3} preserves the synchronous motion of electron and ion, and the quasi-neutrality of plasma
\begin{equation}
\begin{aligned}
  (\mathbf{j}_i+\mathbf{j}_e)^{n+1}&=
  -r_{L_i}\lambda_D^2\frac{\mathbf{E}^{n+1}-\mathbf{E}^*}
  {\Delta t}\\
   \xrightarrow{r_{L_i},\hat{\lambda_D}\to 0}
   \mathbf{V}^{n+1}_i&=\mathbf{V}^{n+1}_e, \nabla\cdot \mathbf{j}=0,n_i^{n+1}=n_e^{n+1}.
\end{aligned}
\end{equation}
As $r_{L_i}\to 0$, Eq.\eqref{implicit2} converges to the ideal Ohm's law
\begin{equation}\label{limit2}
\begin{aligned}
  (n_e \mathbf{E}-\mathbf{j}_e\times \mathbf{B})^{n+1}&=r_{L_i}\frac{m_e\mathbf{j}_e^{n+1}-
  m_e\mathbf{j}_e^*}{\Delta t}\\
  \xrightarrow{r_{L_i}\to 0} \mathbf{E}^{n+1}+&\mathbf{V}^{n+1}\times \mathbf{B}^{n+1}=0.
\end{aligned}
\end{equation}
The summation of Eq.\eqref{implicit1}, \eqref{implicit2}, and \eqref{implicit3} converges to the MHD momentum equation as $r_{L_i}\to0$ and $\hat{\lambda}_D\sim c^{-1}$
\begin{equation}
\begin{aligned}
    \sum_\alpha (-1)^\alpha\frac{m_\alpha \mathbf{j}^{n+1}_\alpha-m_\alpha \mathbf{j}_\alpha^*}{\Delta t}&=\frac{\mathbf{E}^{n+1}}{r_{L_i}}(n_i-n_e)^{n+1}
    -\lambda_D^2\frac{\mathbf{E}^{n+1}-\mathbf{E}^*}{\Delta t}\times \mathbf{B}^{n+1}\\
    \xrightarrow[\hat{\lambda}_D\sim c^{-1}]{r_{L_i}\to0}
     \frac{m_i \mathbf{j}-m_i \mathbf{j}^*}{\Delta t}&\approx \mathbf{B}^{n+1}\times(\nabla\times \mathbf{B}^n).
\end{aligned}
\end{equation}
Eq.\eqref{limit2} indicates that UGKS provides a smooth transition from the Amp$\grave{\text{e}}$re's Law to the ideal Ohm's law with a decreasing of the Larmor radius $r_{L_i}$,
which can remove the light speed constraint on time step
in the continuum MHD regime.
For large $r_{L_i}$, the time step constraint of UGKS is
\begin{equation}\label{cfl1}
  \Delta t=\frac{\text{CFL}\Delta x}{\max(U+4c_s,c)}=\text{CFL}\frac{\Delta x}{c}.
\end{equation}
The CFL number increases as $r_{L_i}$ decreases.
For example, in the Brio-Wu shock test case,
$\text{CFL}=0.3$ for $r_{L_i}<1$,
$\text{CFL}=0.4$ for $r_{L_i}=10^{-1}$,
$\text{CFL}=0.5$ for $r_{L_i}=10^{-2}$,
$\text{CFL}=1.2$ for $r_{L_i}=10^{-3}$.
When $r_{L_i}$ is set to be zero for ideal MHD solutions,
the time step will not be limited by the speed of light,
and is determined by
\begin{equation}\label{cfl2}
  \Delta t=\text{CFL}\frac{\Delta x}{U+\max(c_m,4c_s)},
\end{equation}
where $c_s$ is the sound speed and $c_m$ is the fast magneto-sound speed.

\section{Numerical experiments}
In this section, the UGKS is tested for the cases from collisionless regime to MHD regime.
In the collisionless Vlasov regime, we consider the classical test cases of Landau damping and two stream instability.
The relaxation parameter $\tau$ in Vlasov limit is set to be $\tau=10^3$
(in program $\exp(-\Delta t/\tau)$ is assigned to $1$ in order to avoid machine error).
For these test cases, we view ions as a fixed background and consider the motion of electrons.
Because no magnetic field is involved,
and the Maxwell equations degenerate to the Poisson equation,
so the FFT-based Poisson solver can be used to calculate the electric field.
In MHD regime, we first calculate the one dimensional Brio-Wu shock tube test case.
With the reduction of normalized Larmor radius, the solution goes from the Euler solution to Hall-MHD solution,
and finally converges to the ideal-MHD solution.
For two dimensional test cases, we first consider the Orszag-Tang MHD turbulence problem which tests the performance of UGKS
in capturing MHD solutions.
After testing UGKS in both limiting regimes, the scheme is used to study the GEM magnetic reconnection problem, which happens on the  Debye length scale.

\subsection{Linear Landau damping}

\begin{figure}
  \centering
  \includegraphics[width=0.3\textwidth]{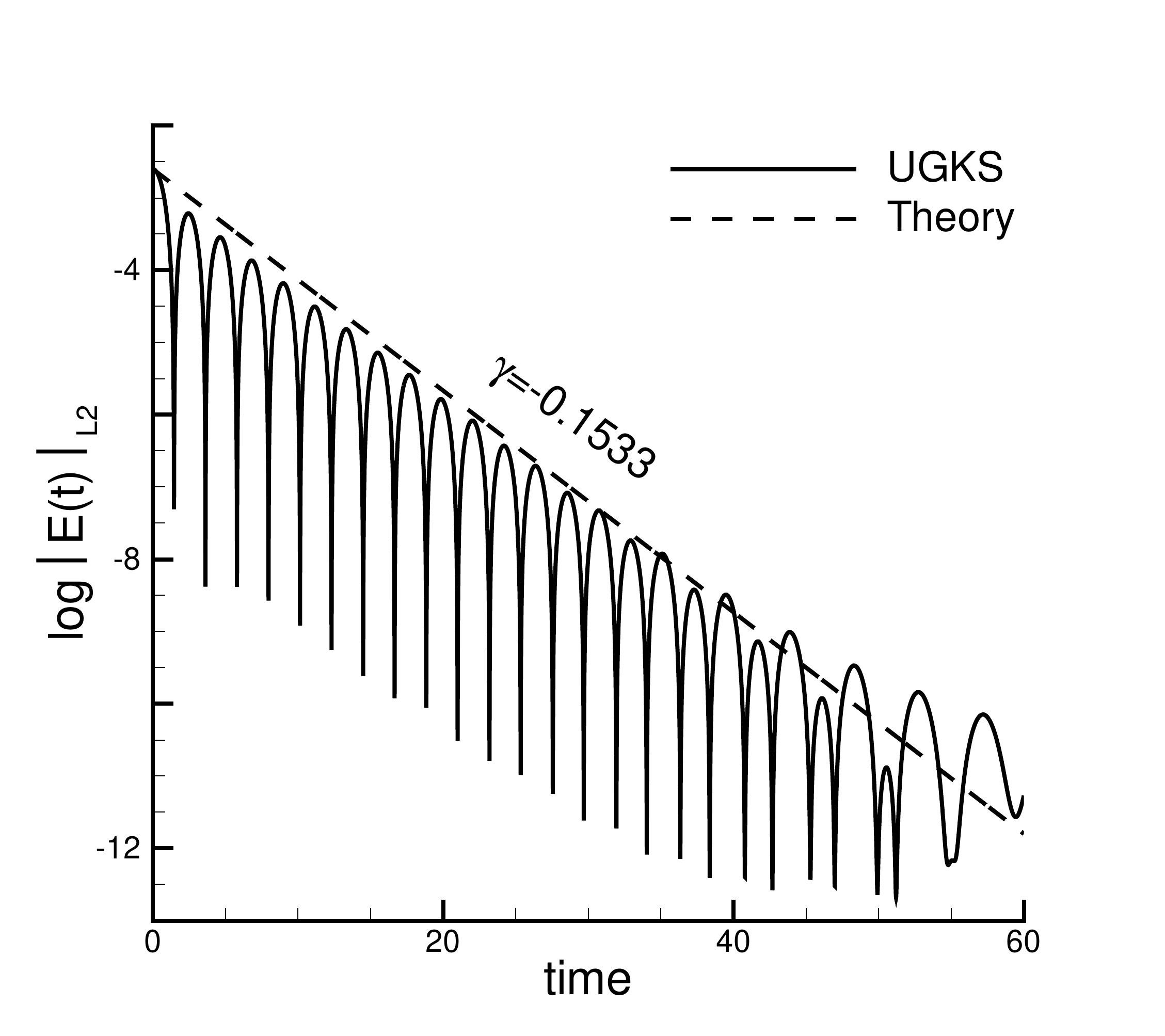}\\
  \includegraphics[width=0.3\textwidth]{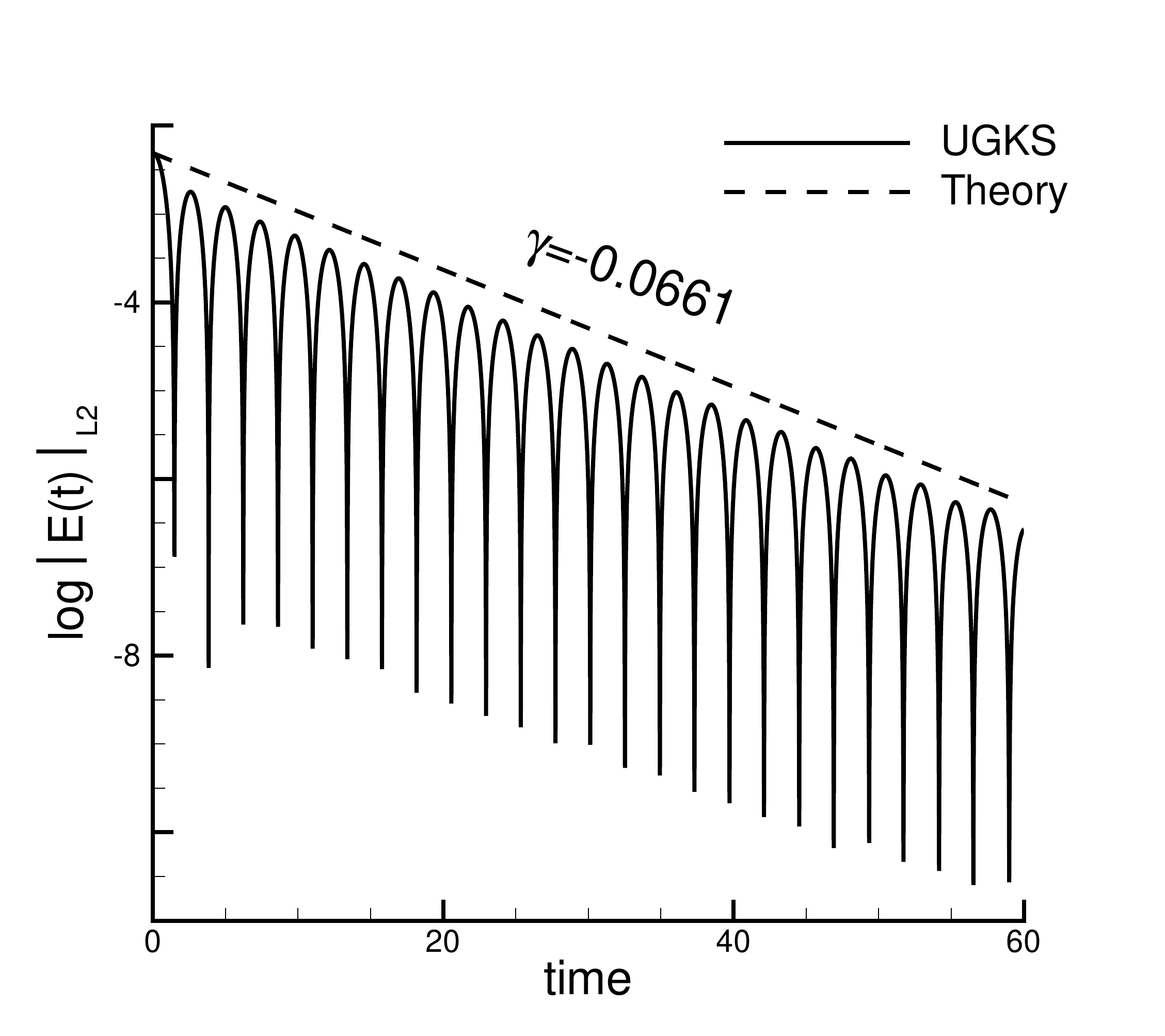}
  \includegraphics[width=0.3\textwidth]{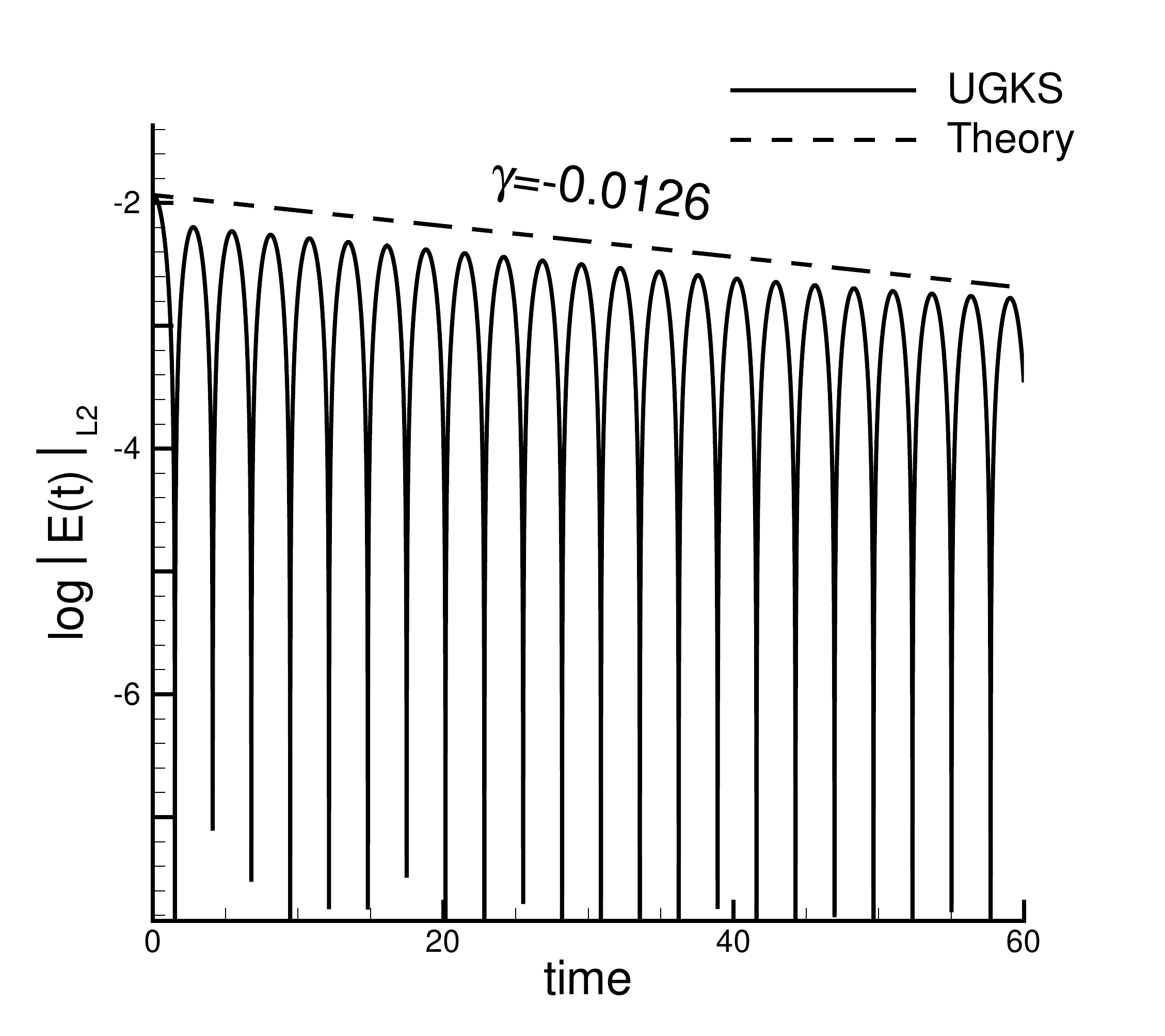}
  \caption{Weak Landau damping. Time evolution of electric field in $L^2$ norm. $k=0.5$ (upper), $k=0.4$ (lower left) and $k=0.3$ (lower right).}
  \label{landau-linear-1}
\end{figure}

\begin{figure}
  \centering
  \includegraphics[width=0.3\textwidth]{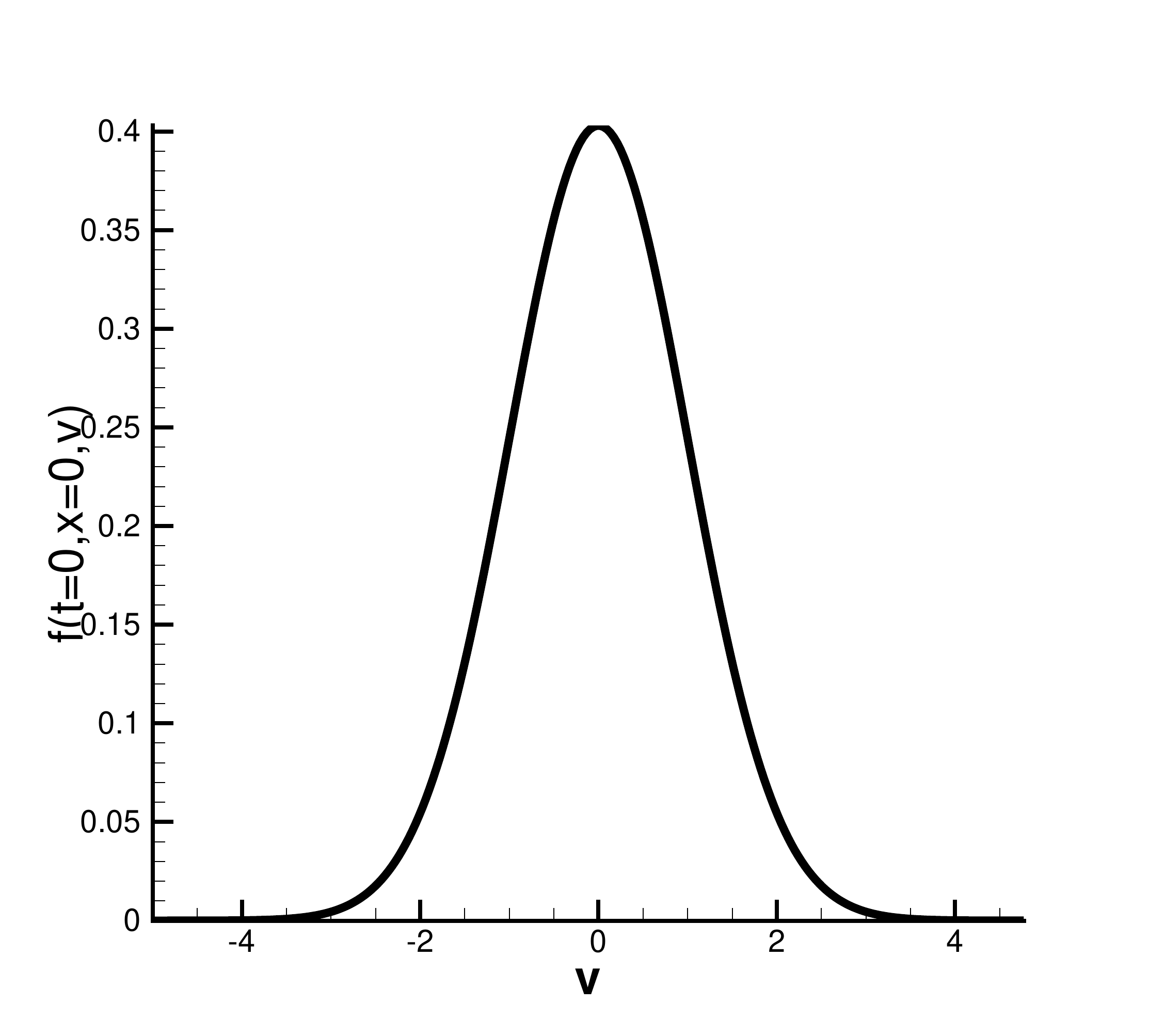}
  \includegraphics[width=0.3\textwidth]{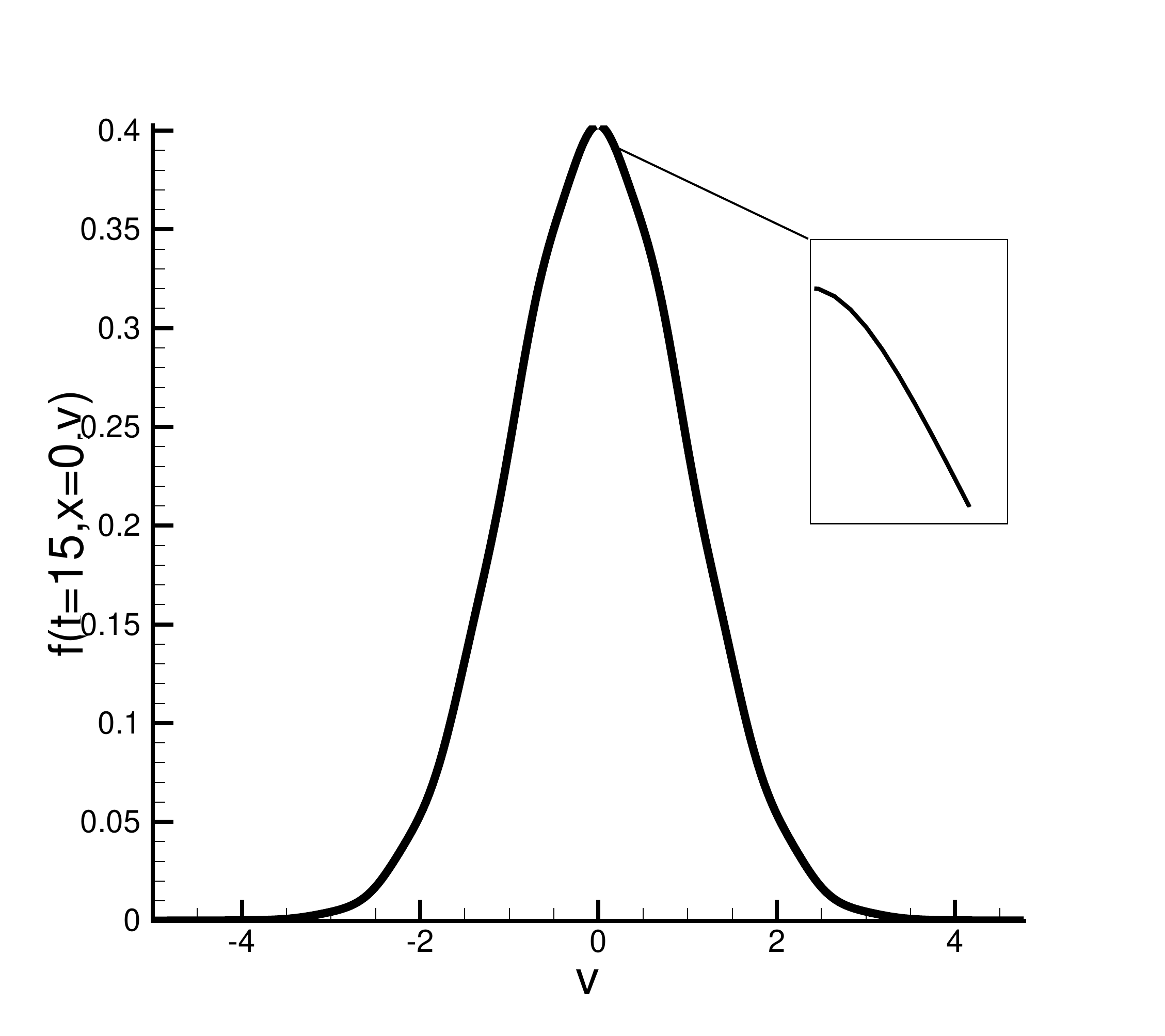}\\
  \includegraphics[width=0.3\textwidth]{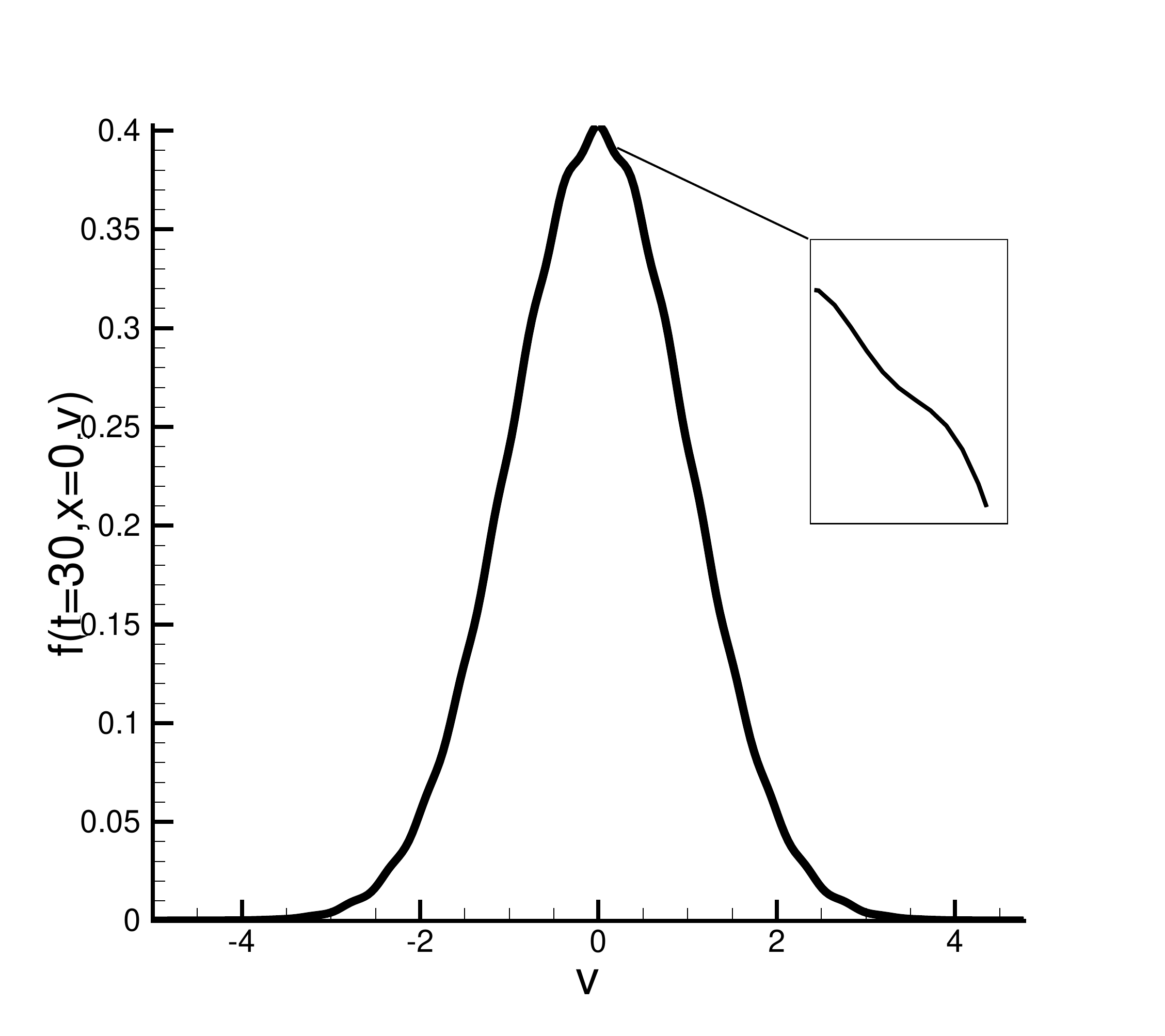}
  \includegraphics[width=0.3\textwidth]{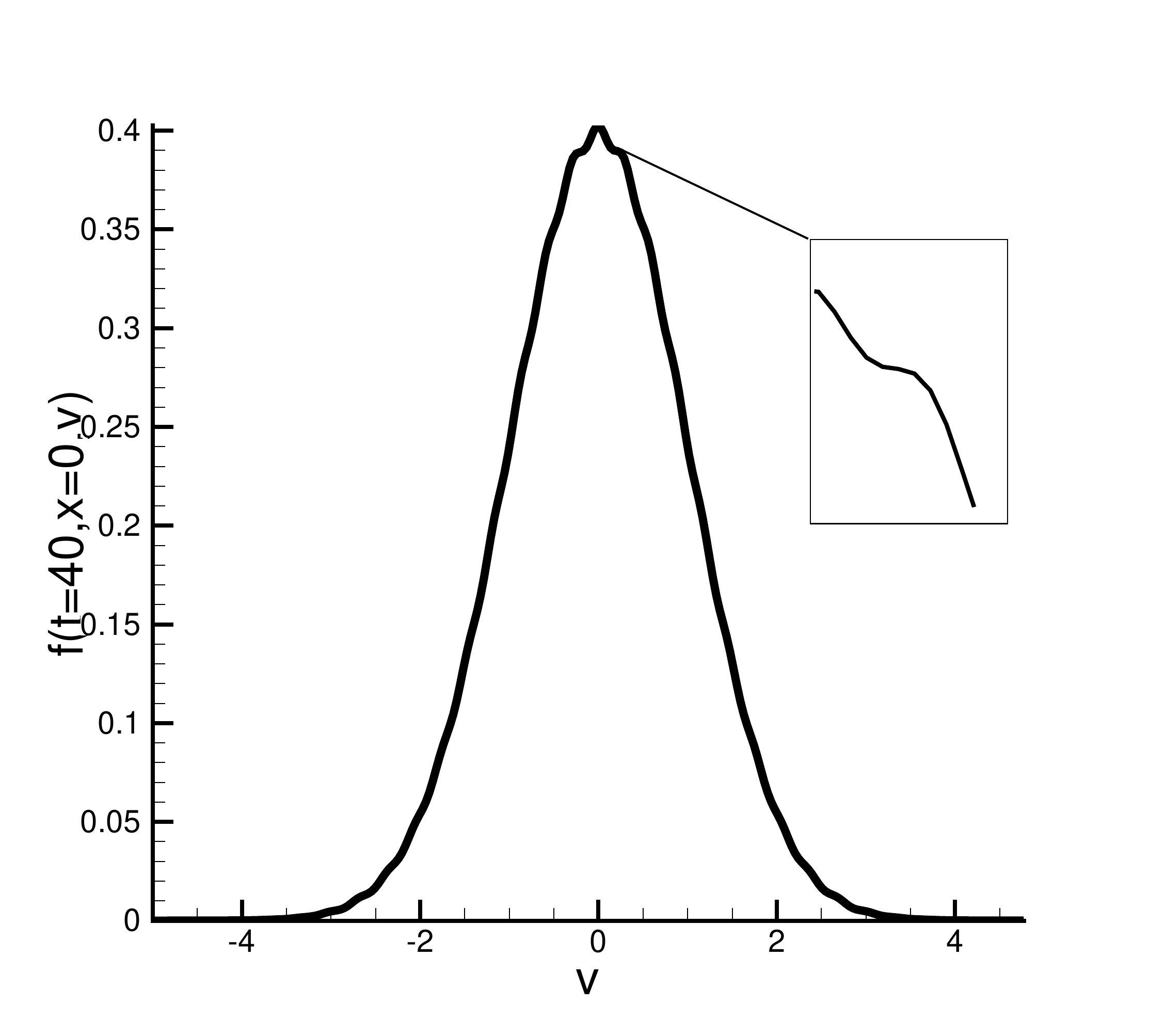}
  \caption{Time development of the distribution function at $x=0$.}
  \label{landau-linear-3}
\end{figure}

A Vlasov-Poisson (VP) system is perturbed by a weak signal.
The linear theory of Landau damping can be applied to predict the linearly decay of electric energy with time \cite{chen1984plasma}.
The initial condition of linear Landau damping for the Vlasov Poisson system is
\begin{equation}\label{landaudamping-initial}
  f_0(x,u)=\frac{1}{\sqrt{2\pi}}\left(1+\alpha\cos(kx)\right)\text{e}^{-\frac{u^2}{2}},
\end{equation}
with $\alpha=0.01$.
The length of the domain in the x direction is $L=2\pi/k$.
The background ion distribution function is fixed,
uniformly chosen so that the total net charge density for the system is zero.
For perturbation parameter $\alpha=0.01$ is small enough,
the Vlasov-Poisson system can be approximated by linearization around the Maxwellian equilibrium.
The analytical damping rate of electric field can be derived accordingly.
We test our scheme with different wave numbers and compare the numerical damping rates with theoretical values.
The phase space is discretized with $N_x\times N_u=128 \times 128$ cells with $u_{max}=5$.
We plot the evolution of electric field in $L^2$ norm in Fig. \ref{landau-linear-1} for $k=0.5$, $k=0.4$, and $k=0.3$.
The correct decay rates of the electric field are observed and are matched with theoretical values.
In addition, the numerical frequencies of oscillation consist with the corresponding theoretical values $\omega=1.41$, $\omega=1.29$, $\omega=1.16$.
The profile of velocity distribution $f(x=0, u, t)$ is plotted in Fig. \ref{landau-linear-3}, which shows
that the  particles with low  velocity absorb energy from the electric wave.

\subsection{Nonlinear Landau damping}

\begin{figure}
\centering
\includegraphics[width=0.5\textwidth]{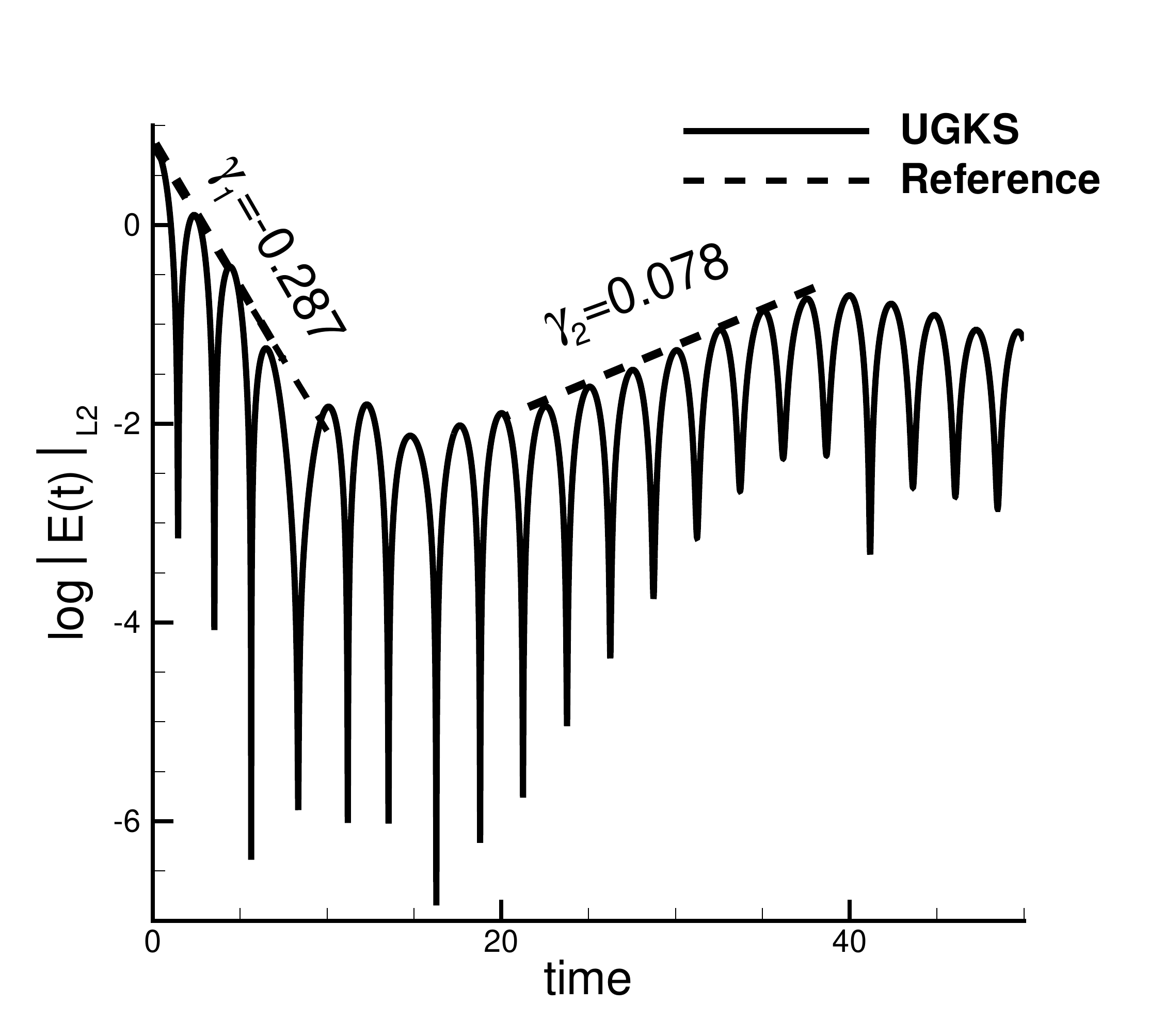}
\caption{Strong Landau damping. Time evolution of electric field in $L^2$ norm.}
  \label{landau-nonlinear-1}
\end{figure}

\begin{figure}
  \centering
  \includegraphics[width=0.3\textwidth]{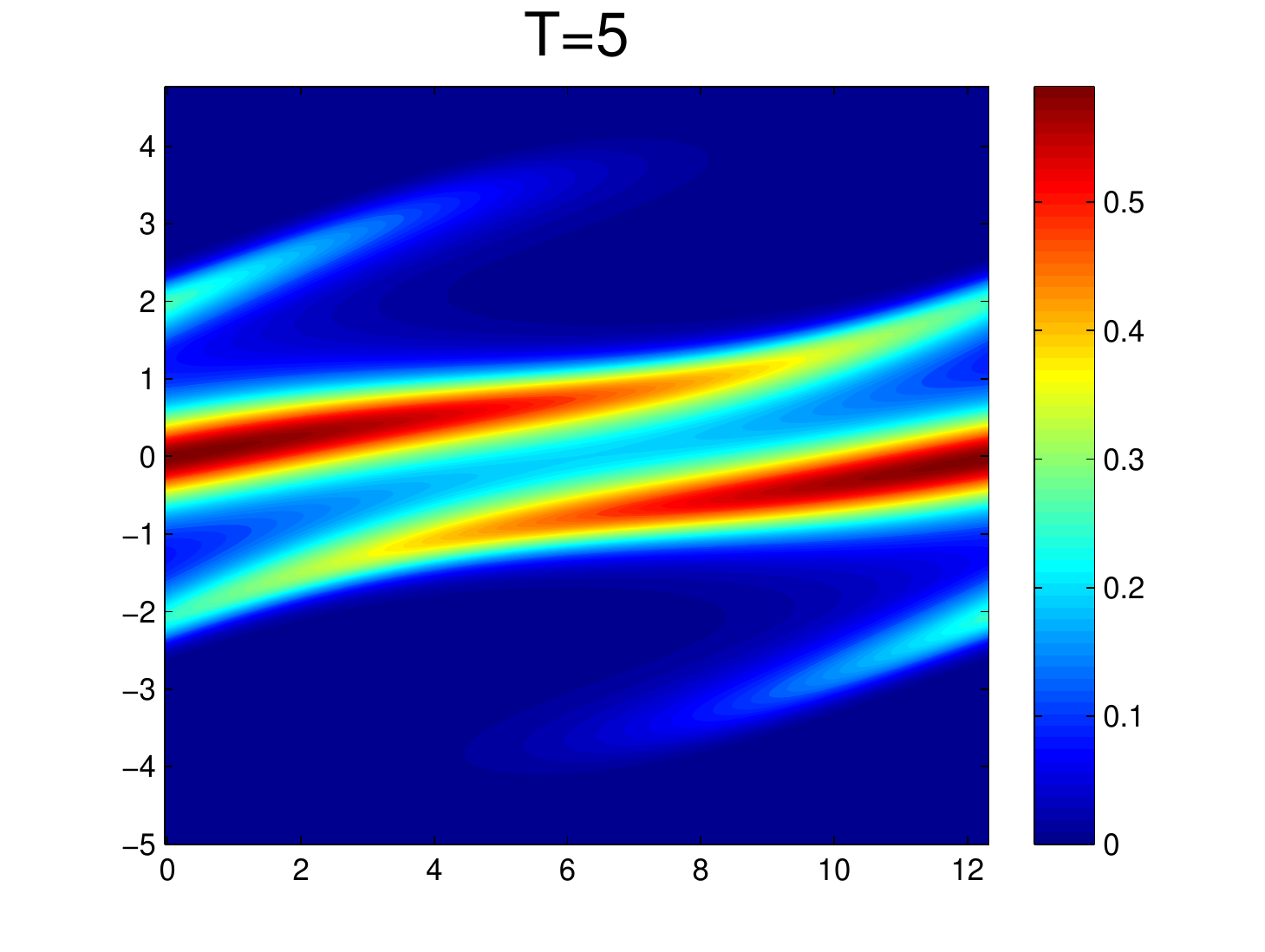}
  \includegraphics[width=0.3\textwidth]{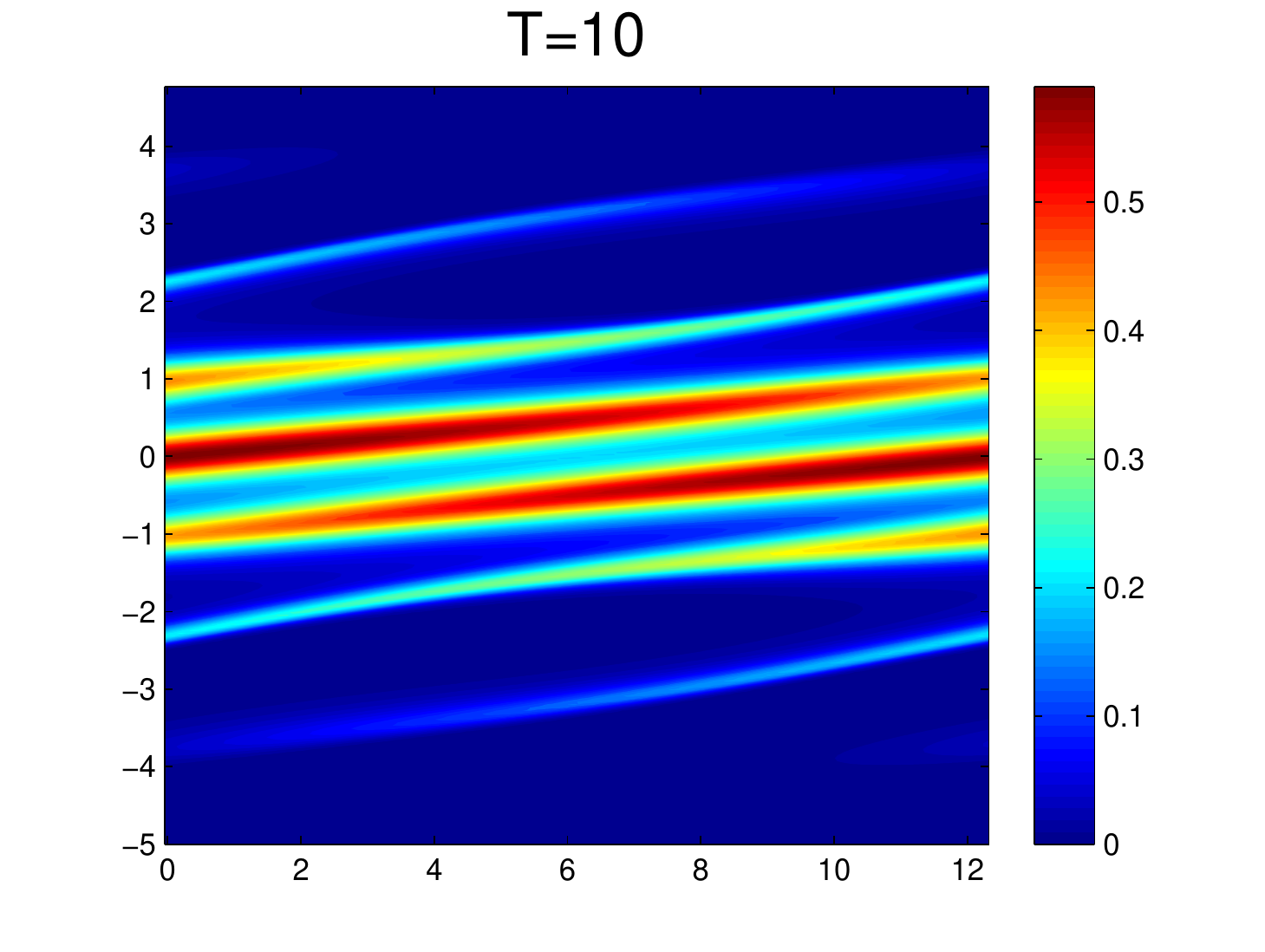}
  \includegraphics[width=0.3\textwidth]{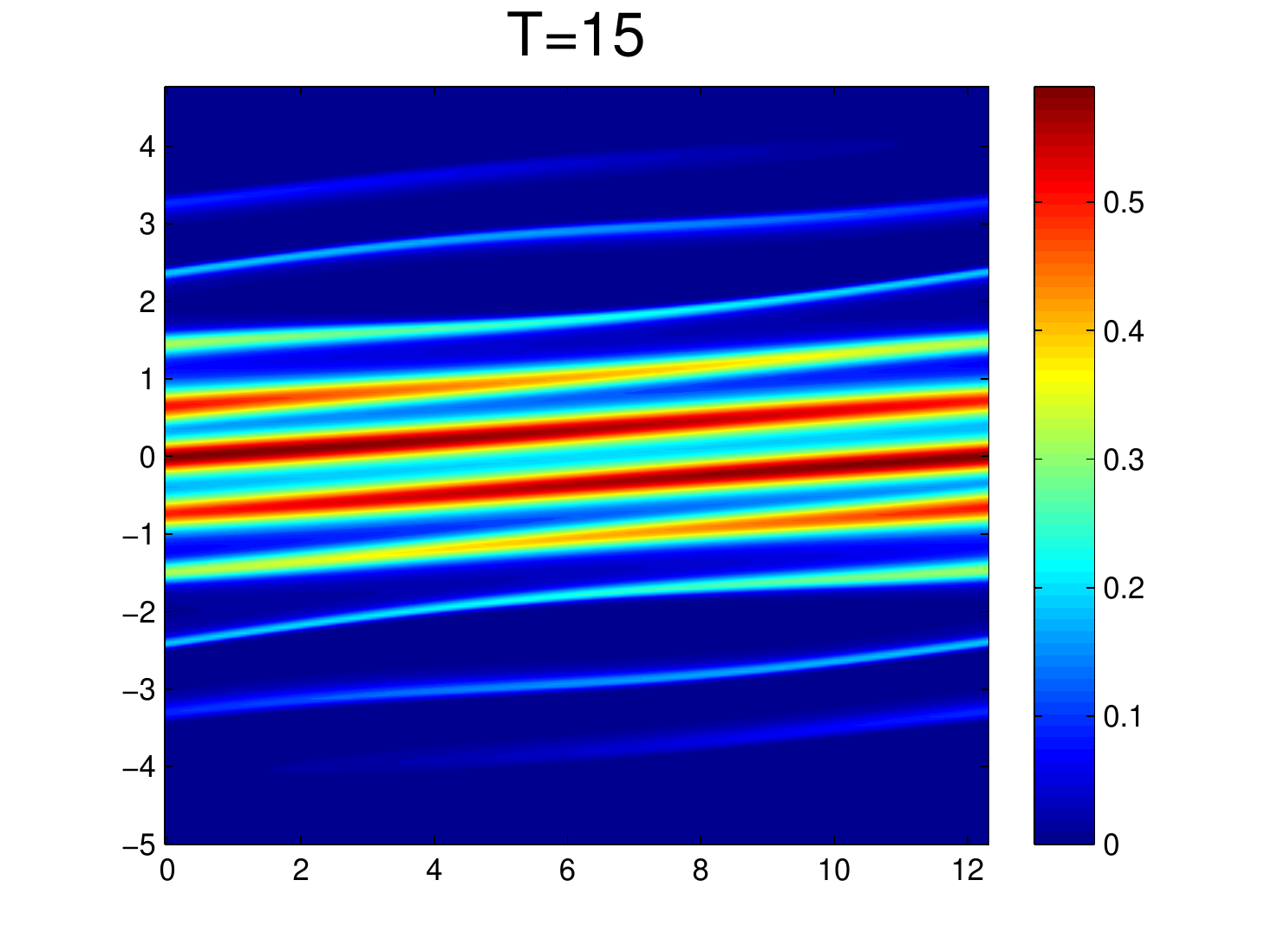}\\
  \includegraphics[width=0.3\textwidth]{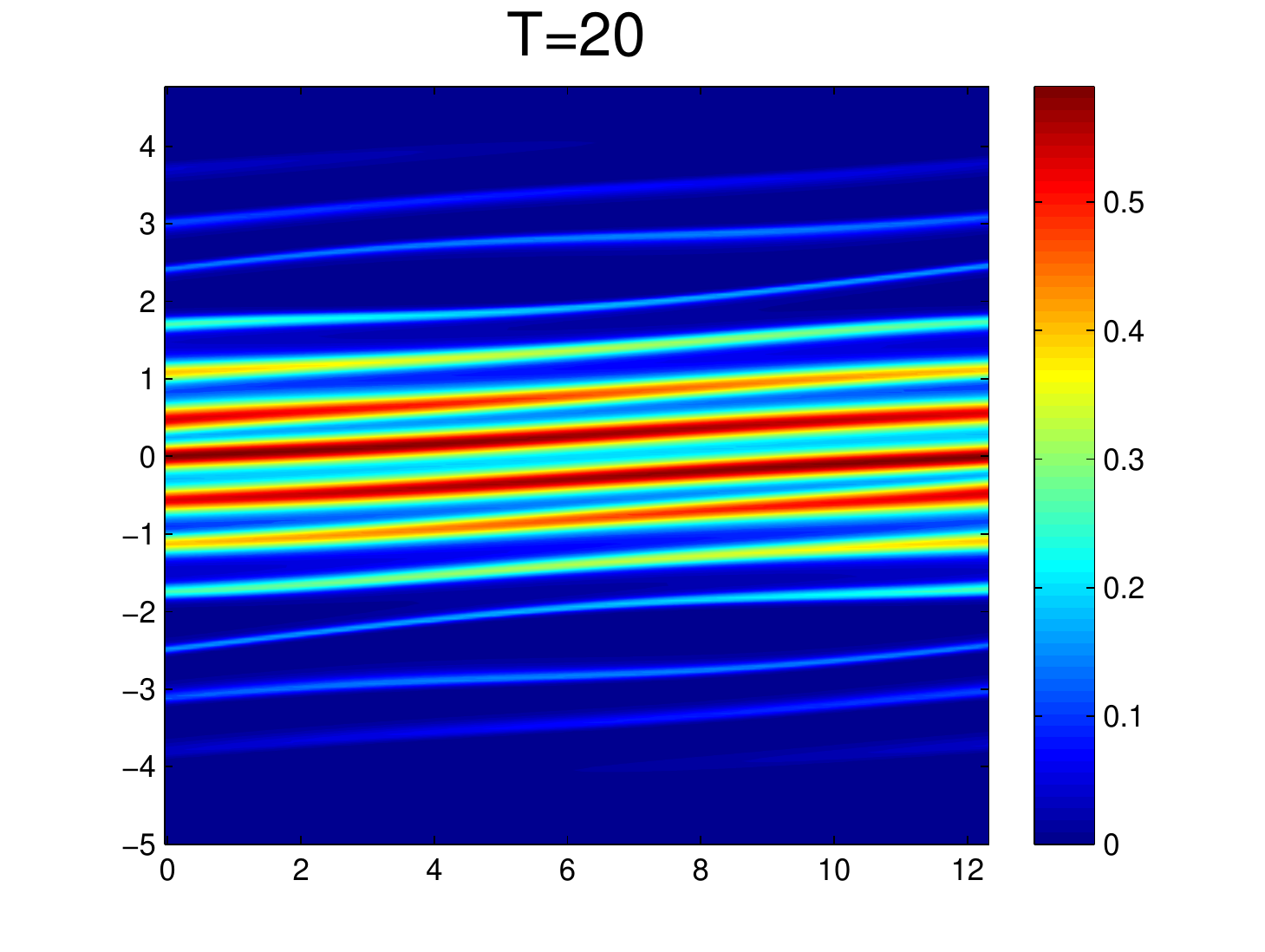}
  \includegraphics[width=0.3\textwidth]{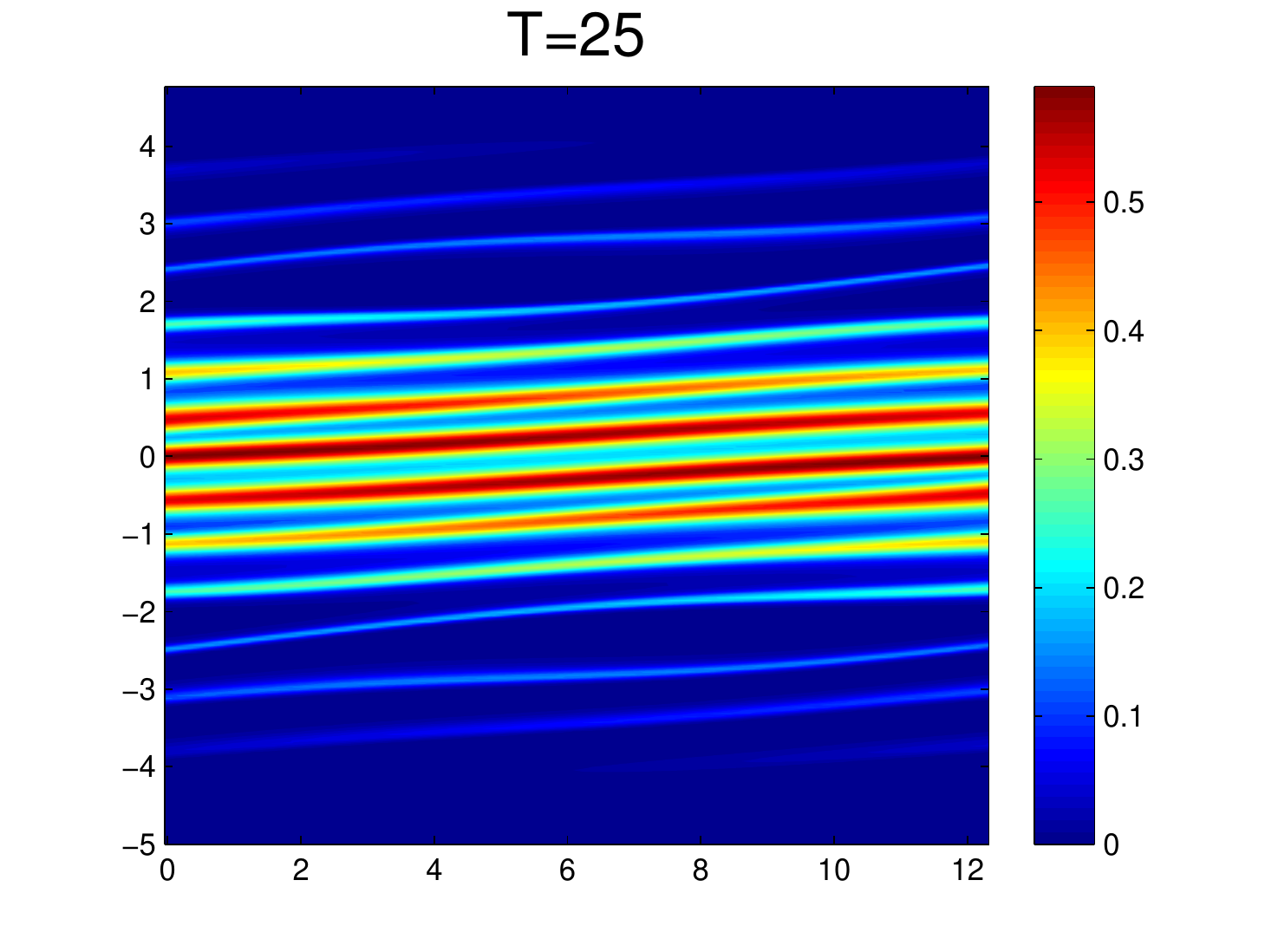}
  \includegraphics[width=0.3\textwidth]{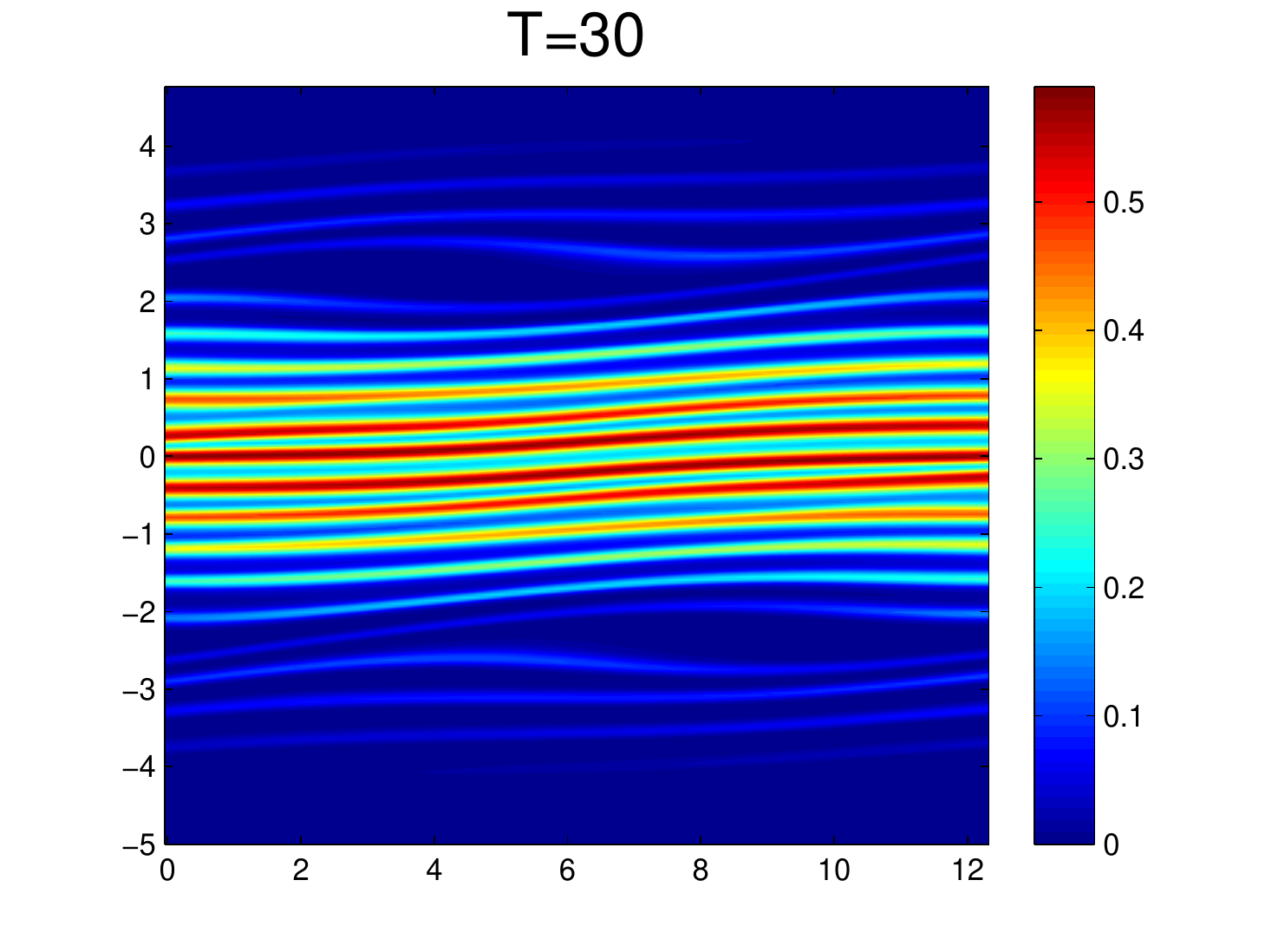}\\
  \includegraphics[width=0.3\textwidth]{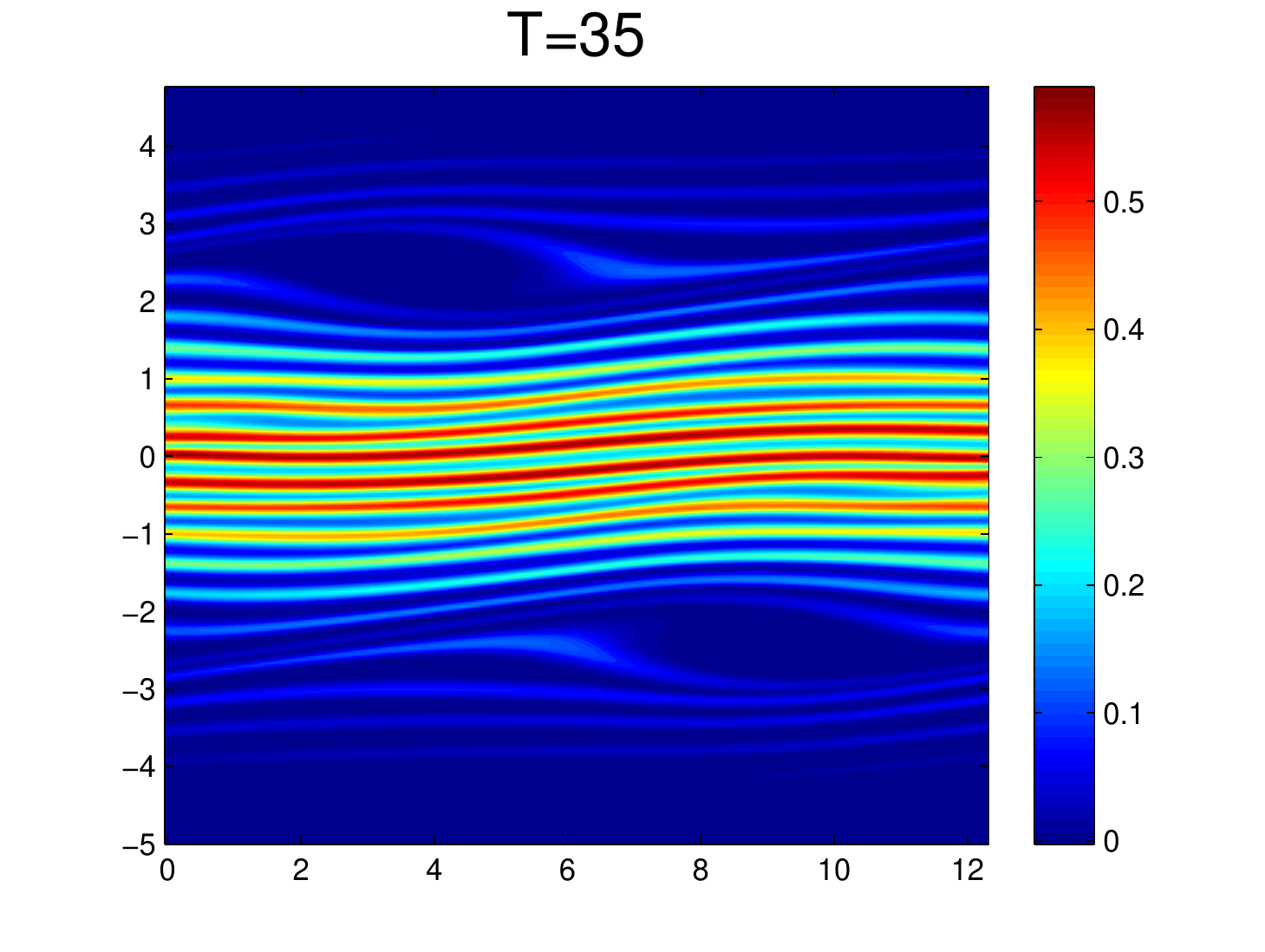}
  \includegraphics[width=0.3\textwidth]{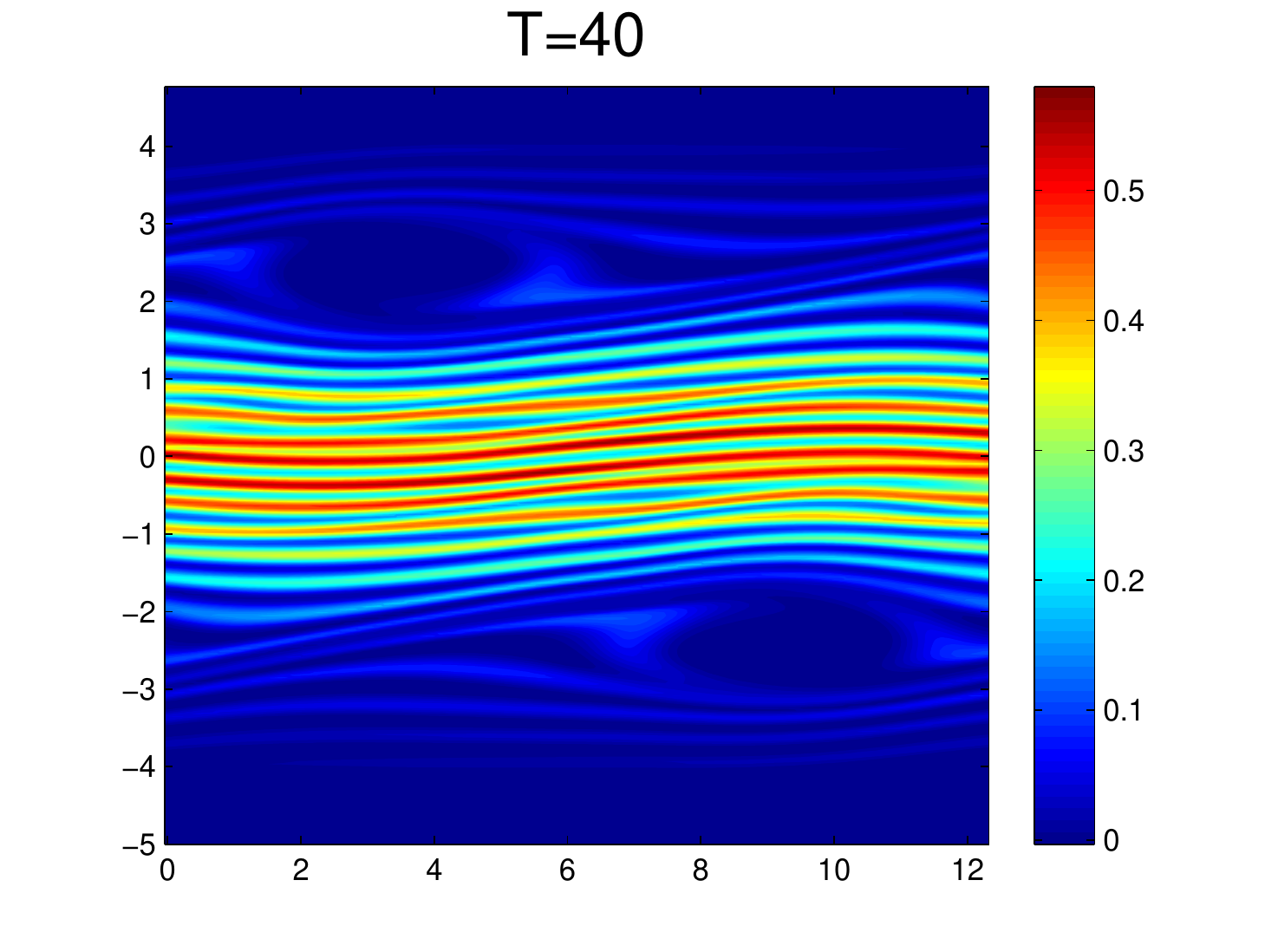}
  \includegraphics[width=0.3\textwidth]{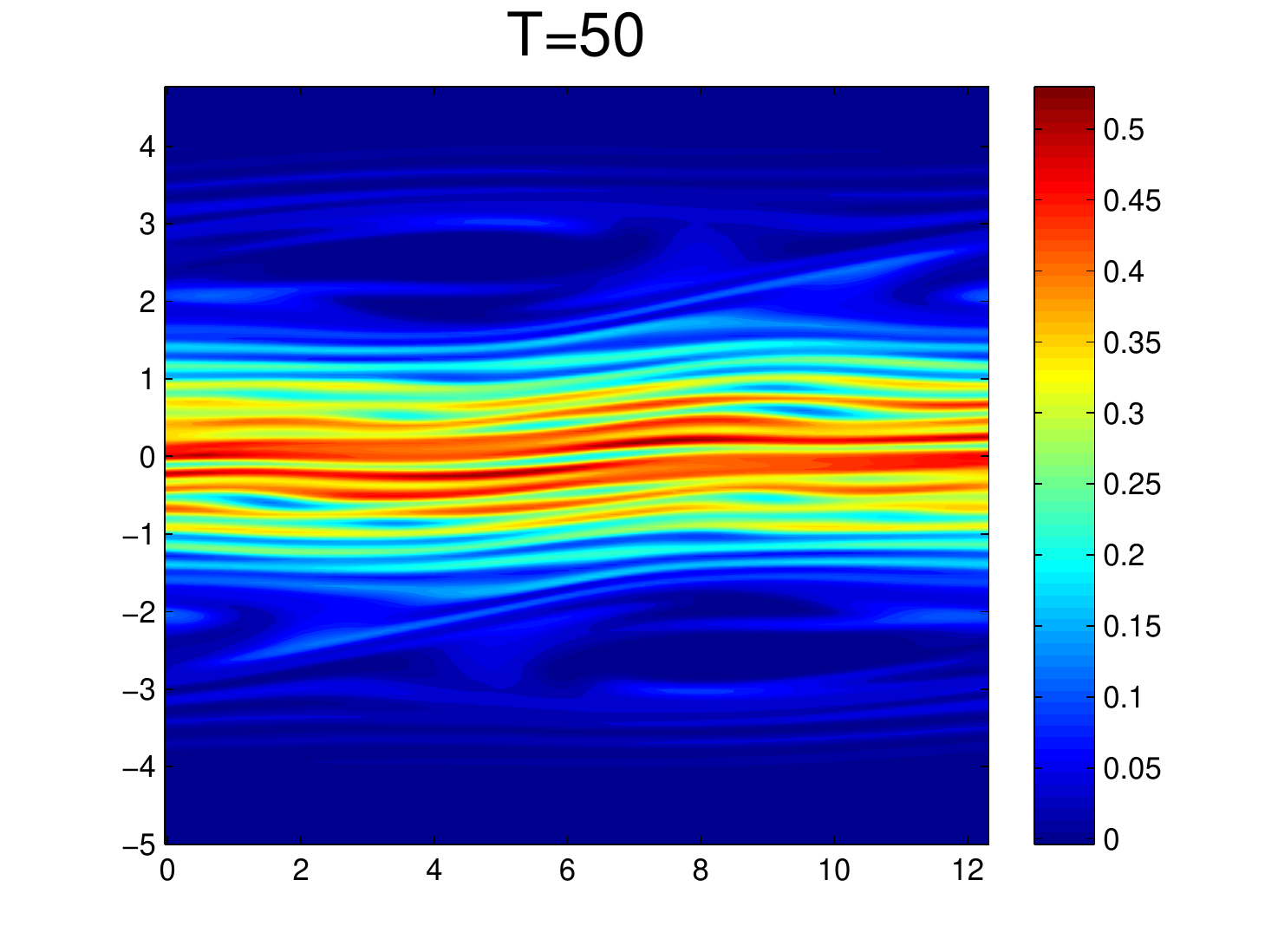}
  \caption{Strong Landau damping. Velocity distribution contours. $N_x\times N_u=256\times 256$}
  \label{landau-nonlinear-2}
\end{figure}

\begin{figure}
  \centering
  \includegraphics[width=0.3\textwidth]{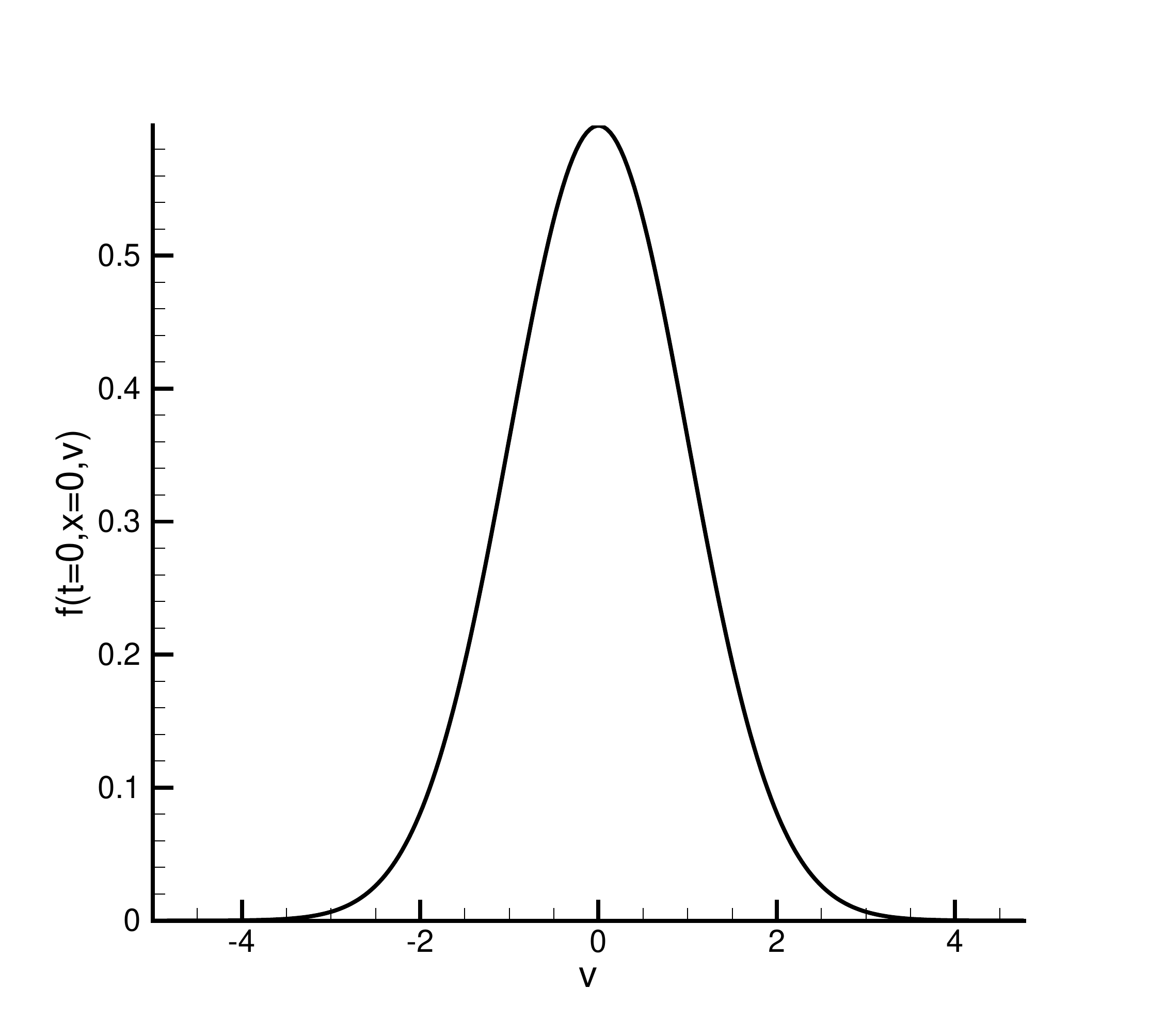}
  \includegraphics[width=0.3\textwidth]{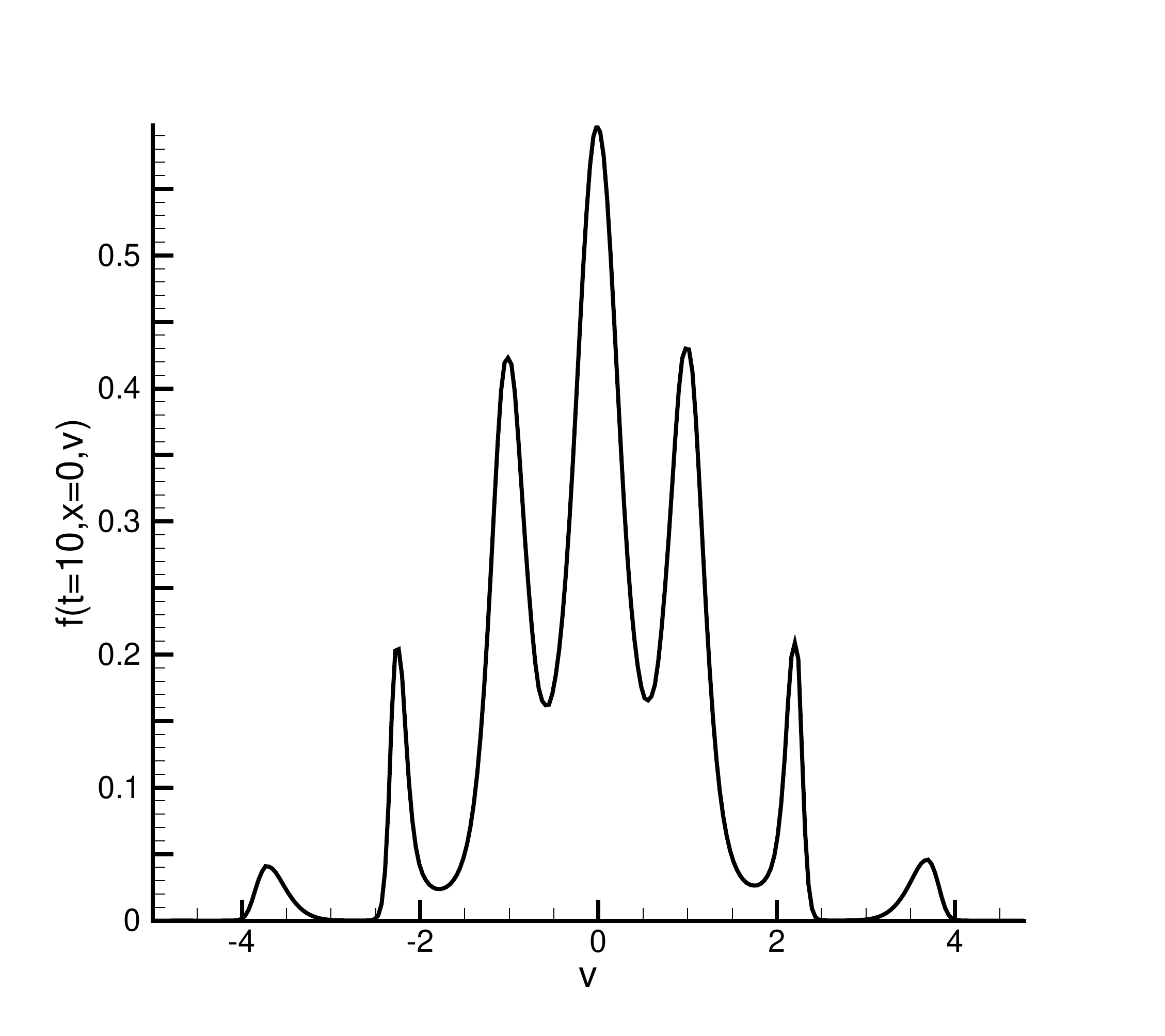}\\
  \includegraphics[width=0.3\textwidth]{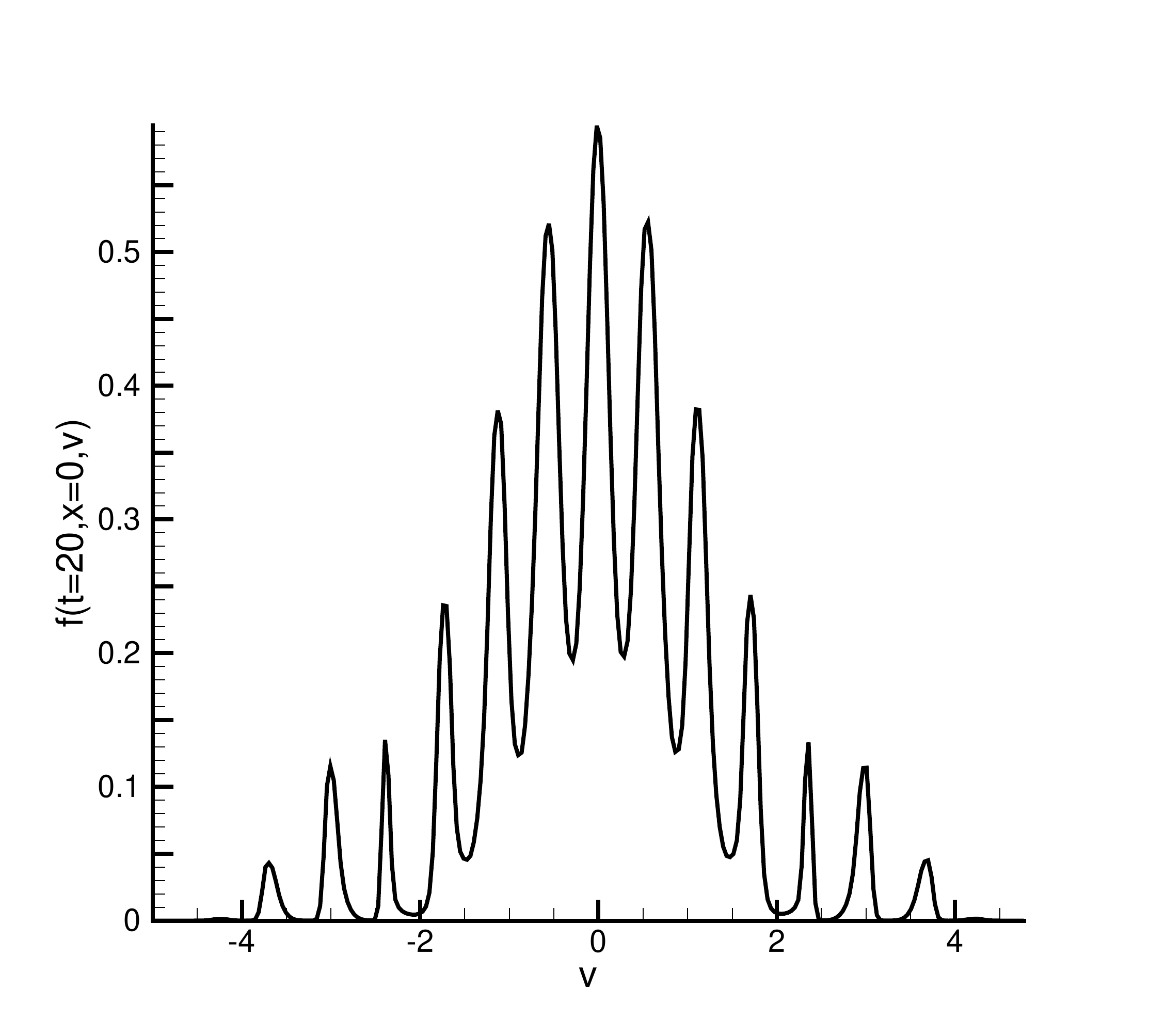}
  \includegraphics[width=0.3\textwidth]{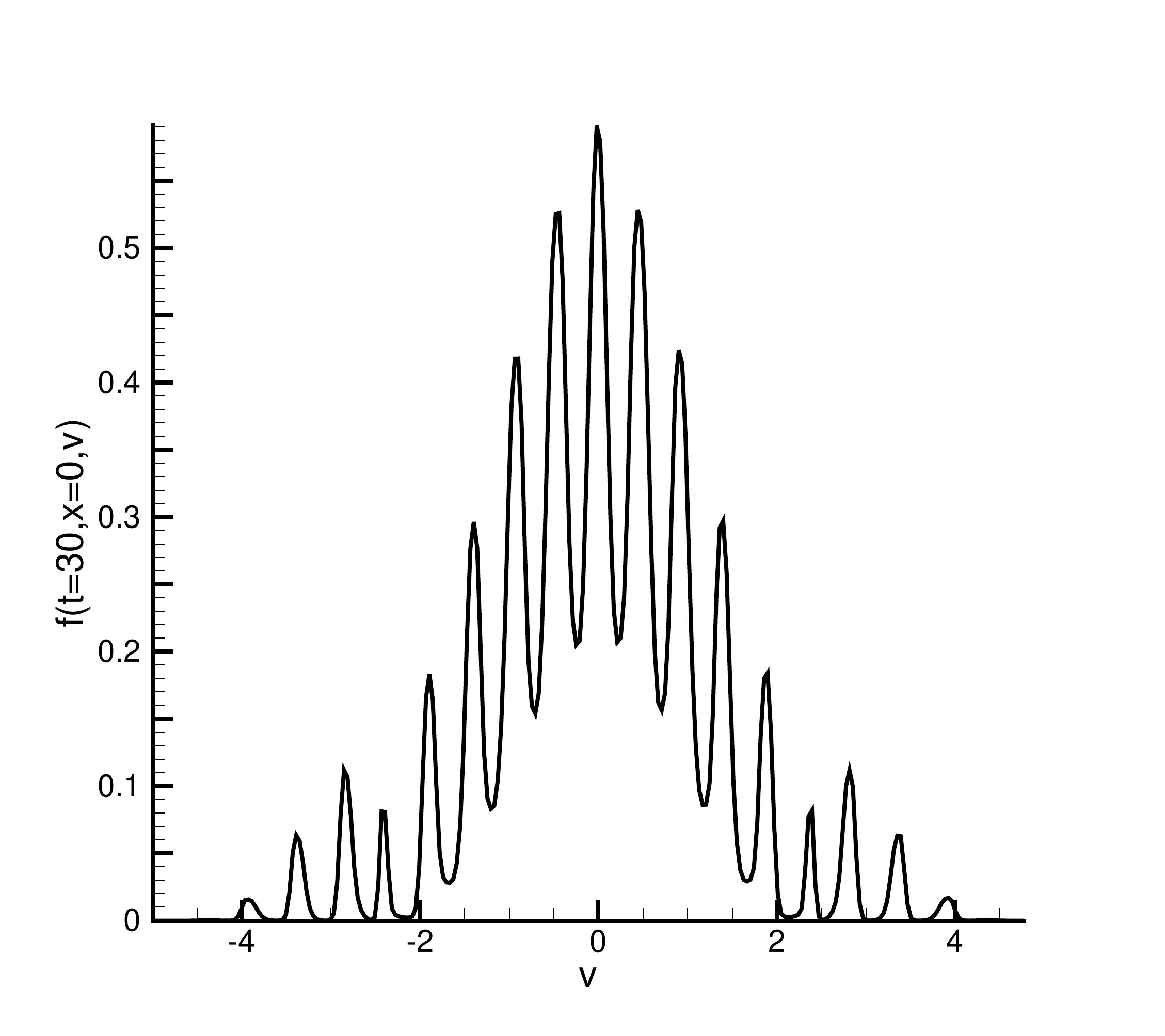}\\
  \caption{Strong Landau damping. Time development of the distribution function at $x=0$.}
  \label{landau-nonlinear-3}
\end{figure}

When the VP system is perturbed by a large amplitude, the nonlinear effects will appear \cite{chen1984plasma}.
For nonlinear landau damping,
the initial condition is similar with the linear case
while the parameters are set as $\alpha=0.5$ and $k=0.5$.
Fig. \ref{landau-nonlinear-1} shows the $L^2$ norms of electric field computed by UGKS.
The linear decay rate of electric energy is approximately $\gamma_1=-0.287$,
which is identical to the values obtained  by Heath \emph{et al.}.
The growth rate provided by UGKS is approximately $\gamma_2=0.078$, which is between
the value of $0.0815$ computed by Rossmanith and Seal and $0.0746$ by Heath \emph{et al.}.
The contours of velocity distribution at different times are shown in Fig. \ref{landau-nonlinear-2}.
The profile of velocity distribution at $x=0$ is plotted in Fig. \ref{landau-nonlinear-3}, from which the nonlinear effect is clearly observed.

\subsection{Linear two stream instability}

\begin{figure}
\centering
\includegraphics[width=0.5\textwidth]{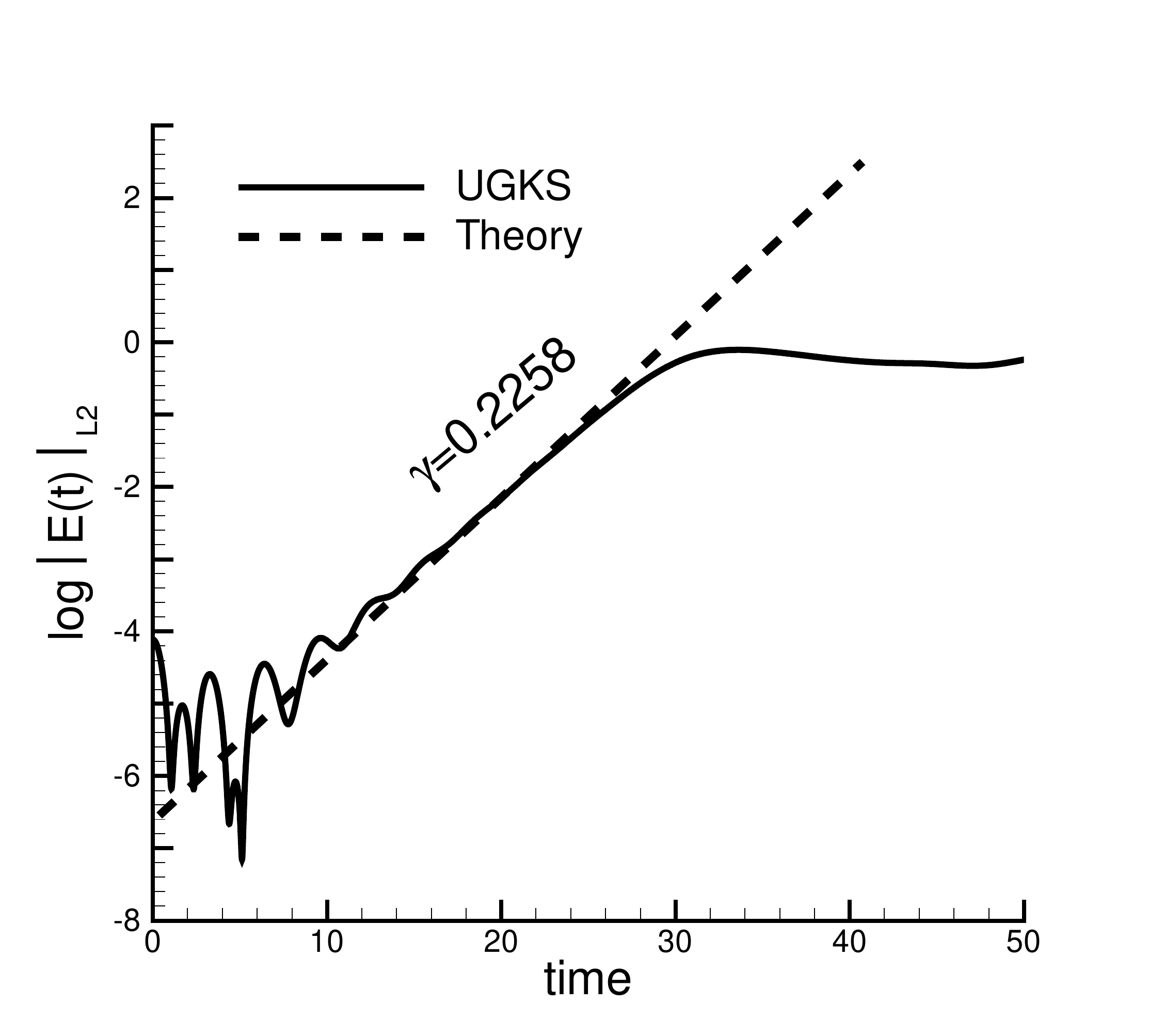}
\caption{Two stream instability. Time evolution of electric field in $L^2$ norm. $\alpha=0.001$, $u_{th}=1$ and $k=0.2$.}
\label{instability-linear-1}
\end{figure}

\begin{figure}
\centering
\includegraphics[width=0.45\textwidth]{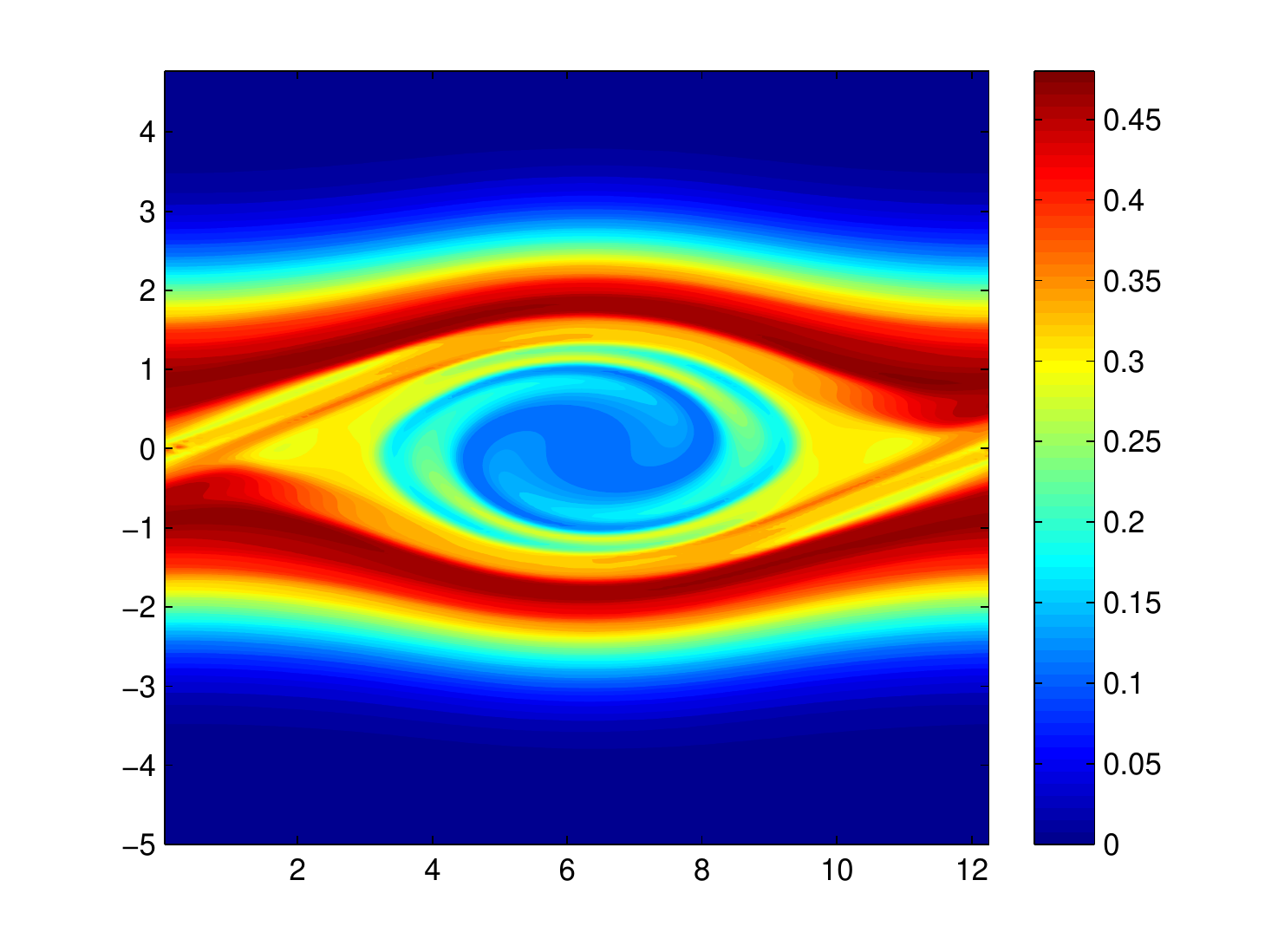}
\includegraphics[width=0.45\textwidth]{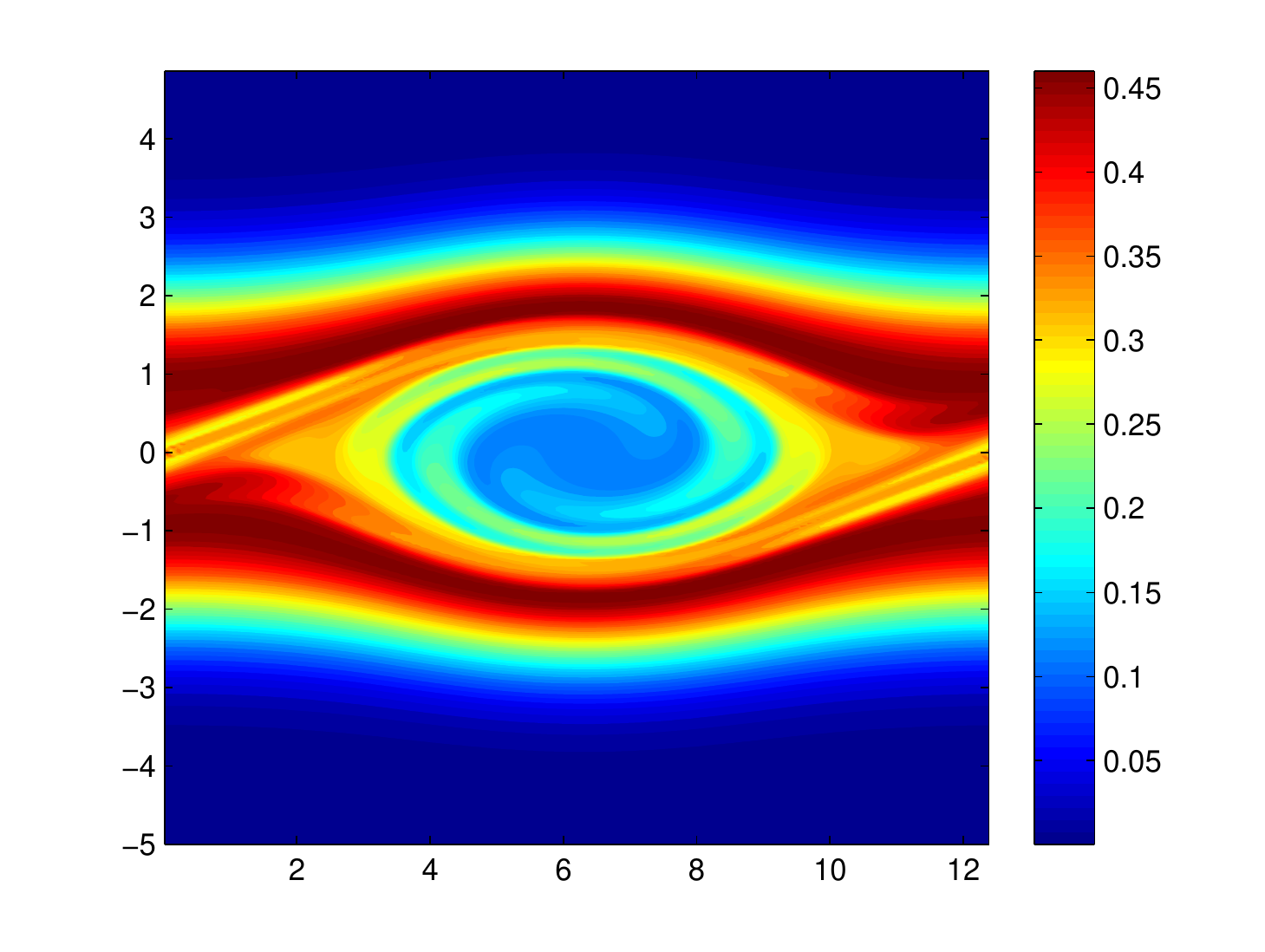}
\caption{Two stream instability (weak). Velocity distribution contours at $t=70$. $N_x\times N_u=256\times 256$ (left); $N_x\times N_u=512\times 512$ (right)}
\label{instability-linear-2}
\end{figure}

Consider linear two stream instability problem with initial distribution function:
\begin{equation}\label{instability-initial1}
  f(x,u,t=0)=\frac{2}{7\sqrt{2\pi}}(1+5v^2)(1+\alpha((\cos(2kx)+\cos(3kx))/1.2+\cos(kx)))\text{e}^{-\frac{u^2}{2}},
\end{equation}
where $\alpha=0.001$ and $k=0.2$.
The length of the domain in the x direction is $L=\frac{2\pi}{k}$.
The background ion distribution function is
fixed, uniformly to balance the charge density of electron.
After a certain amount of time, a linear growth rate of electric field can be found, and the value can be theoretically calculated \cite{chen1984plasma}. In Fig. \ref{instability-linear-1}, we plot the evolution of electric field in $L^2$ norm, and the growth
rate predicted by UGKS is the same as the theoretical one.
The velocity distribution contours at time $t=70$ with different mesh sizes are presented in Fig. \ref{instability-linear-2}.

\subsection{Nonlinear two stream instability}

\begin{figure}
\centering
\includegraphics[width=0.45\textwidth]{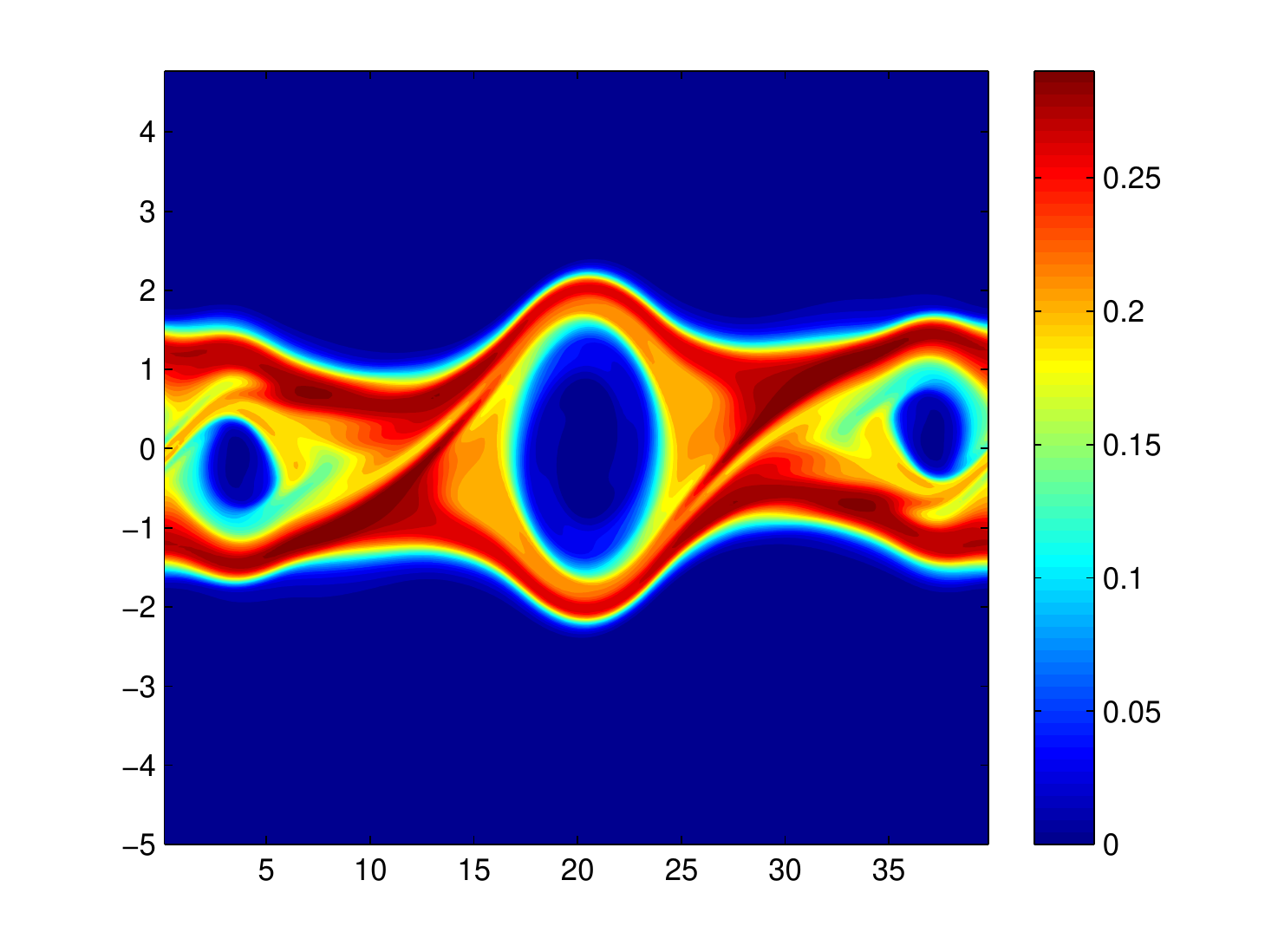}
\includegraphics[width=0.45\textwidth]{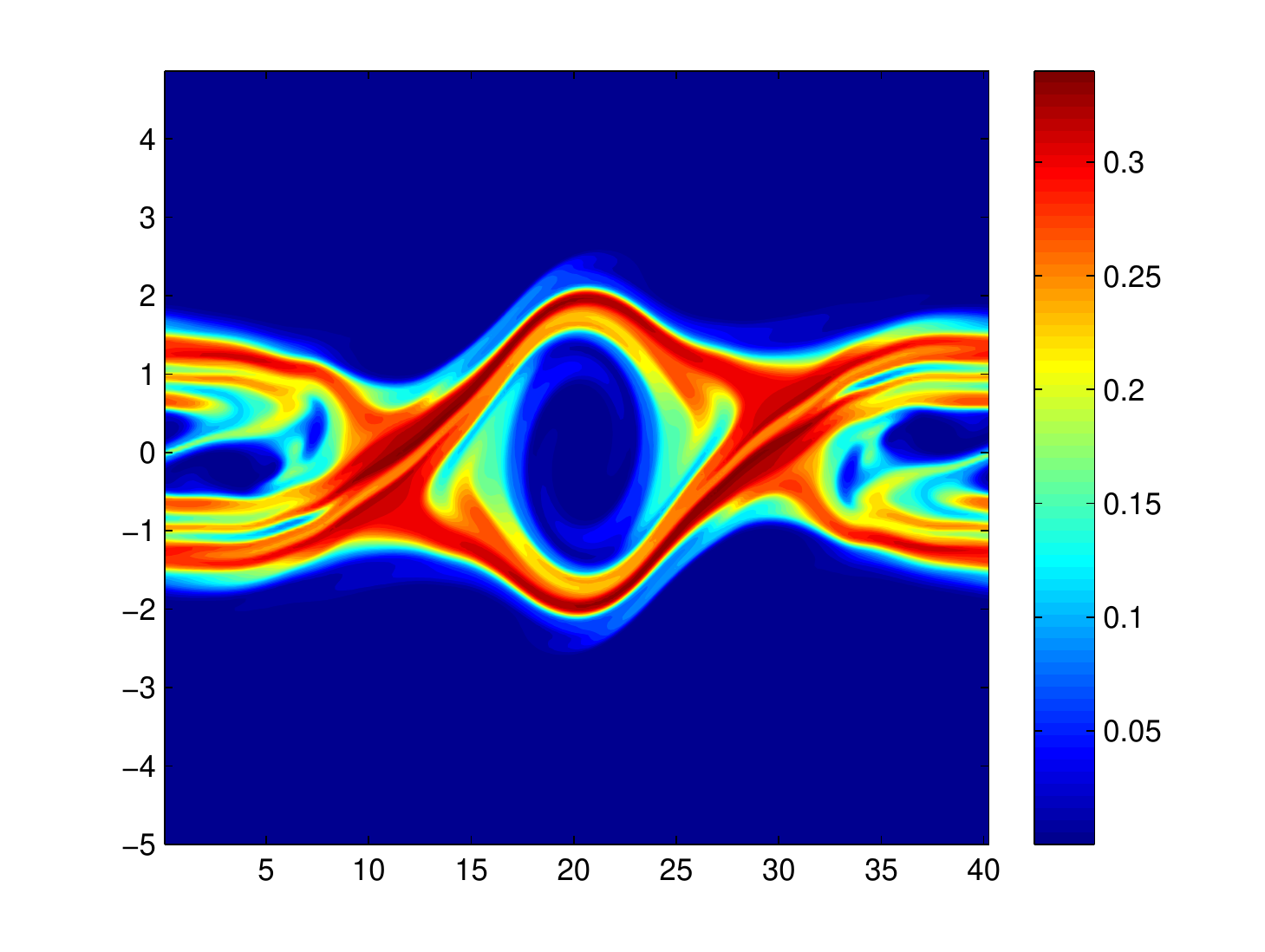}
\caption{Two stream instability (nonlinear). Velocity distribution contours at $t=70$. $N_x\times N_u=256\times 256$ (left); $N_x\times N_u=512\times 512$ (right)}
\label{instability-nonlinear-1}
\end{figure}

For the nonlinear case of two stream instability, we use a symmetric initial condition
\begin{equation}\label{instability-initial2}
  f(x,u,t=0)=\frac{1}{2v_{th}\sqrt{2\pi}}\left[\exp\left(-\frac{(u-U)^2}{2u_{t}^2}\right)+
  \exp\left(-\frac{(u+U)^2}{2u_{t}^2}\right)\right](1+\alpha \cos(kx)),
\end{equation}
with $\alpha=0.05$, $U=0.99$, $u_{t}=0.3$, and $k=\frac{2}{13}$.
In Fig. \ref{instability-nonlinear-1}, we show
the numerical results of the contours of distribution function at $t=70$.
The computations show that the detailed structures of $f$ can only be captured with very fine mesh for a second order scheme.

\begin{figure}
  \centering
  \includegraphics[width=0.8\textwidth]{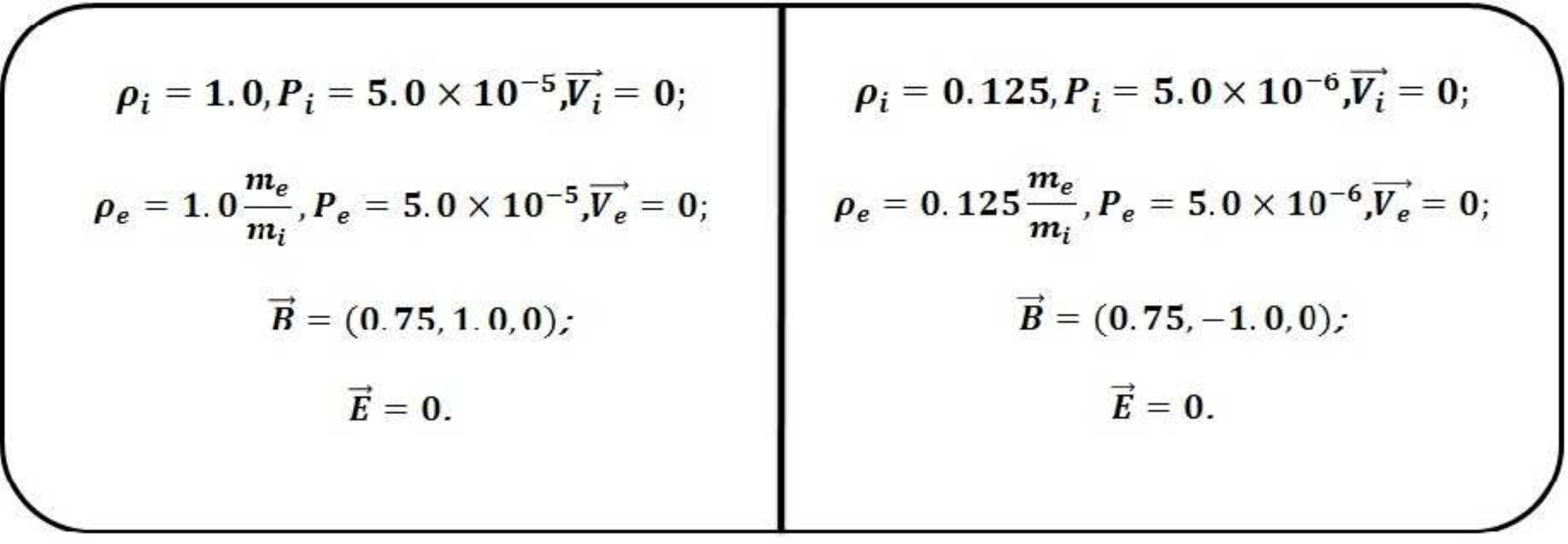}
\caption{Initial condition for Brio-Wu shock tube problems.}
\label{Brio-initial}
\end{figure}

\subsection{Brio-Wu shock tube}

The Brio-Wu shock tube is a standard test case for ideal MHD solvers in continuum regime \cite{brio1988upwind,hakim2006high}.
The same initial condition as the Brio-Wu one is shown in Fig. \ref{Brio-initial}.
The ion to electron mass ratio is set to be 1836, and the ionic charge state is set to be unity. $\nu_{ie}$ is defined the same as the one
 in the original AAP model \cite{AAP}. The normalized Debye length is $0.01$, and the normalized speed of light is $100$.
The ion Larmor radius takes different values normalized by the length of the domain,
i.e. $r_{L_i}=10, 1, 0.01, 0.003$. The relaxation parameter $\tau$ of the following test case is set to be $10^{-5}$ and the grid points in physical space are 1000. The velocity space is $[-5,5]$ for ion and $[-5\sqrt{1836},5\sqrt{1836}]$ for electron with $32$ grid points.

In this test case, different time steps for electron and ion are used to reduce the computational cost.
Specifically, based on the CFL number, the time step $\Delta t_i$
for ions is chosen to be five times of the time step $\Delta t_e$ for electrons. During one $\Delta t_e$ the ions are supposed to be fixed, after evolving the electrons five times with $\Delta t_e$ each, we update ions for one $\Delta t_i$ and couple them with electrons through electromagnetic field.

\begin{figure}
\centering
\includegraphics[width=0.23\textwidth]{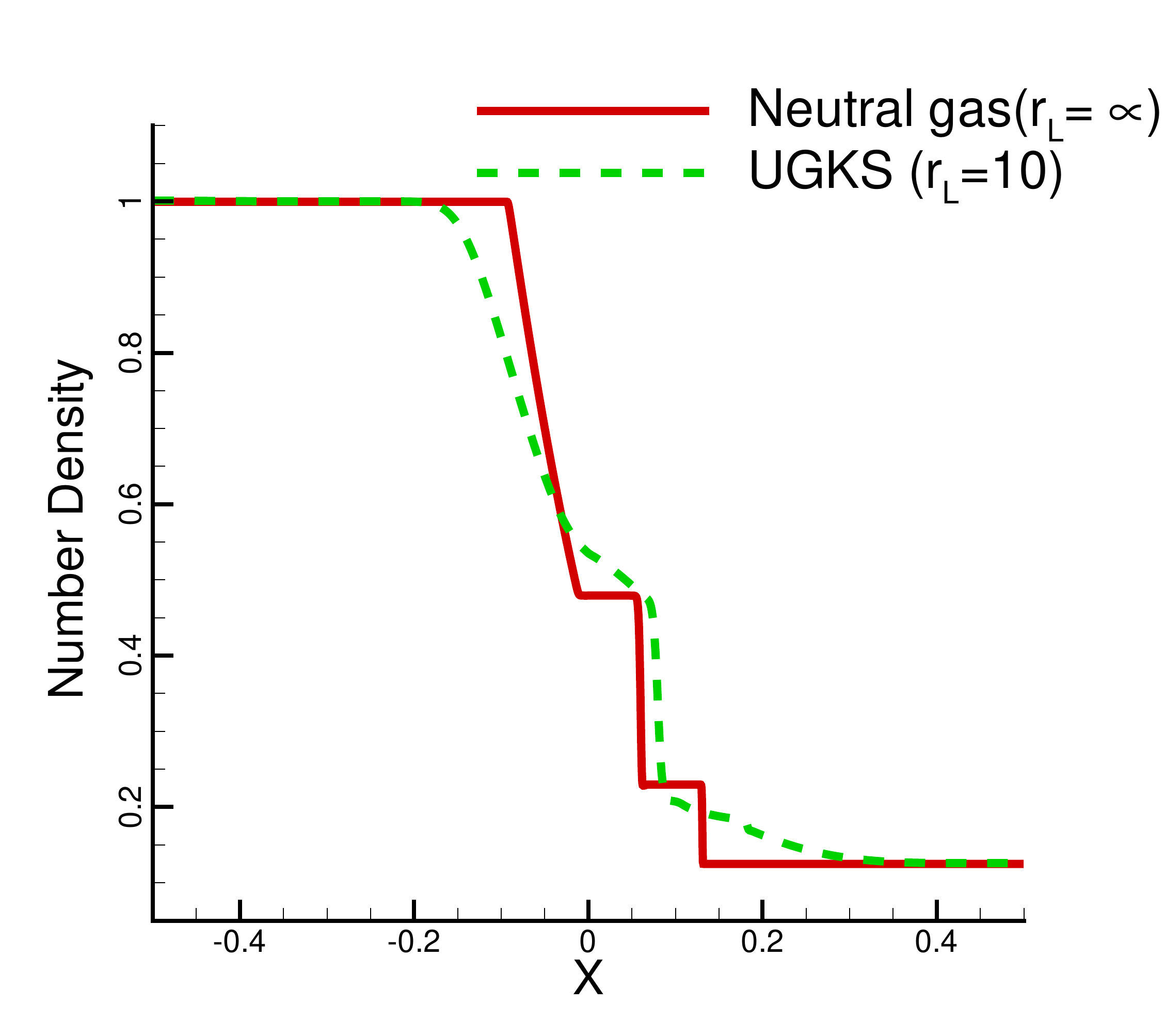}
\includegraphics[width=0.23\textwidth]{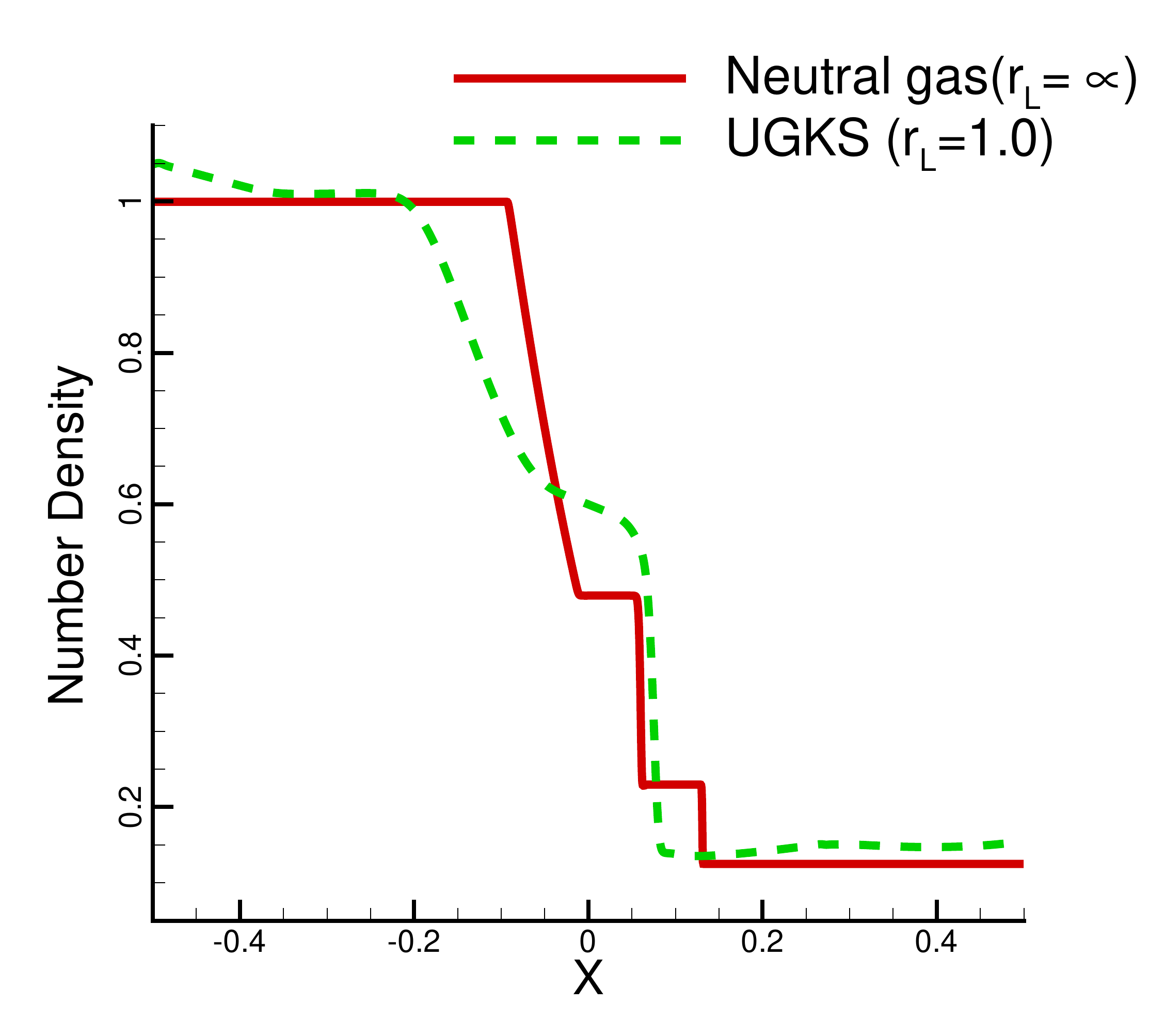}
\includegraphics[width=0.23\textwidth]{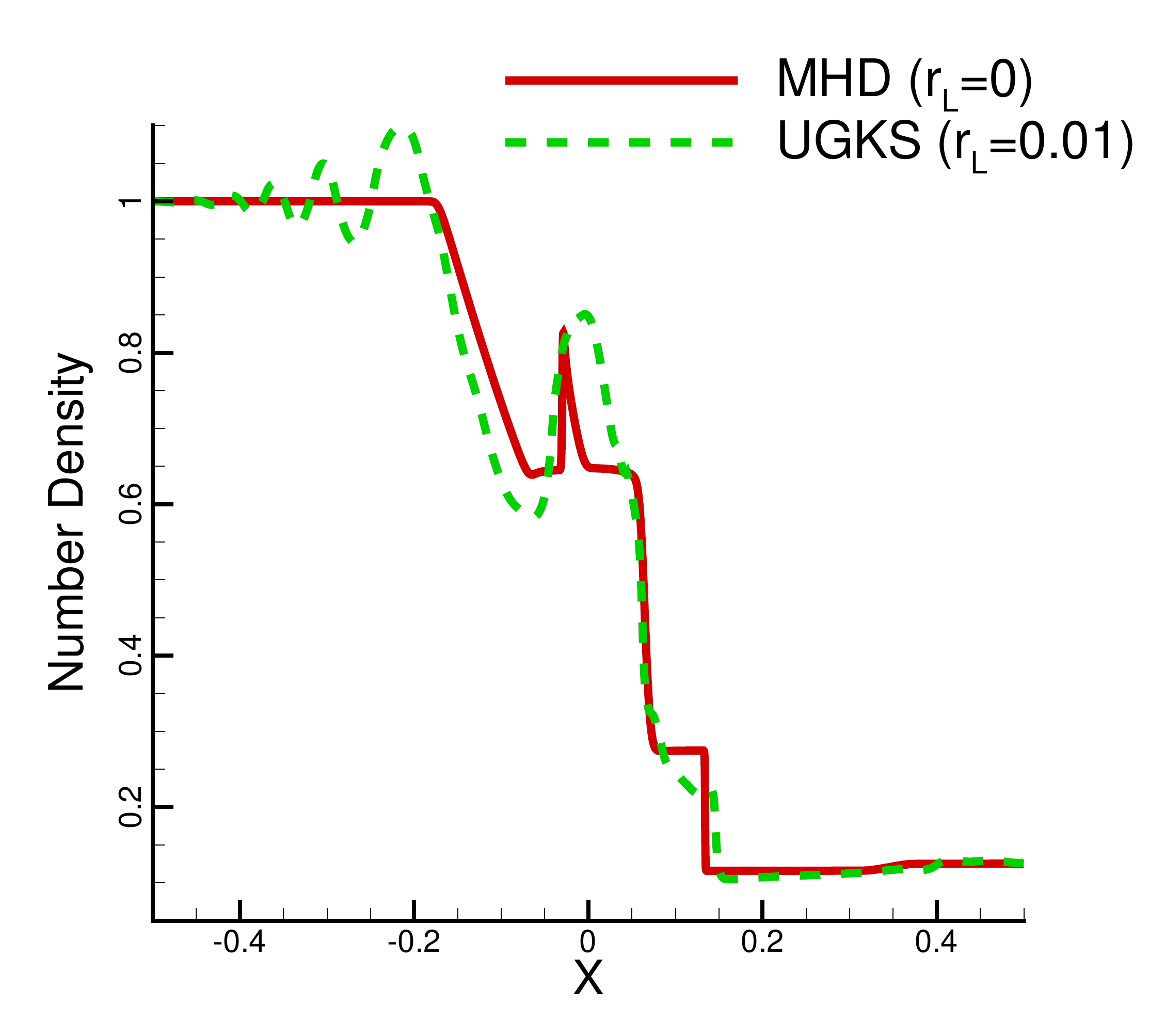}
\includegraphics[width=0.23\textwidth]{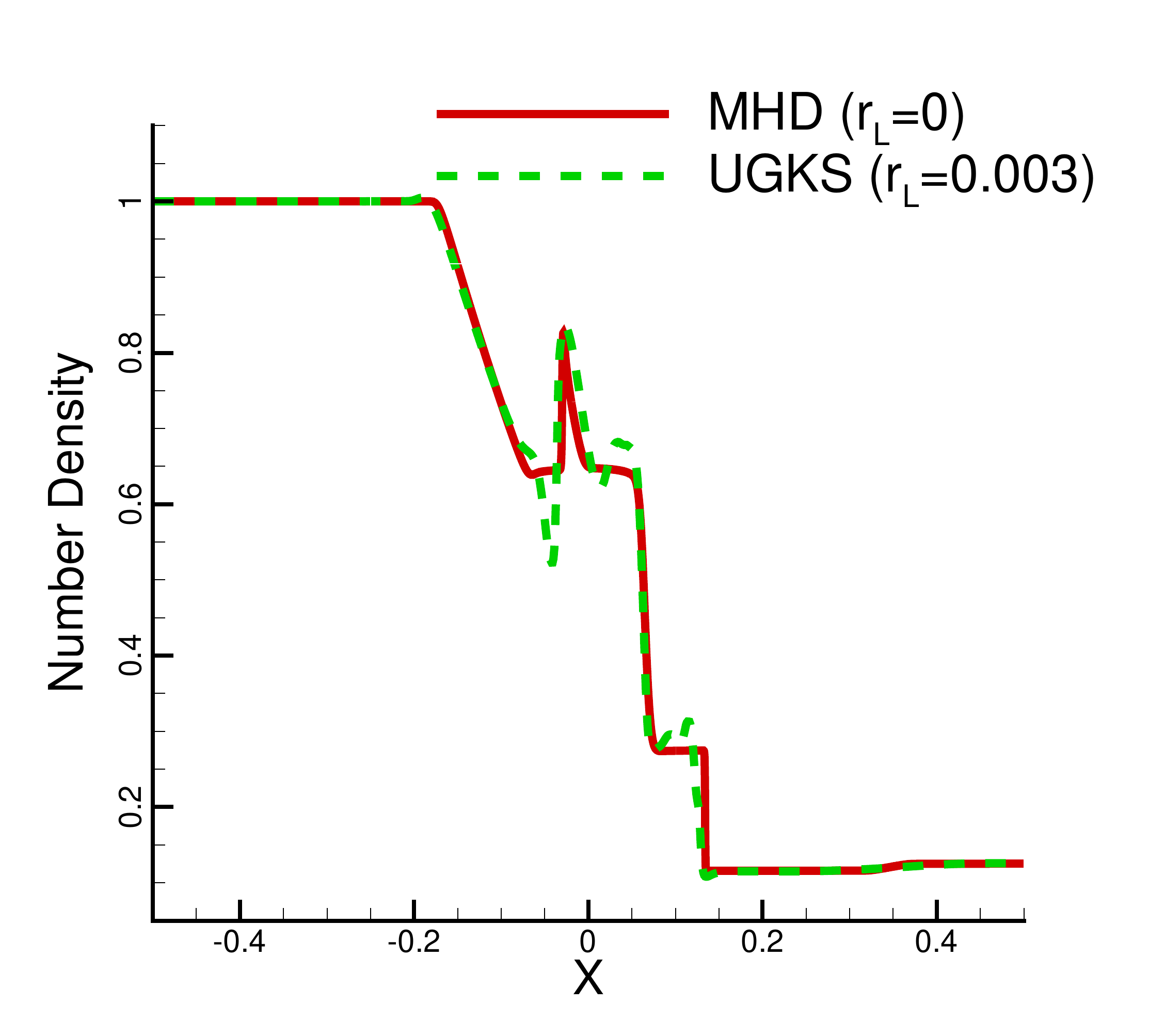}\\
\includegraphics[width=0.23\textwidth]{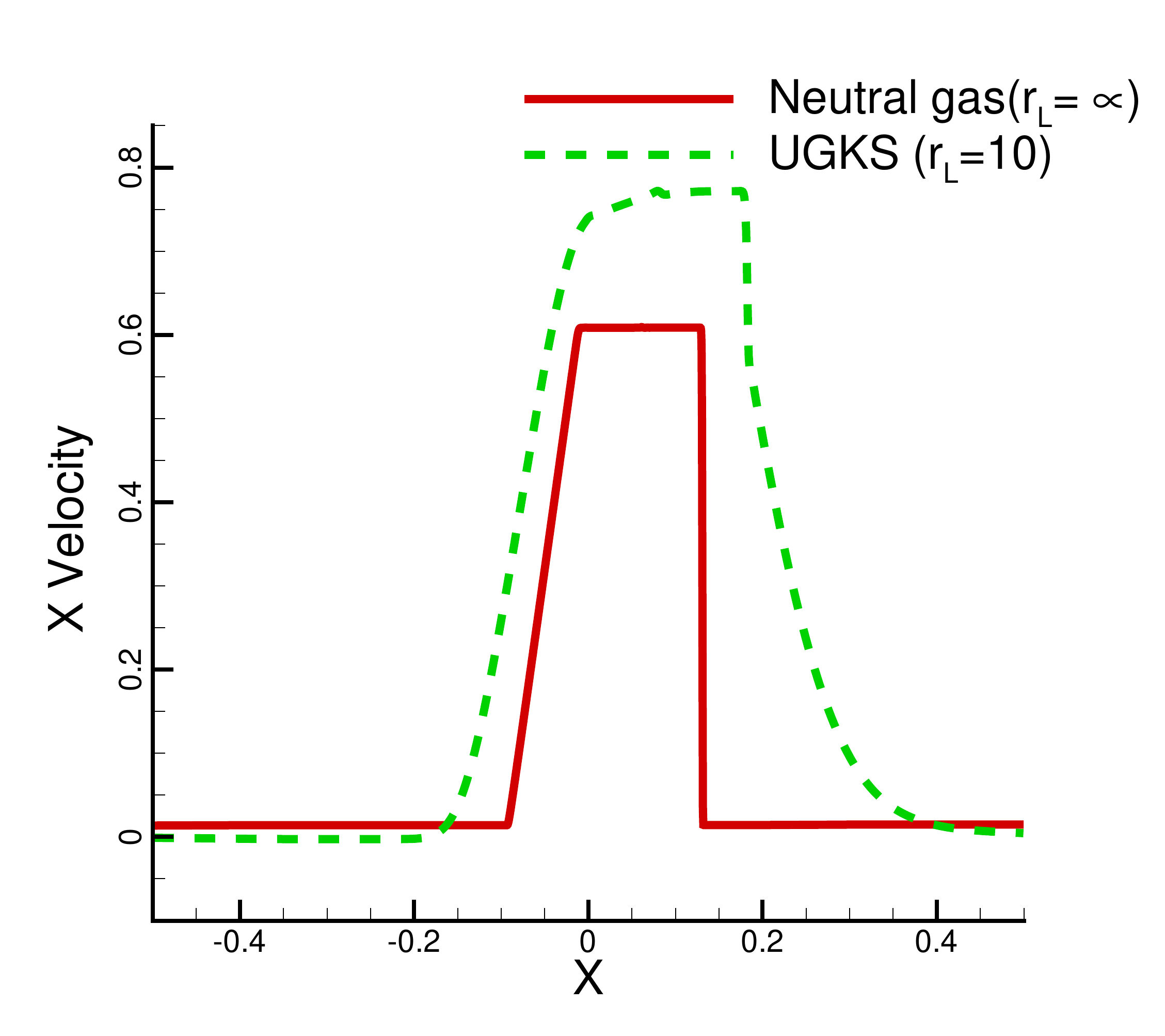}
\includegraphics[width=0.23\textwidth]{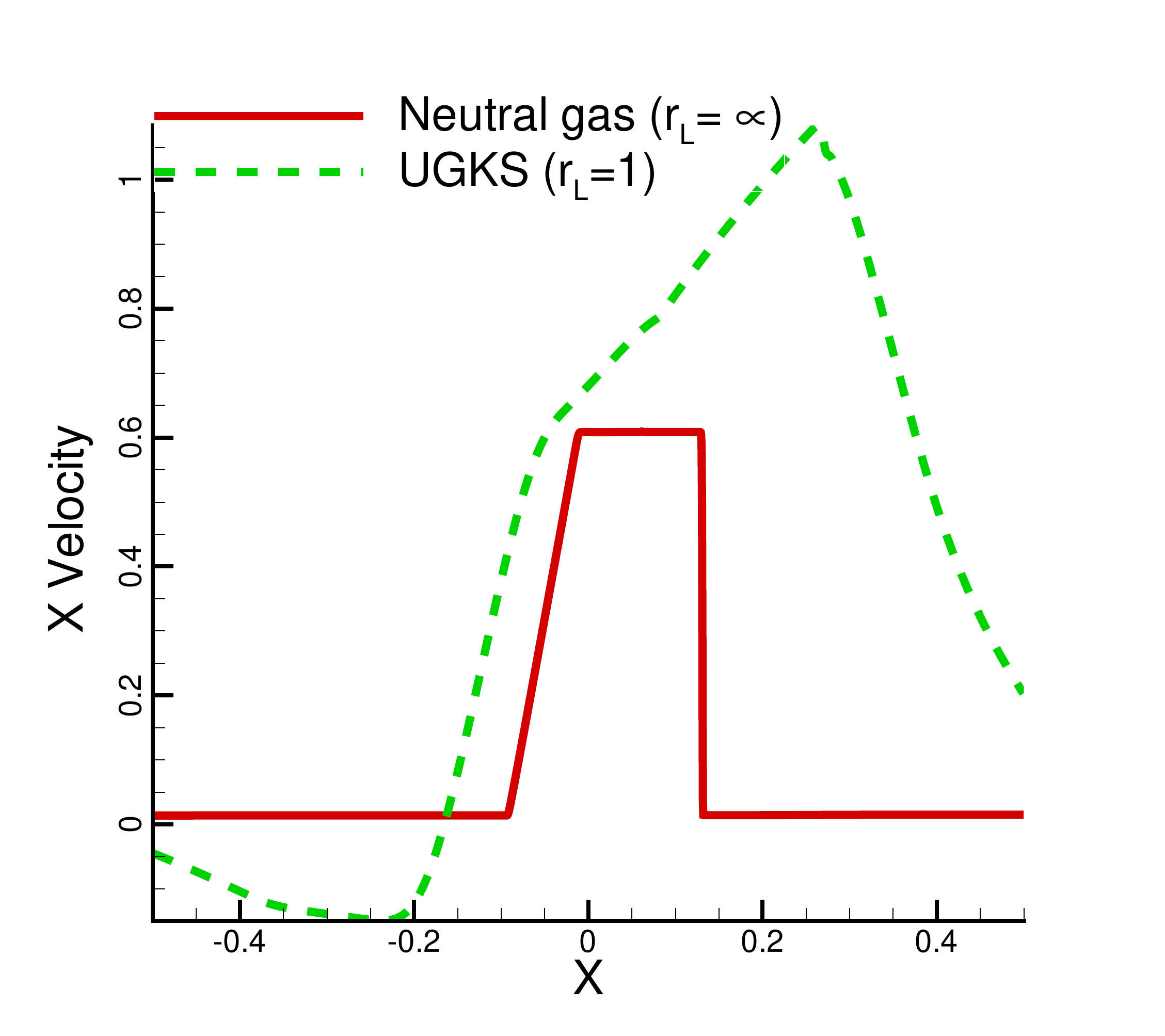}
\includegraphics[width=0.23\textwidth]{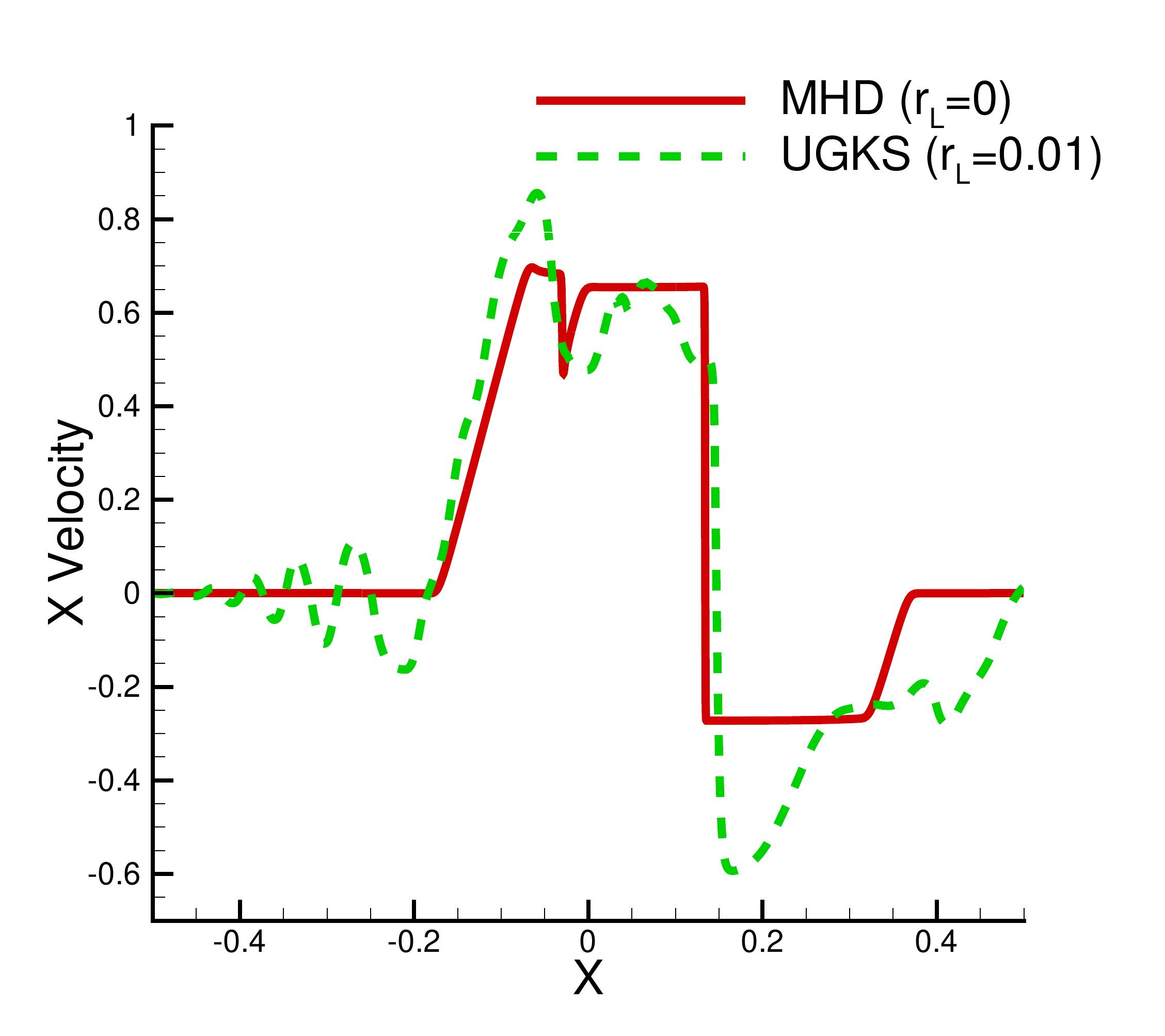}
\includegraphics[width=0.23\textwidth]{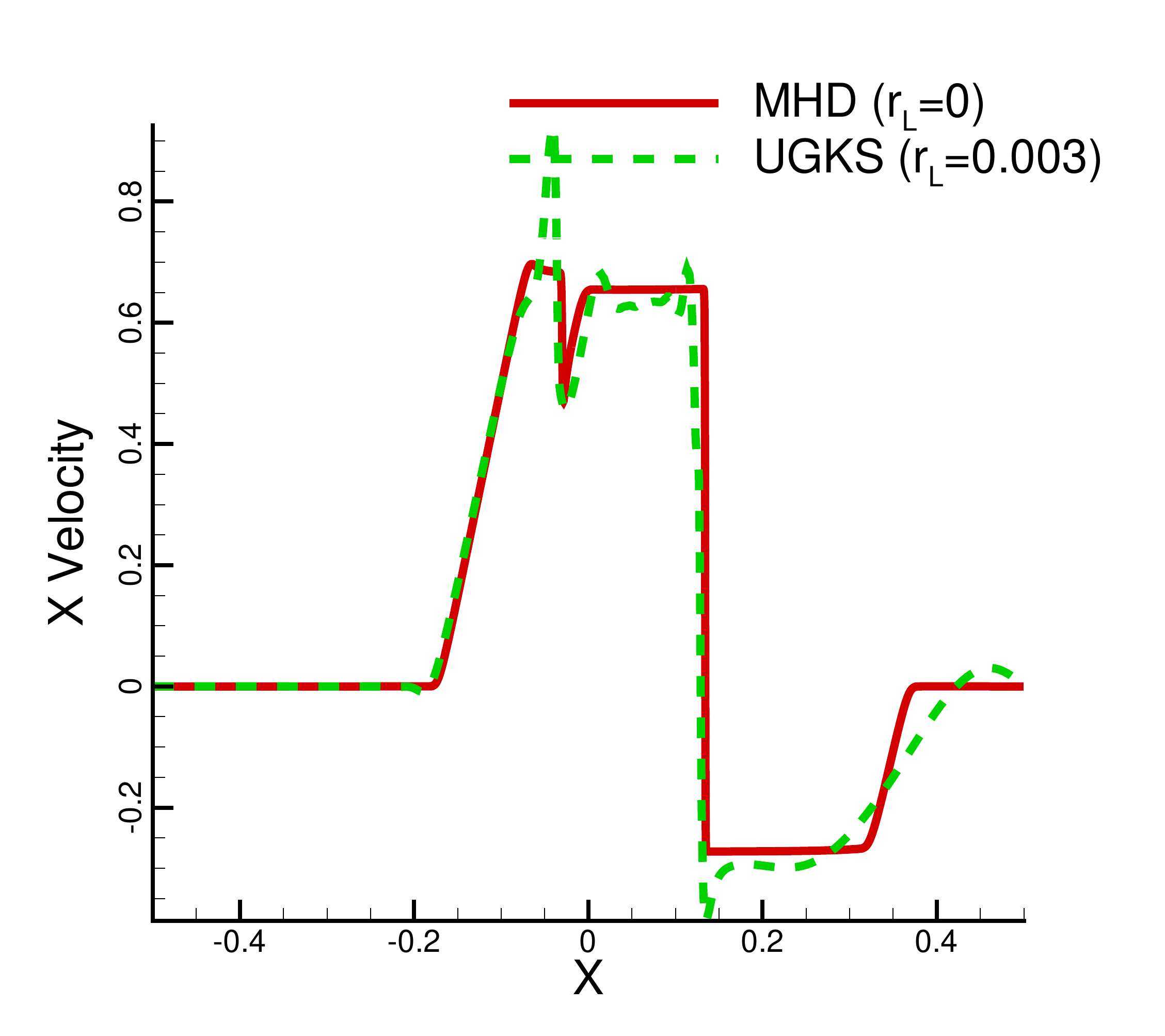}\\
\includegraphics[width=0.21\textwidth,angle=90]{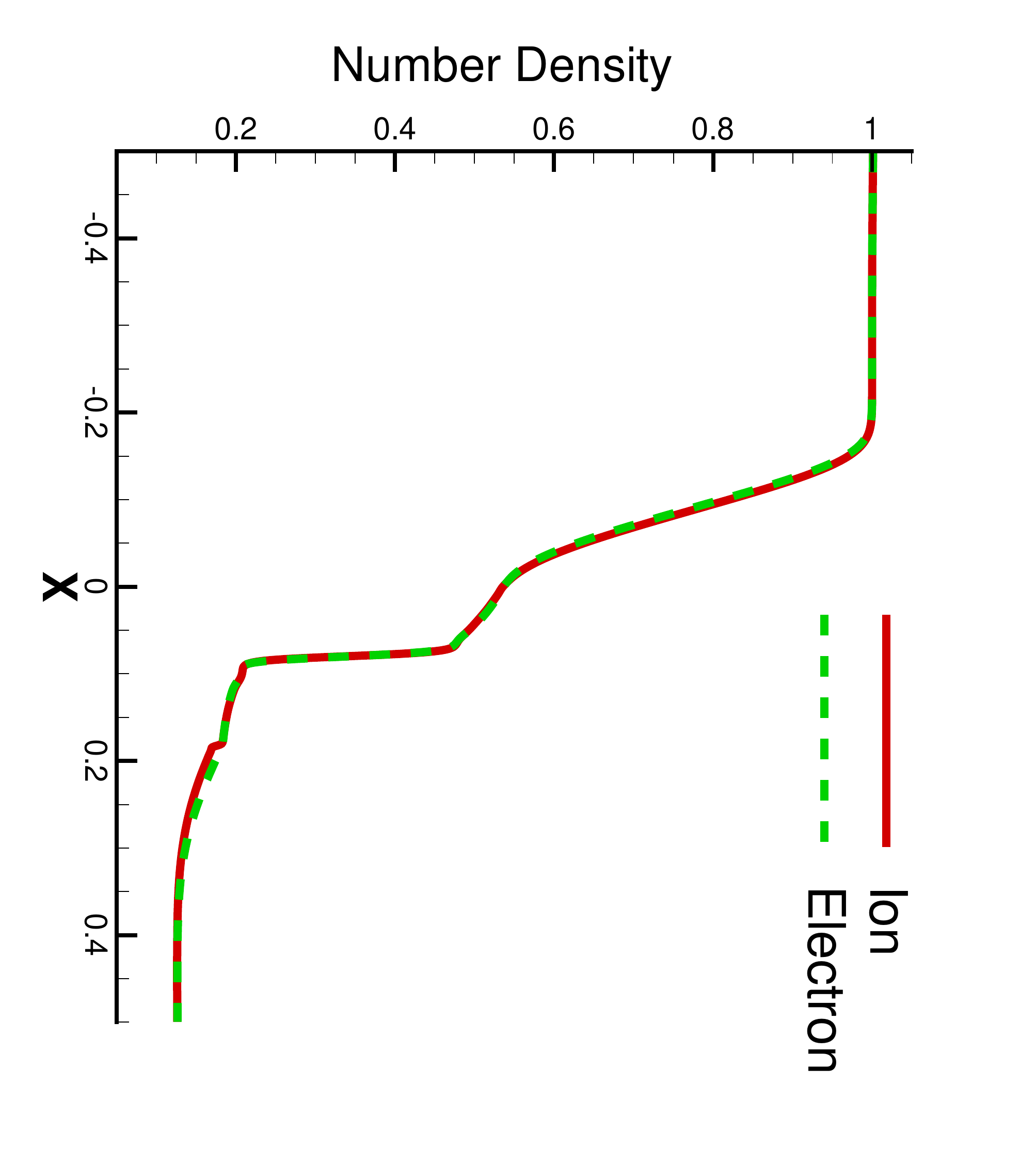}
\includegraphics[width=0.21\textwidth,angle=90]{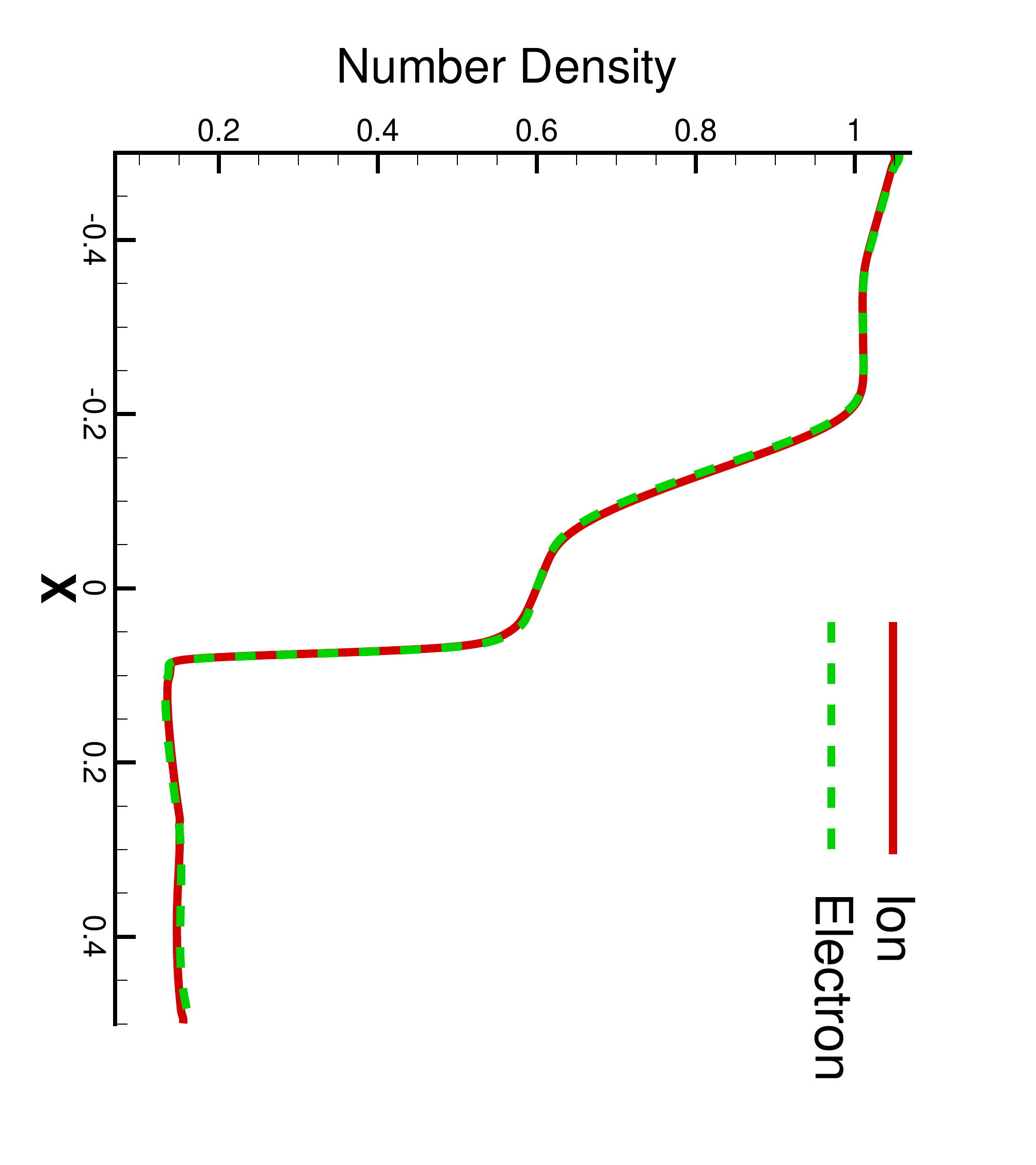}
\includegraphics[width=0.23\textwidth]{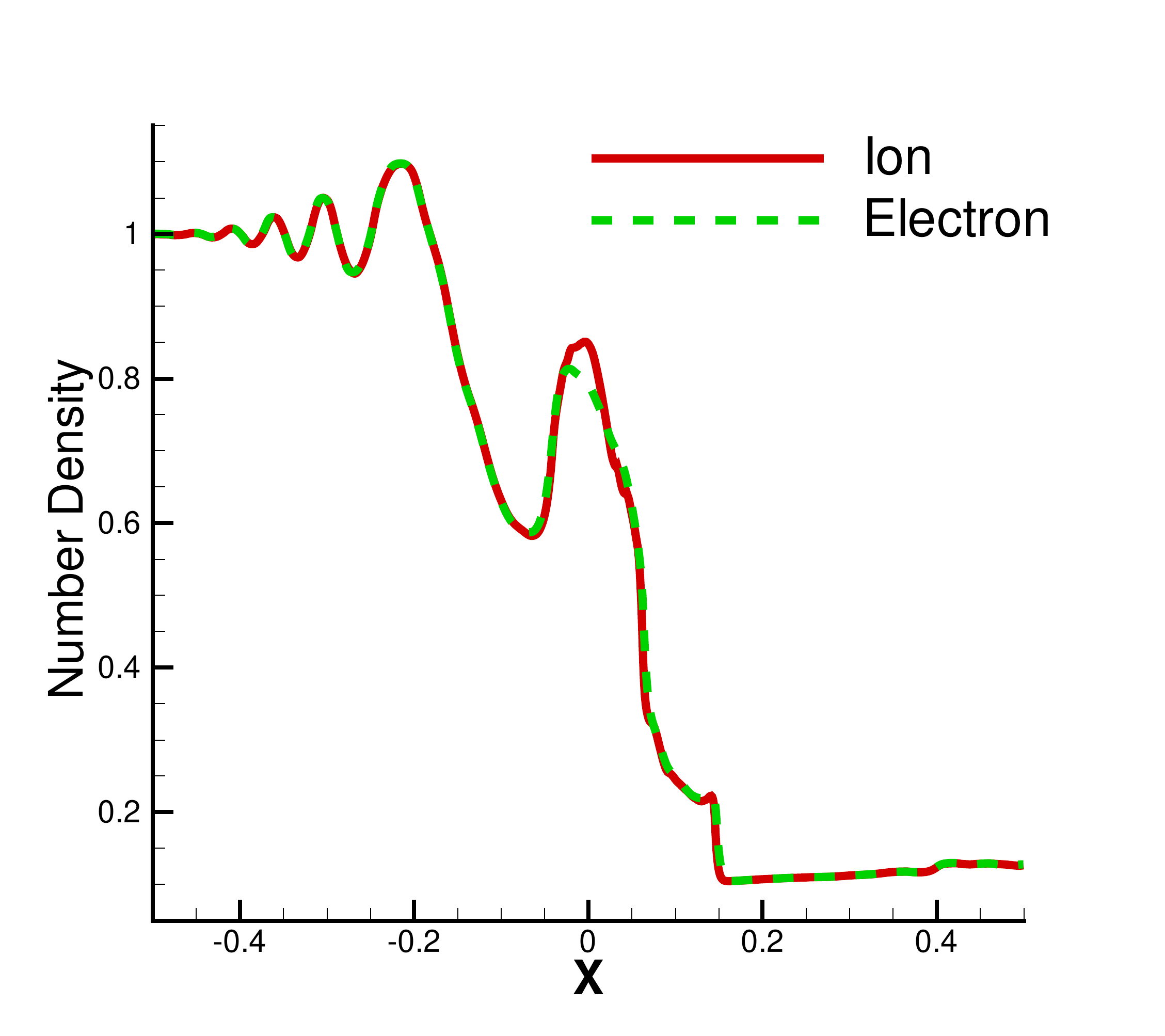}
\includegraphics[width=0.21\textwidth,angle=90]{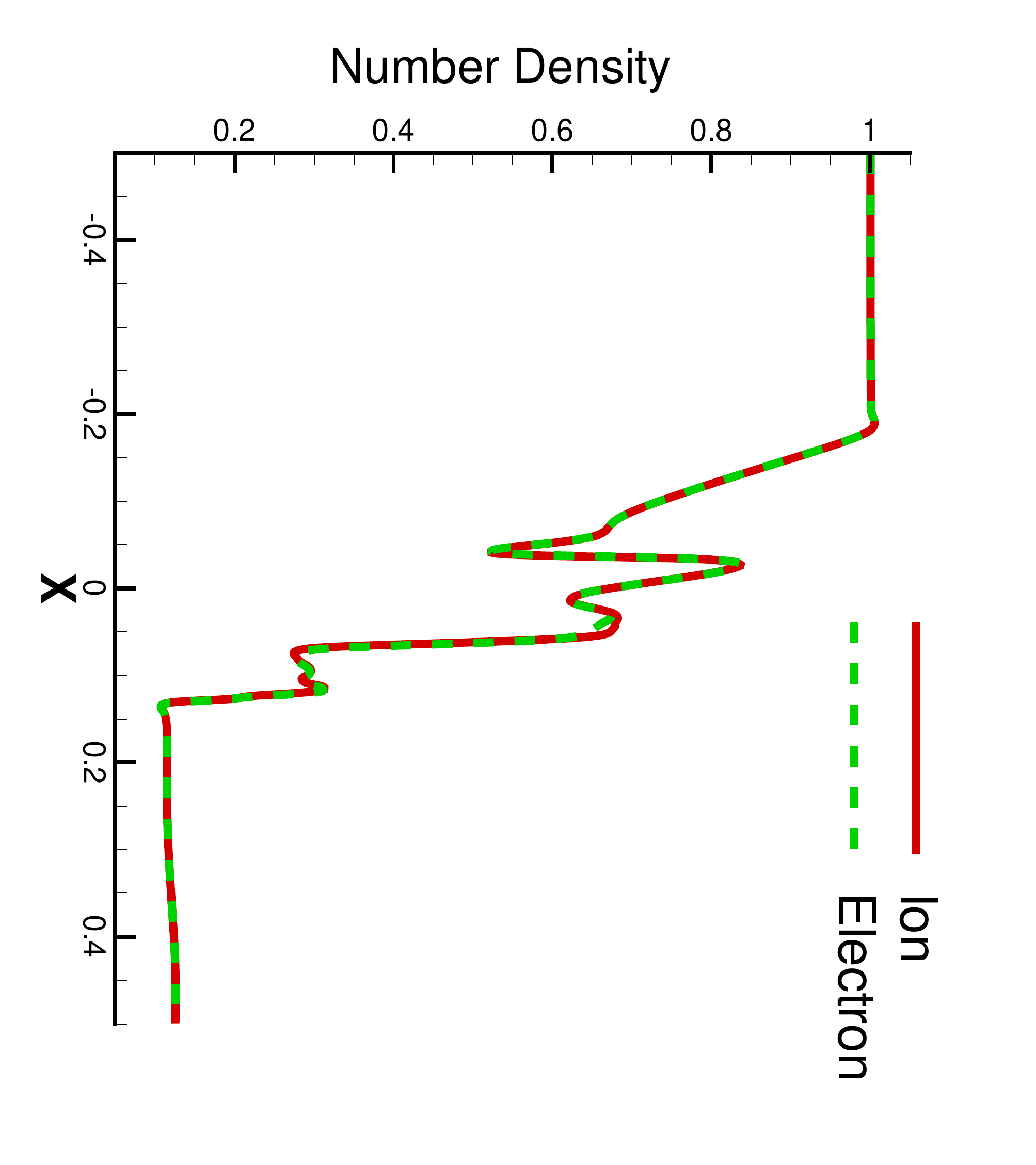}\\
\includegraphics[width=0.21\textwidth,angle=90]{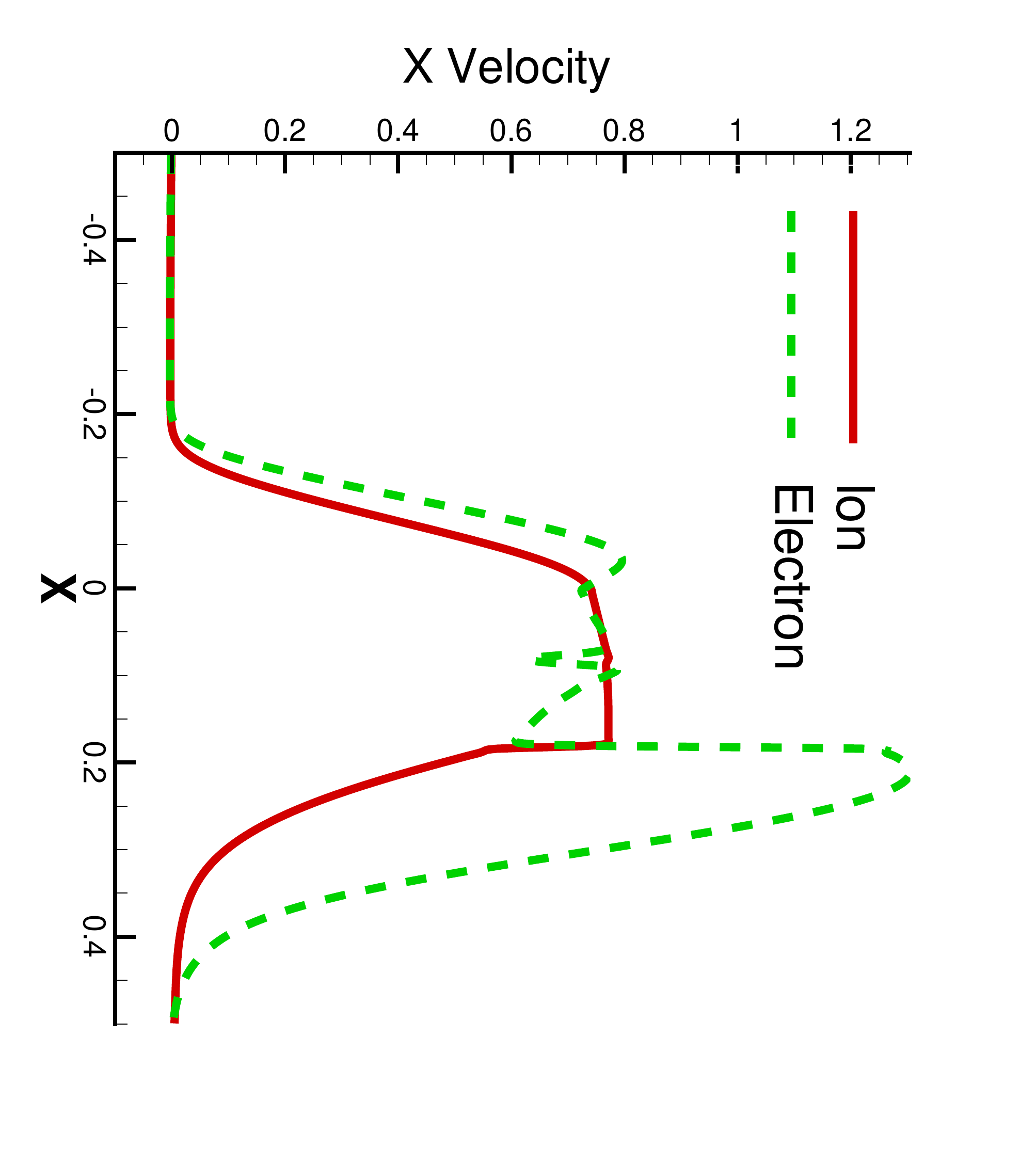}
\includegraphics[width=0.21\textwidth,angle=90]{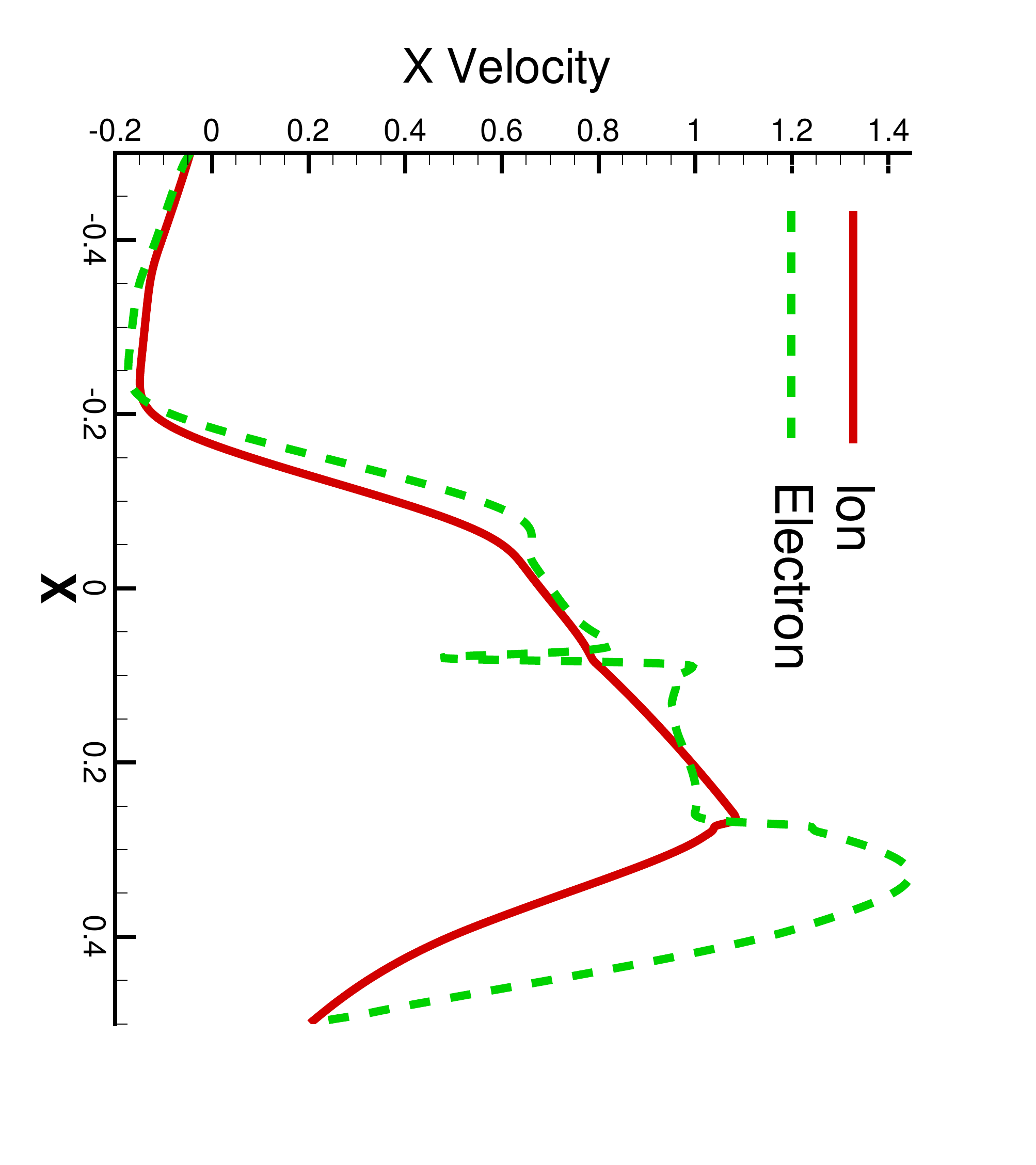}
\includegraphics[width=0.21\textwidth,angle=90]{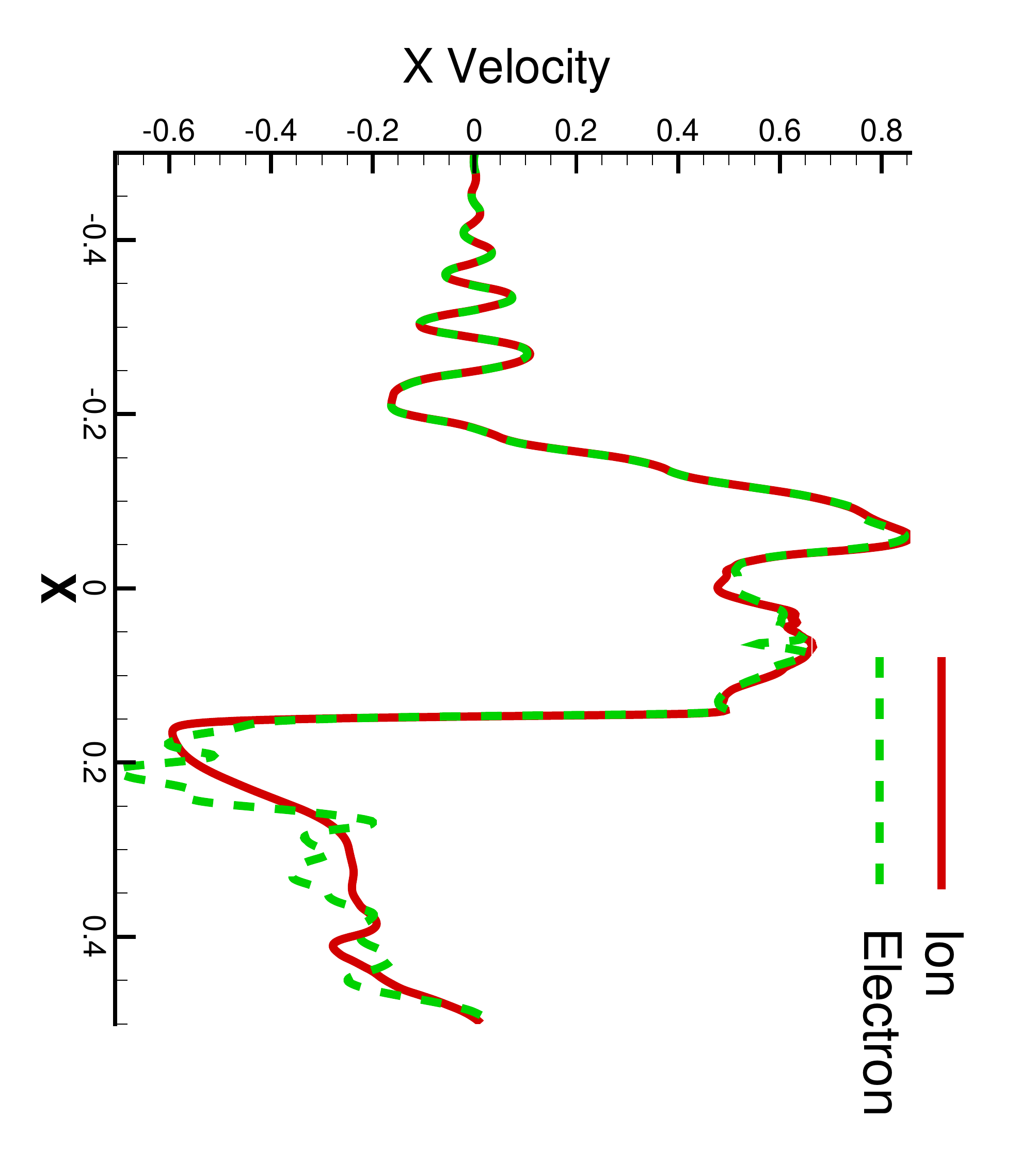}
\includegraphics[width=0.21\textwidth,angle=90]{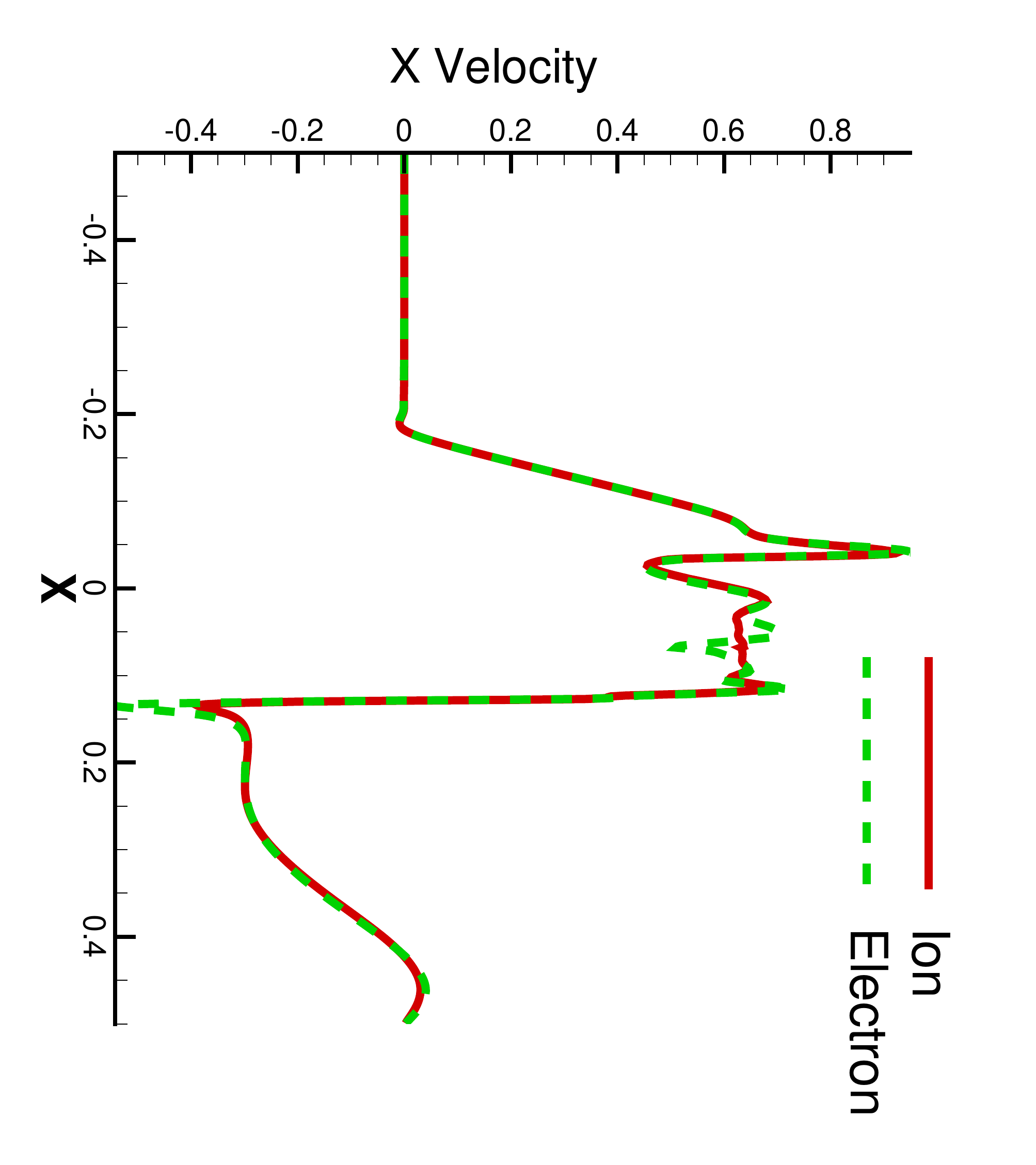}\\
\includegraphics[width=0.21\textwidth,angle=90]{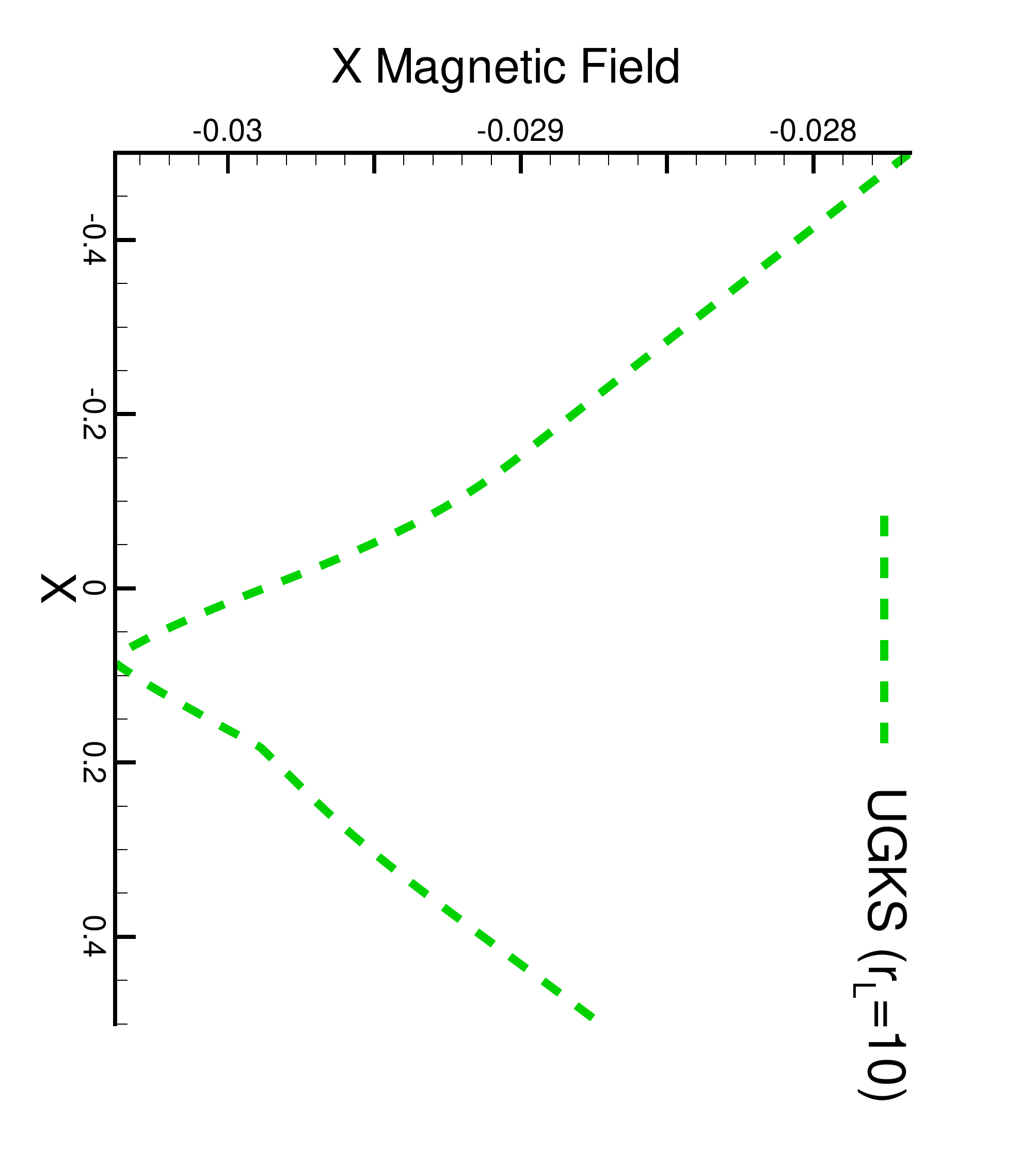}
\includegraphics[width=0.21\textwidth,angle=90]{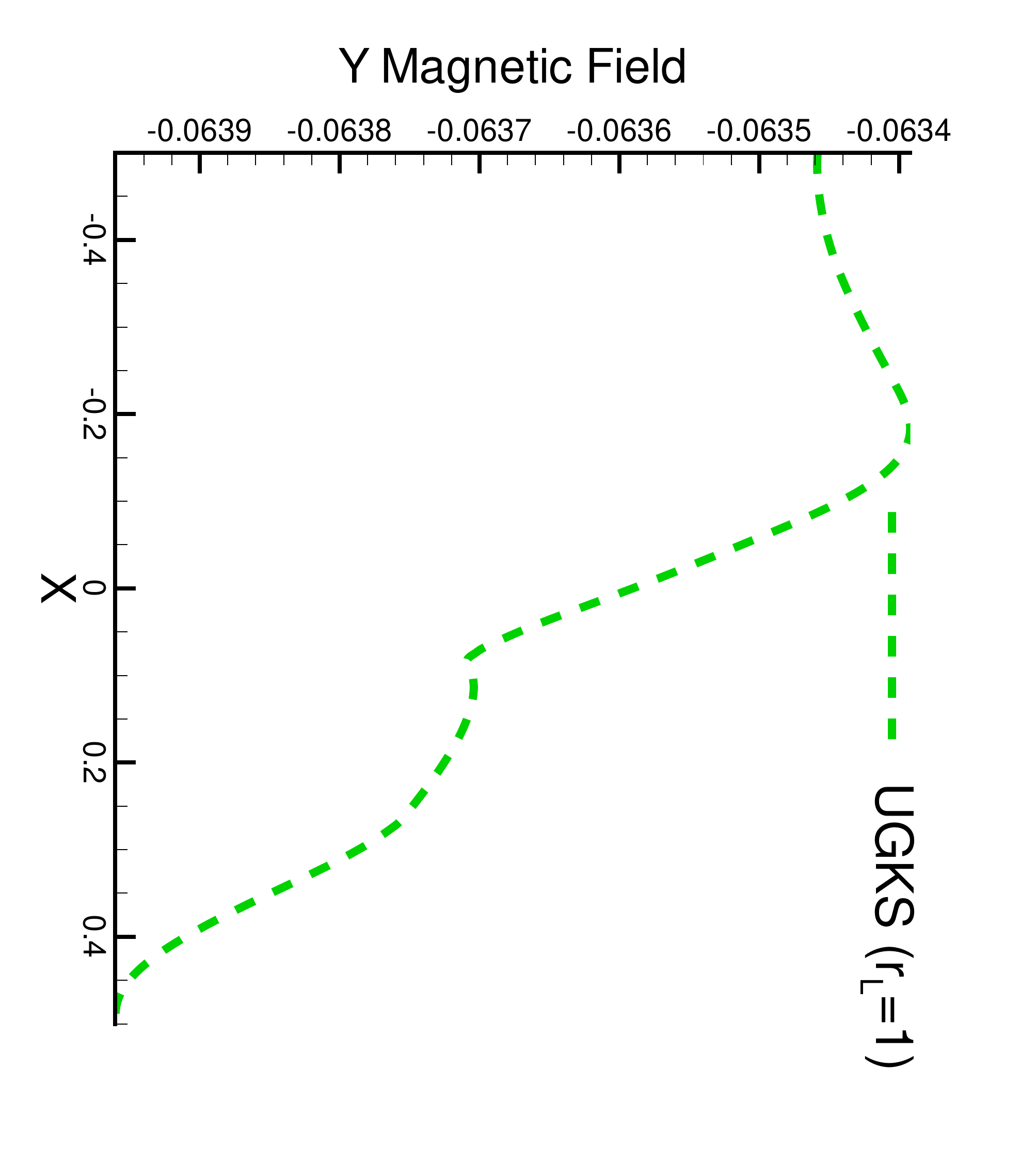}
\includegraphics[width=0.21\textwidth,angle=90]{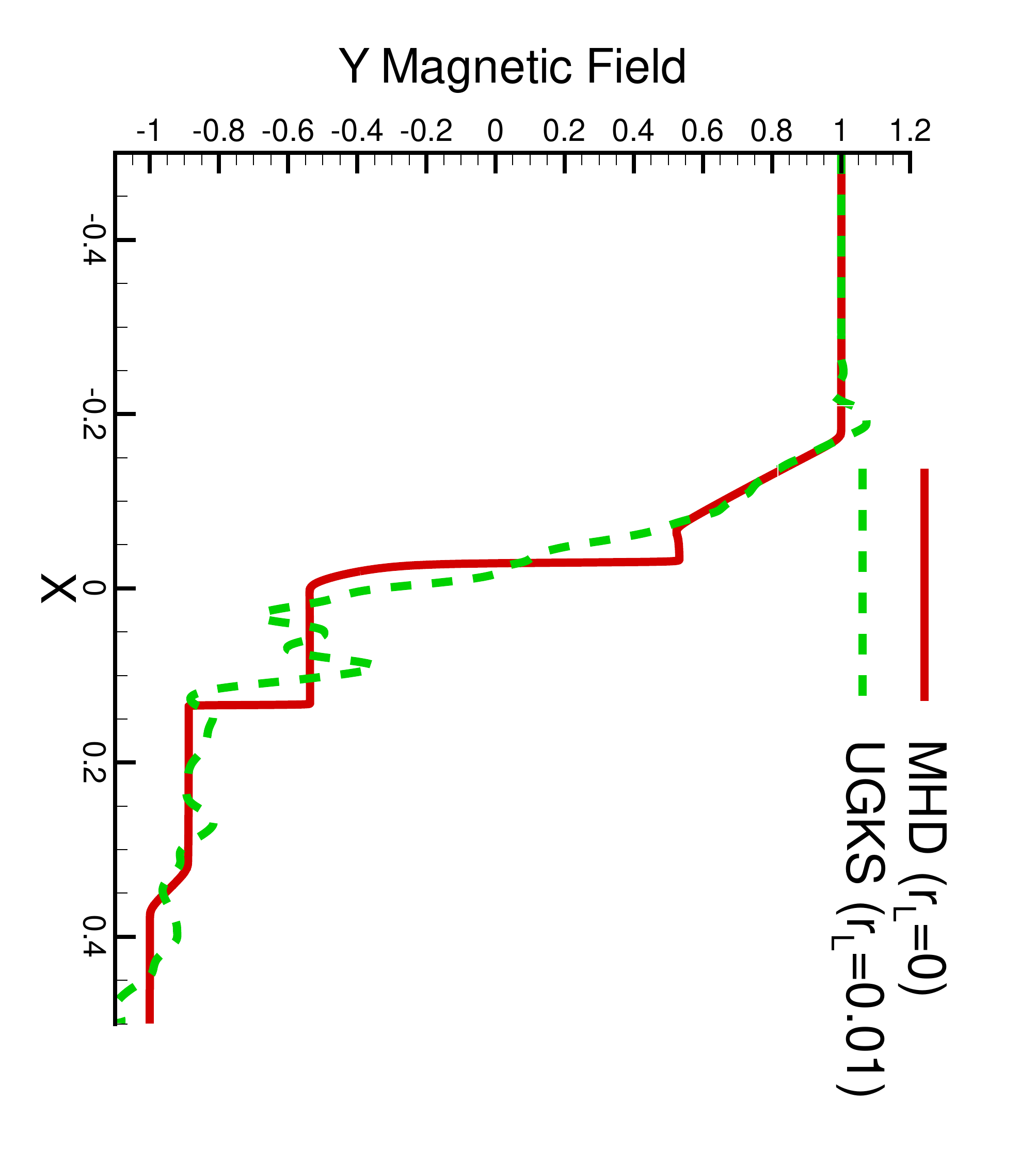}
\includegraphics[width=0.23\textwidth]{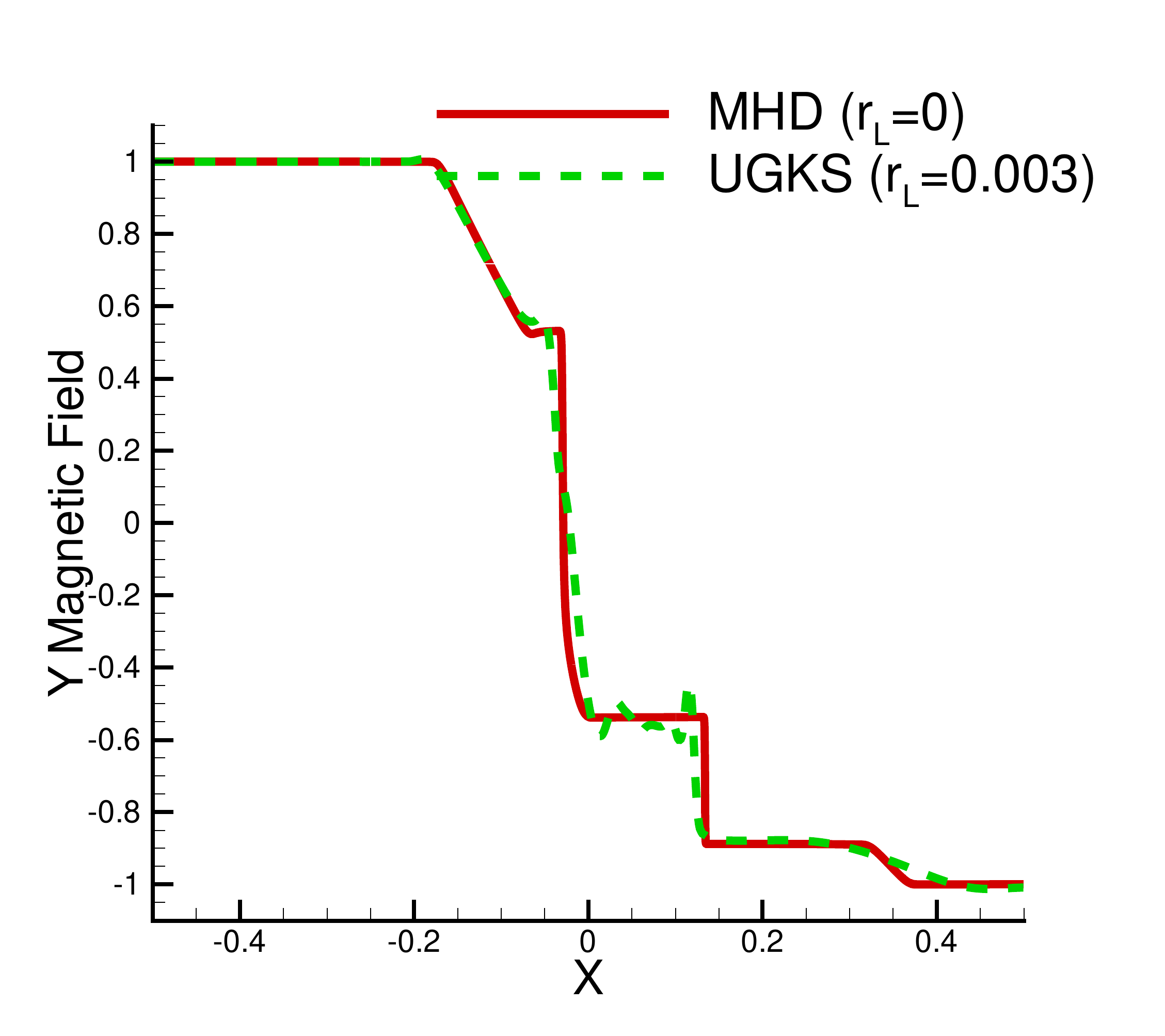}
\caption{Results of the averaged density, averaged velocity, ion-electron density, ion-electron velocity, and
magnetic field profiles from top to bottom, and at $r_{L_i}=10, 1, 0.01, 0.003$ from left to right.}
\label{Brio-1}
\end{figure}

The averaged density and velocity are calculated by $$\rho=\frac{\rho_i m_i+\rho_e m_e}{m_i+m_e},~~ \mathbf{U}=\frac{\mathbf{U}_i m_i+\mathbf{U}_e m_e}{m_i+m_e}.$$
The results of the density, velocity, and magnetic field profiles, which are compared with the gas dynamic results, and MHD results are shown in Fig. \ref{Brio-1}. It can be observed that the solutions behave like Euler solutions at large normalized Larmor radius. When normalized Larmor radius gets small, the electrons and ions are coupled together and the behavior of plasma fluid follows the Hall-MHD and
towards to ideal MHD solutions.

\subsection{Multiple scale shock tube problem}

\begin{figure}
  \centering
  \includegraphics[width=0.45\textwidth]{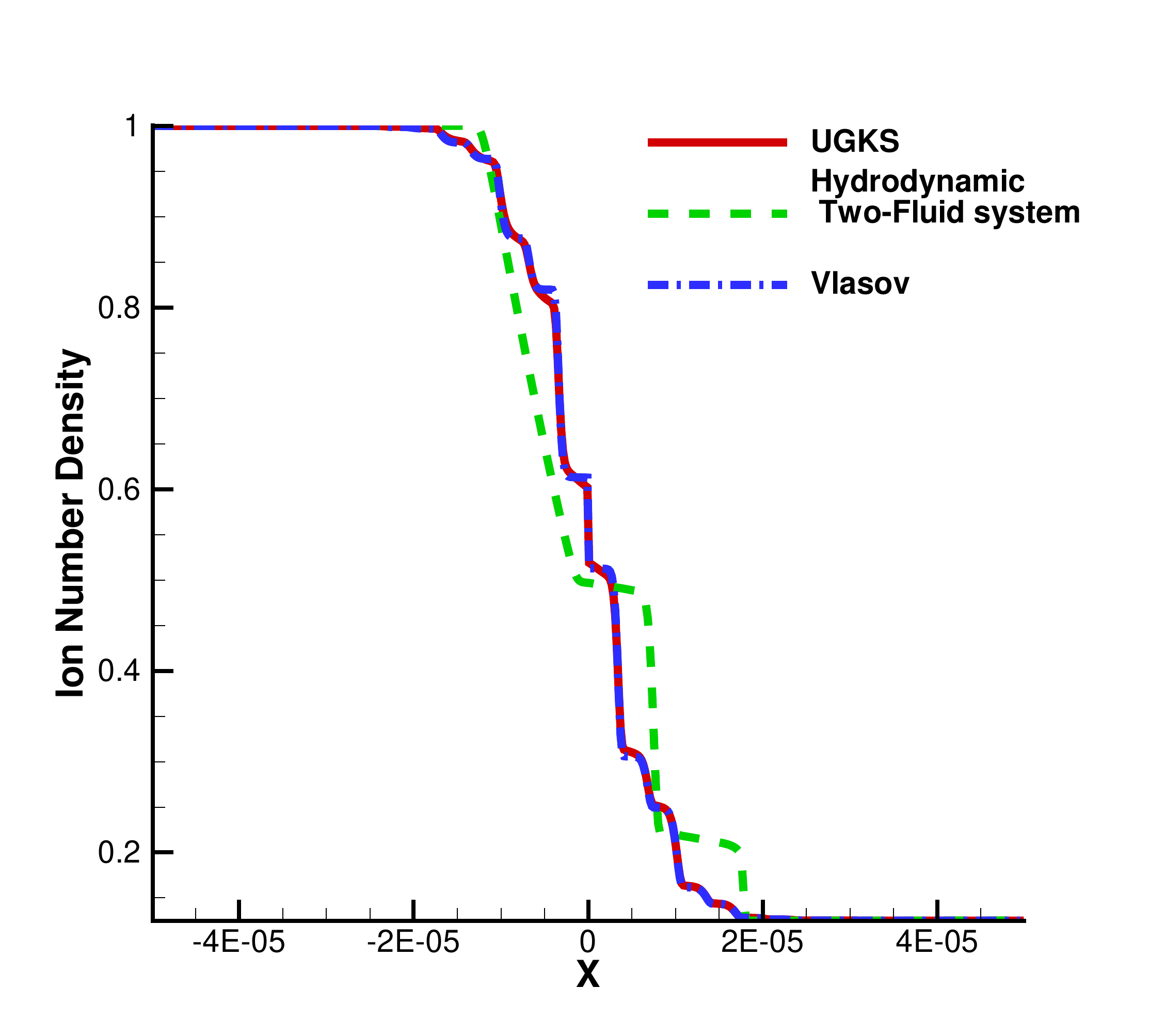}{a}
  \includegraphics[width=0.45\textwidth]{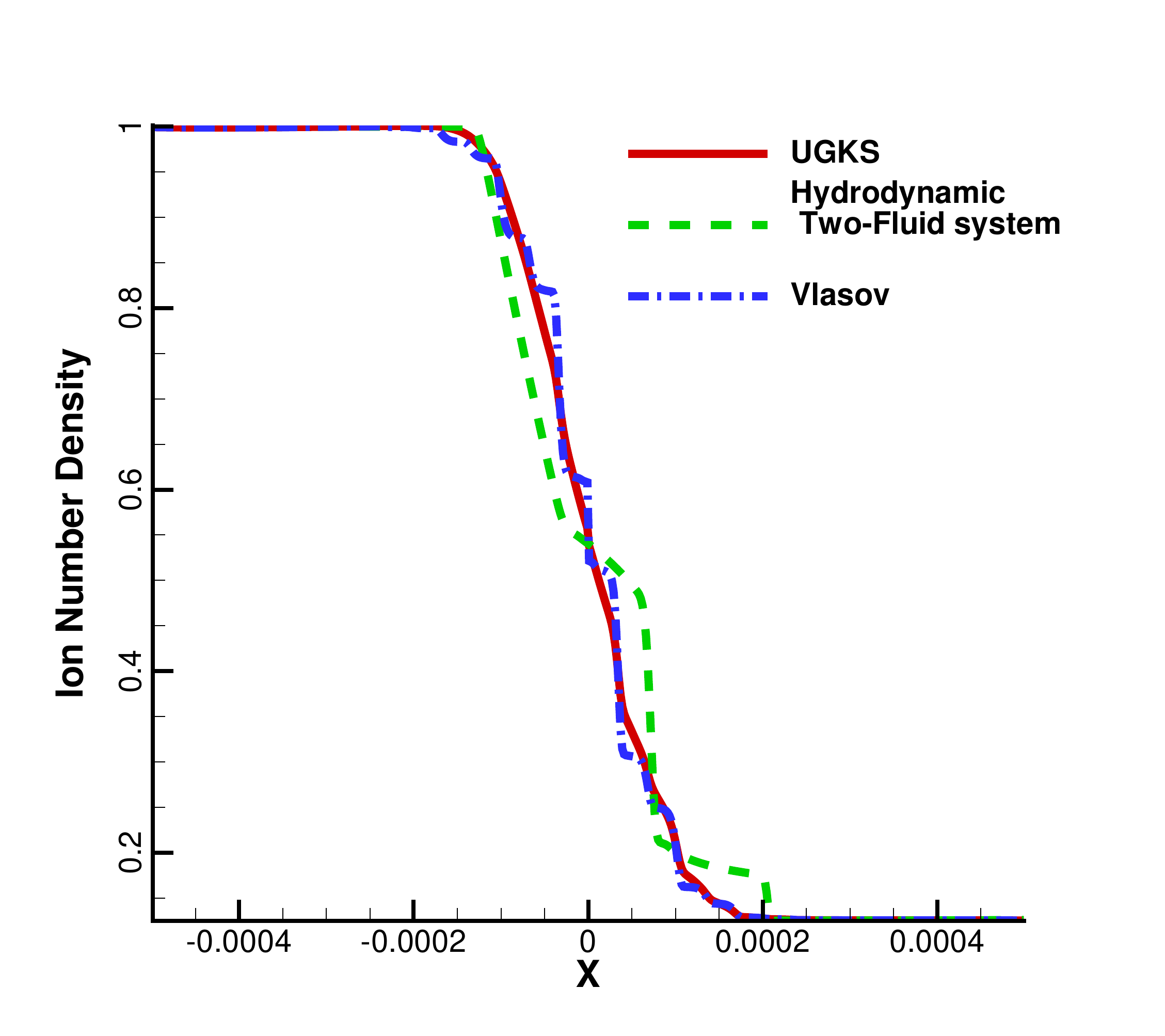}{b}
  \includegraphics[width=0.45\textwidth]{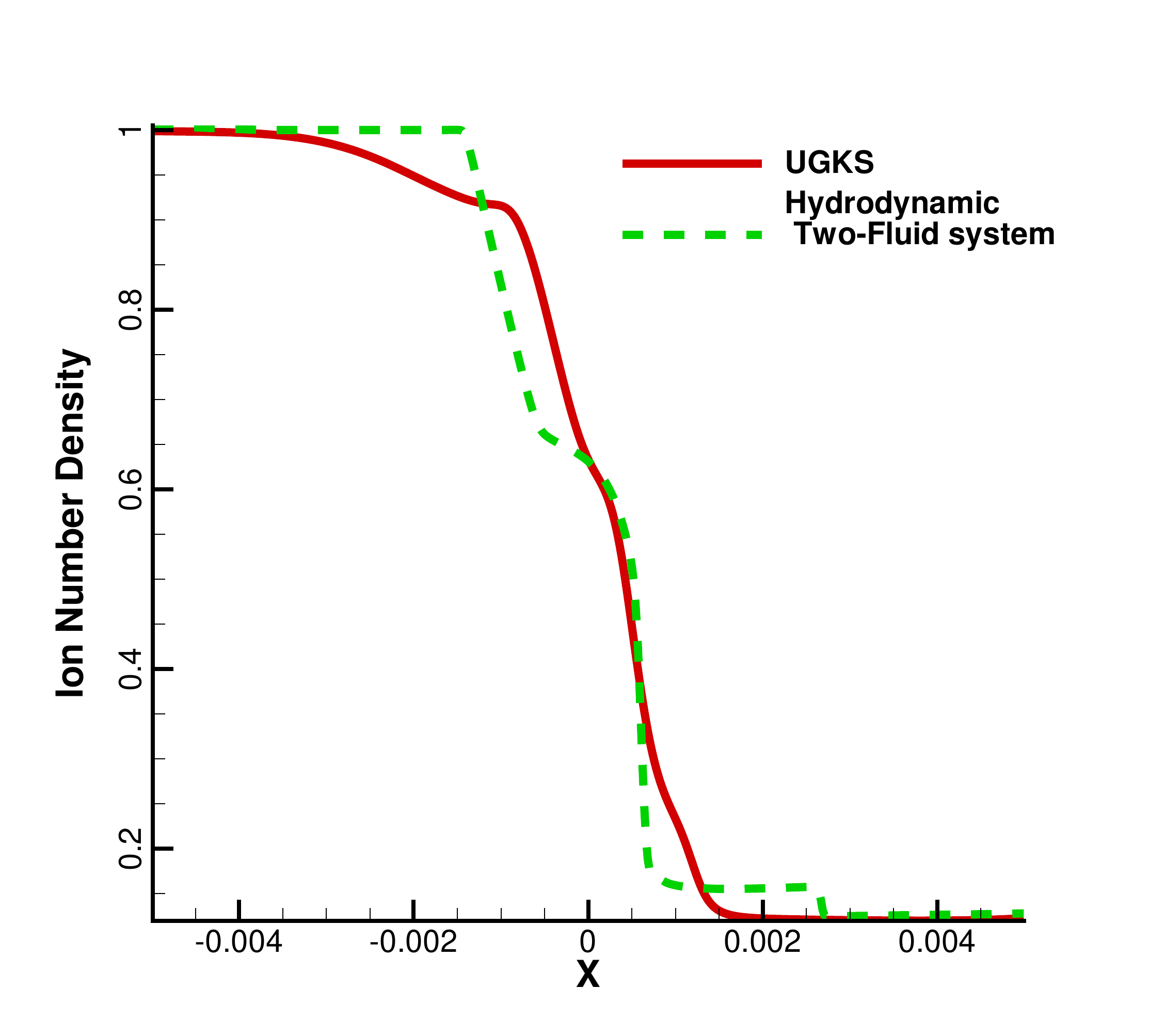}{c}
  \includegraphics[width=0.45\textwidth]{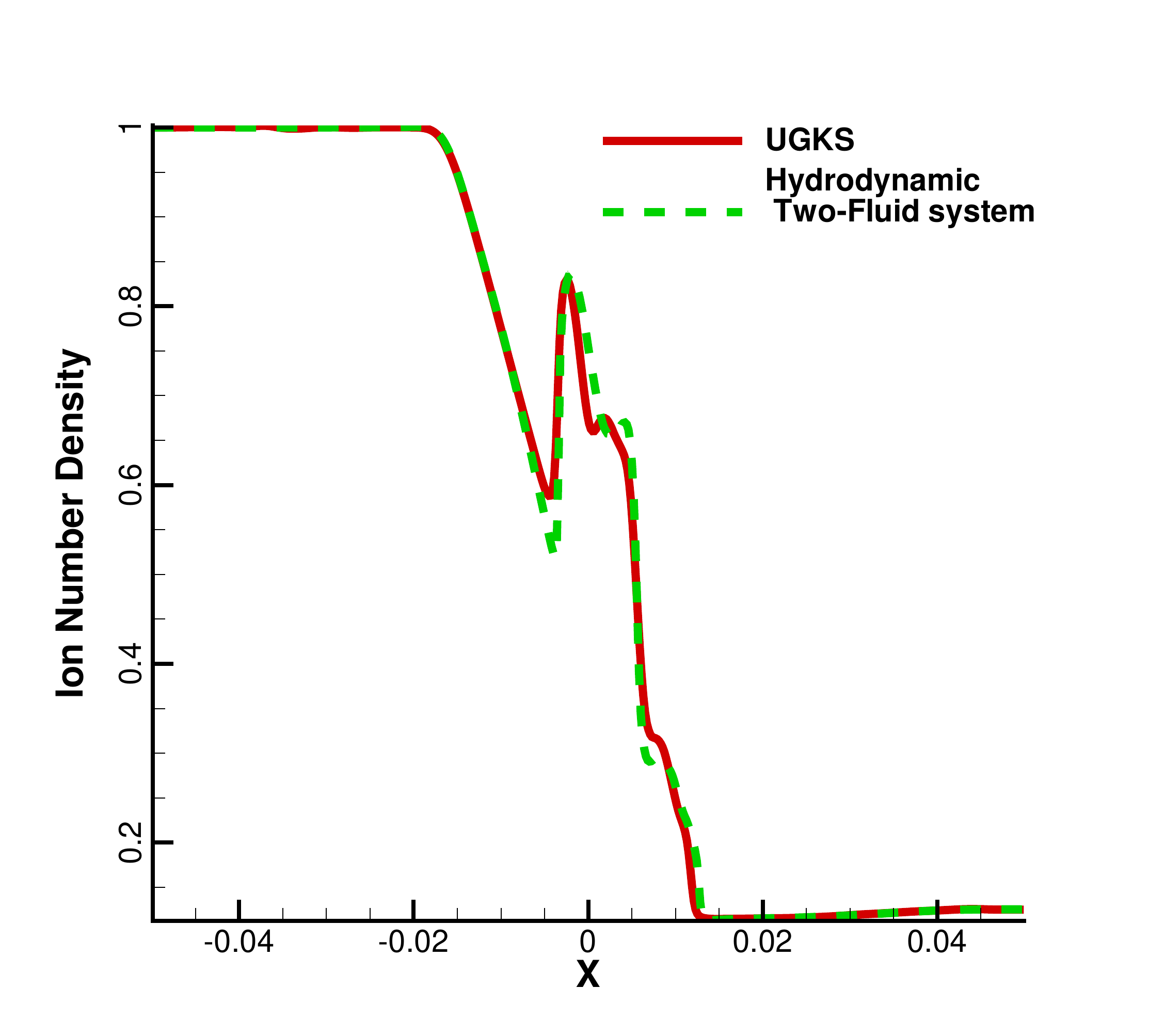}{d}
  \includegraphics[width=0.45\textwidth]{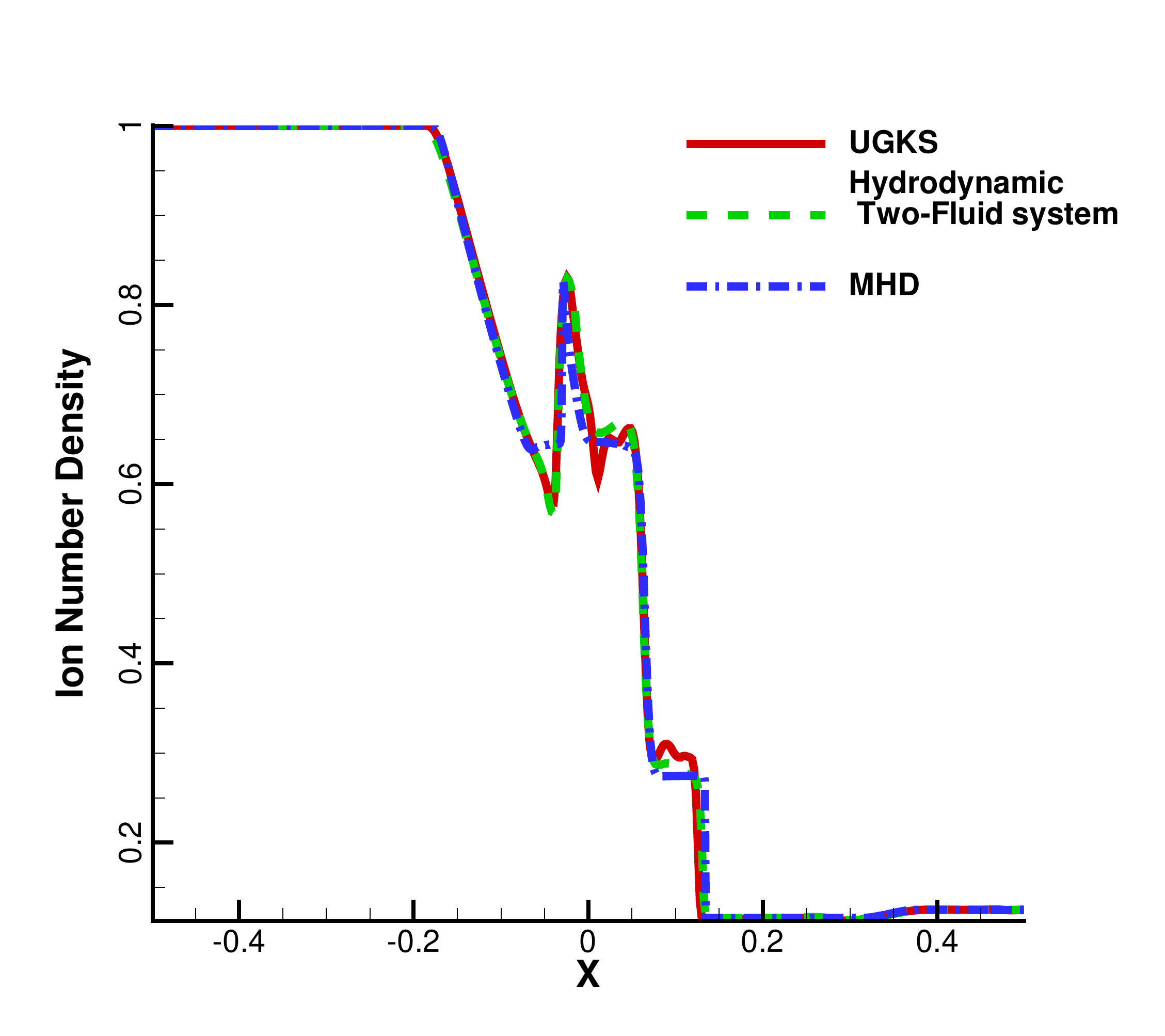}{e}
  \caption{Solutions of the Brio-Wu shock tube at output time $t=10^{-5}$ (a), $10^{-4}$ (b), $10^{-3}$ (c), $10^{-2}$ (d), $10^{-1}$ (e).
  The UGKS solutions are compared with the two fluid system solutions,
  Vlasov solutions (for $t=10^{-5}, 10^{-4}$) and MHD solution (for $t=10^{-1}$).}
  \label{compare1}
\end{figure}

In this calculation, we study how the solution is developed from the initial condition to the final MHD one. Multi-scale solutions can be observed at different output times. The ion to electron mass ratio is set to be 25 and $\nu_{ie}$ is set to be zero.
The characteristic length is fixed to be $1000$ ion Larmor radius.
The upstream Debye length is $0.01 r_{L_i}$.
The relaxation parameter $\tau$ is set to be $10^{-6}$ and the velocity space is $[-3,3]$ for ion and $[-15,15]$ for electron with 16 grid points.
For different output times, the cell size is fixed to different value.
The ion number density at times $t=10^{-5}, 10^{-4}, 10^{-3}, 10^{-2}, 10^{-1}$ by UGKS are shown in Fig. \ref{compare1}, which are compared
with the Vlasov solutions and hydrodynamic two-fluid solutions.
For the solution at $t=10^{-5}$, the cell size is $2.0\times 10^{-7}$,
which is much less than the ion Larmor radius $\Delta x=2\times 10^{-4} r_{L_i}$,
and the time step is $\Delta t =3\times10^{-3}\tau_i=2\times 10^{-3}\omega_{pi}^{-1}$.
At this output time, the UGKS solution goes to Vlasov one due to this
collisionless limit, which is different from the hydrodynamic two-fluid solutions.
The cell sizes used for output times $t=10^{-4}, 10^{-3}, 10^{-2}$ are $\Delta x=2.0\times 10^{-6}, 2.0\times 10^{-5}, 2.0\times 10^{-4}$.
In this transition regime, the solutions from the UGKS, Vlasov, and hydrodynamic two-fluid deviate,
and the solution of UGKS should be physically reliable.
A large cell size $\Delta x=2\times 10^{-3}$ is used for the solution at the output time $t=0.1$,
and this cell size is two times as large as the ion Larmor radius,
and the time step for this case is
$\Delta t=30 \tau_i=25\omega_{pi}^{-1}$.
It is shown that at the output time $t=0.1$,
both the UGKS and two fluid system solutions converge to the MHD solution in this hydrodynamic regime.
The computational time for UGKS to get solution at $t=0.1$ is $126$ seconds on a 3.40GHz 4-core CPU.

\subsection{Orszag-Tang Vortex}

\begin{figure}
  \centering
  \includegraphics[width=0.45\textwidth]{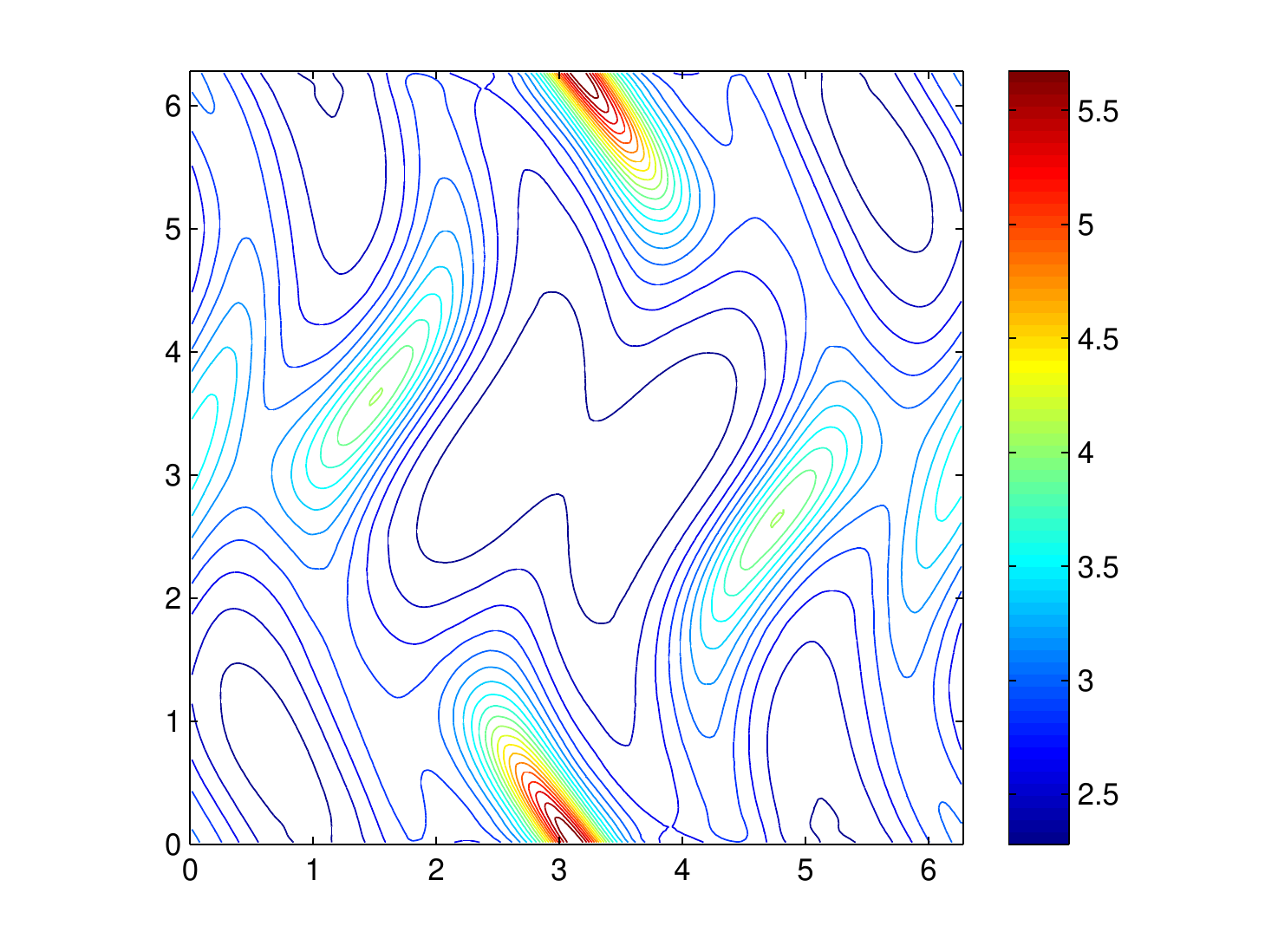}{a}
  \includegraphics[width=0.45\textwidth]{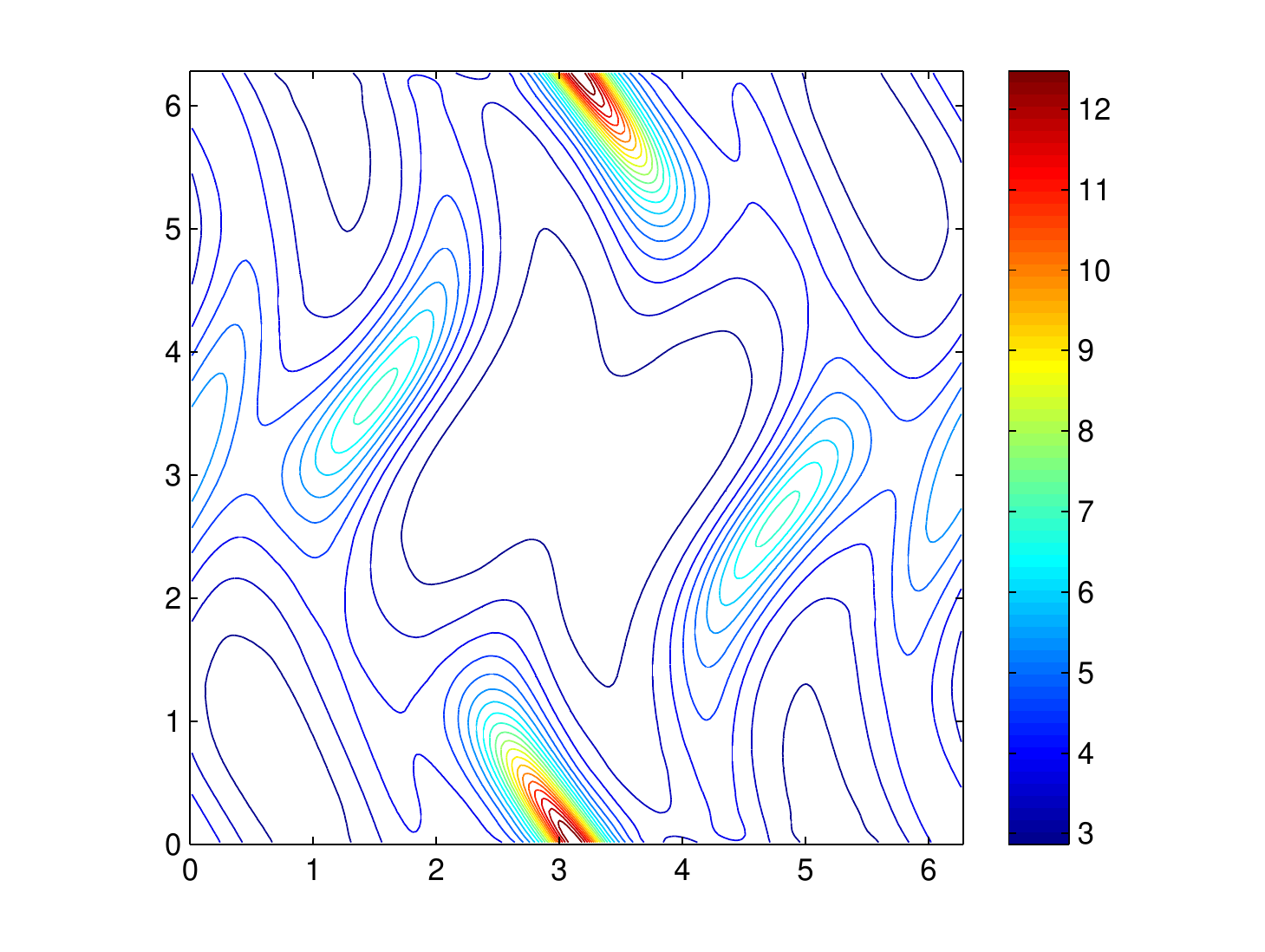}{b}
  \includegraphics[width=0.45\textwidth]{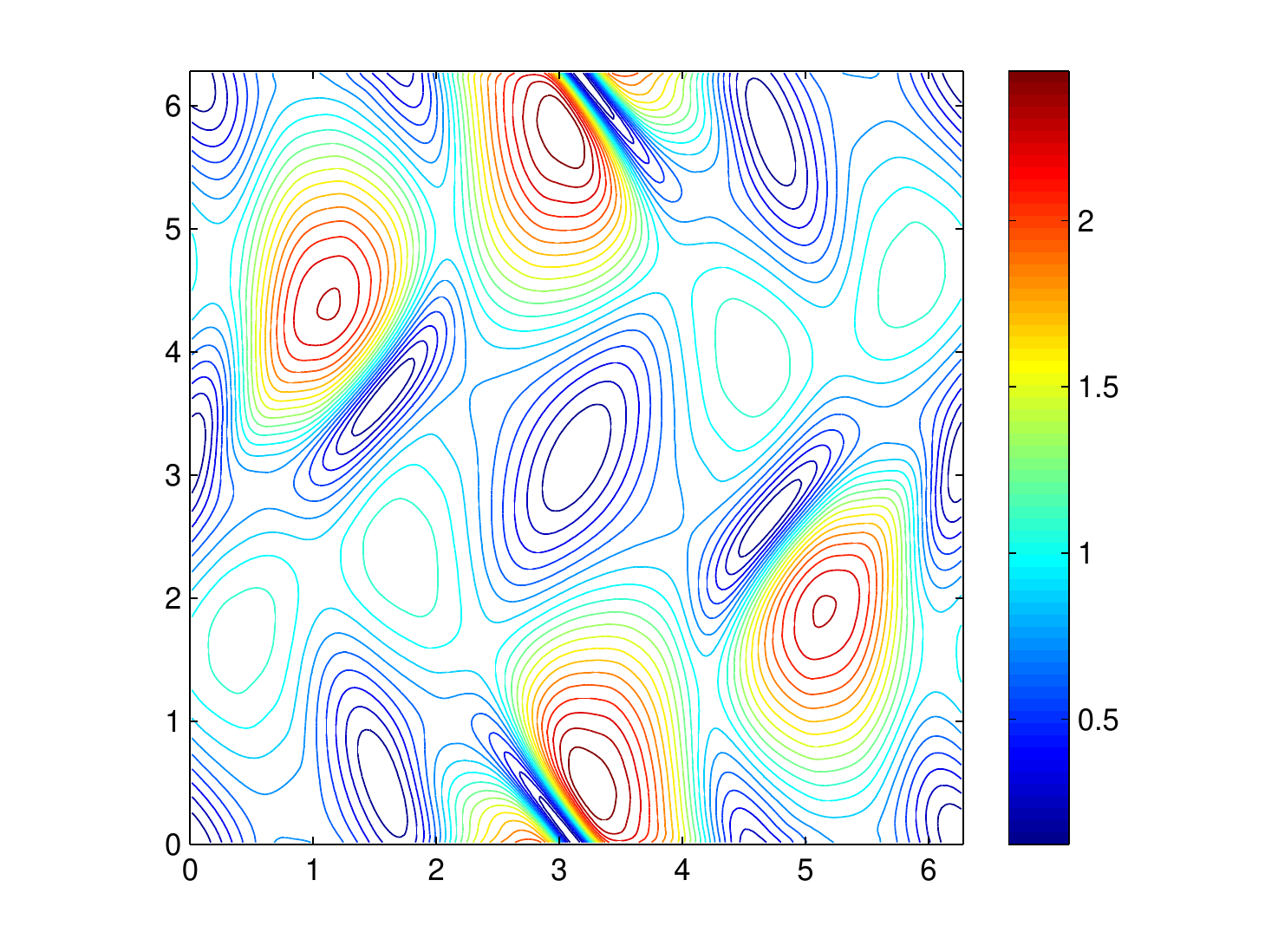}{c}
  \includegraphics[width=0.45\textwidth]{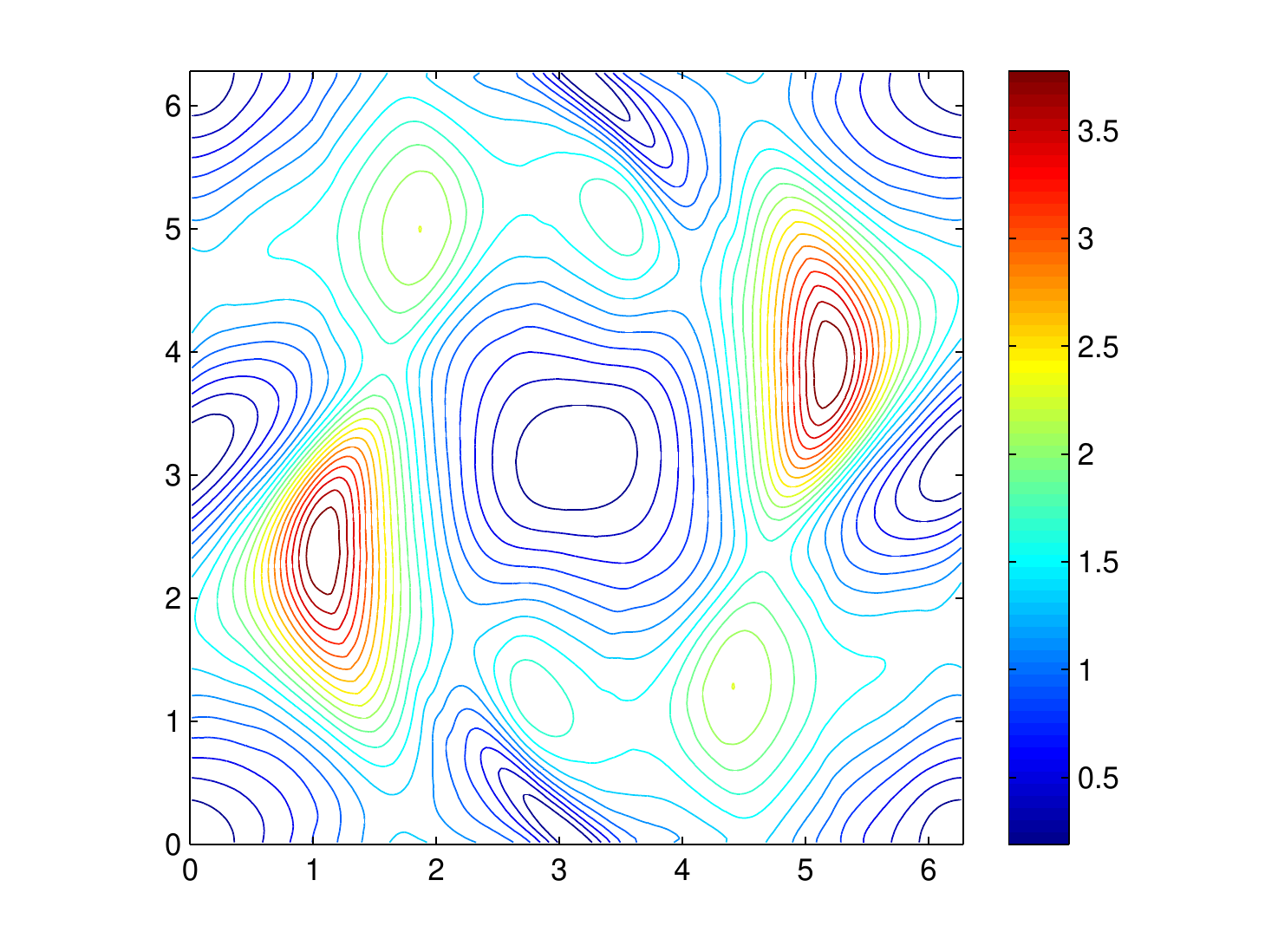}{d}
  \caption{The Orszag-Tang MHD turbulence problem ($r_{L_i}=0$) with a uniform mesh of $192 \times 192$ grid points. The
output time is $t=0.5$. (a) density; (b) gas pressure; (c) magnetic pressure; (d) kinetic energy.}
\label{orszag1}
\end{figure}

\begin{figure}
  \centering
  \includegraphics[width=0.45\textwidth]{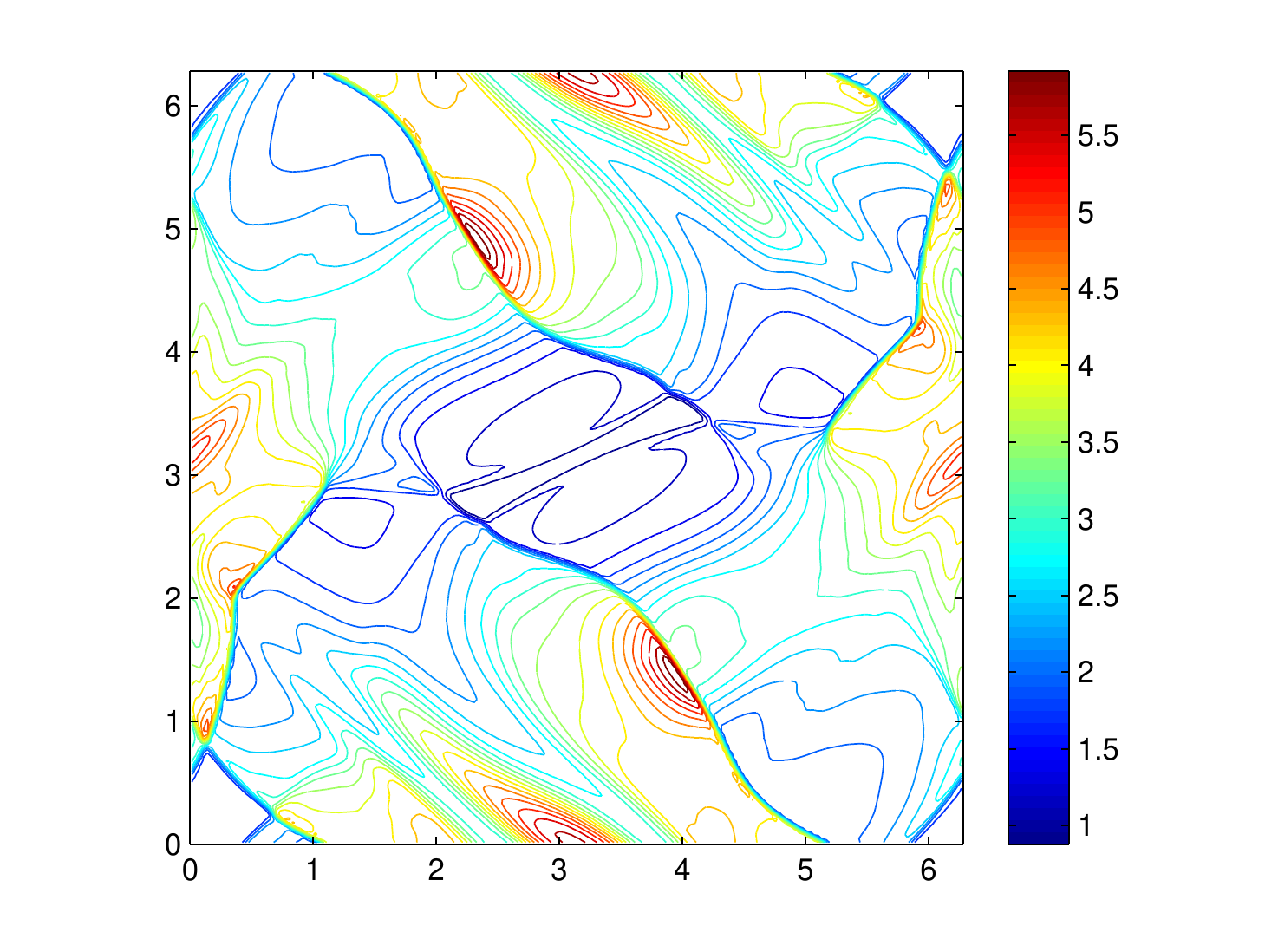}{a}
  \includegraphics[width=0.45\textwidth]{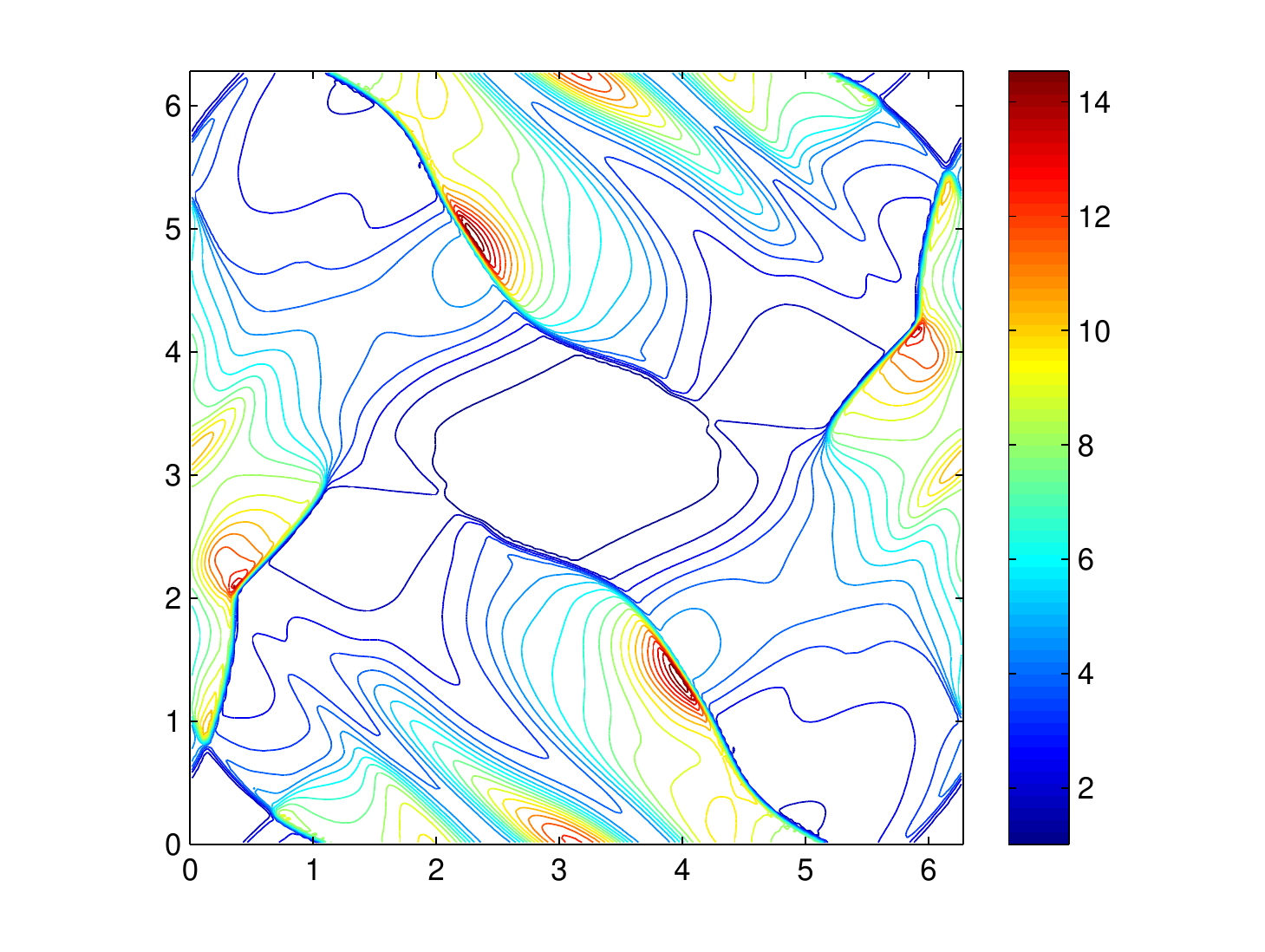}{b}
  \includegraphics[width=0.45\textwidth]{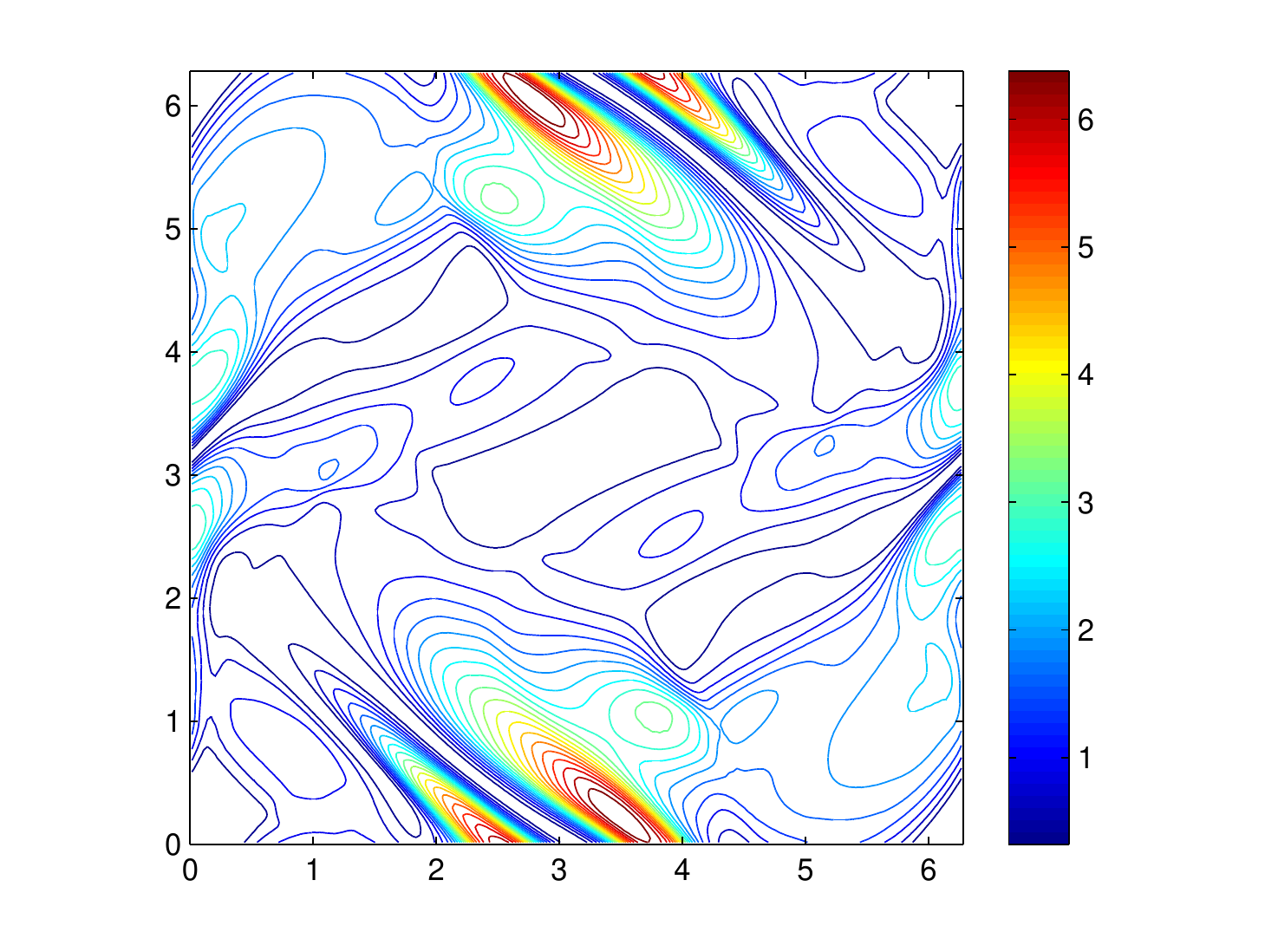}{c}
  \includegraphics[width=0.45\textwidth]{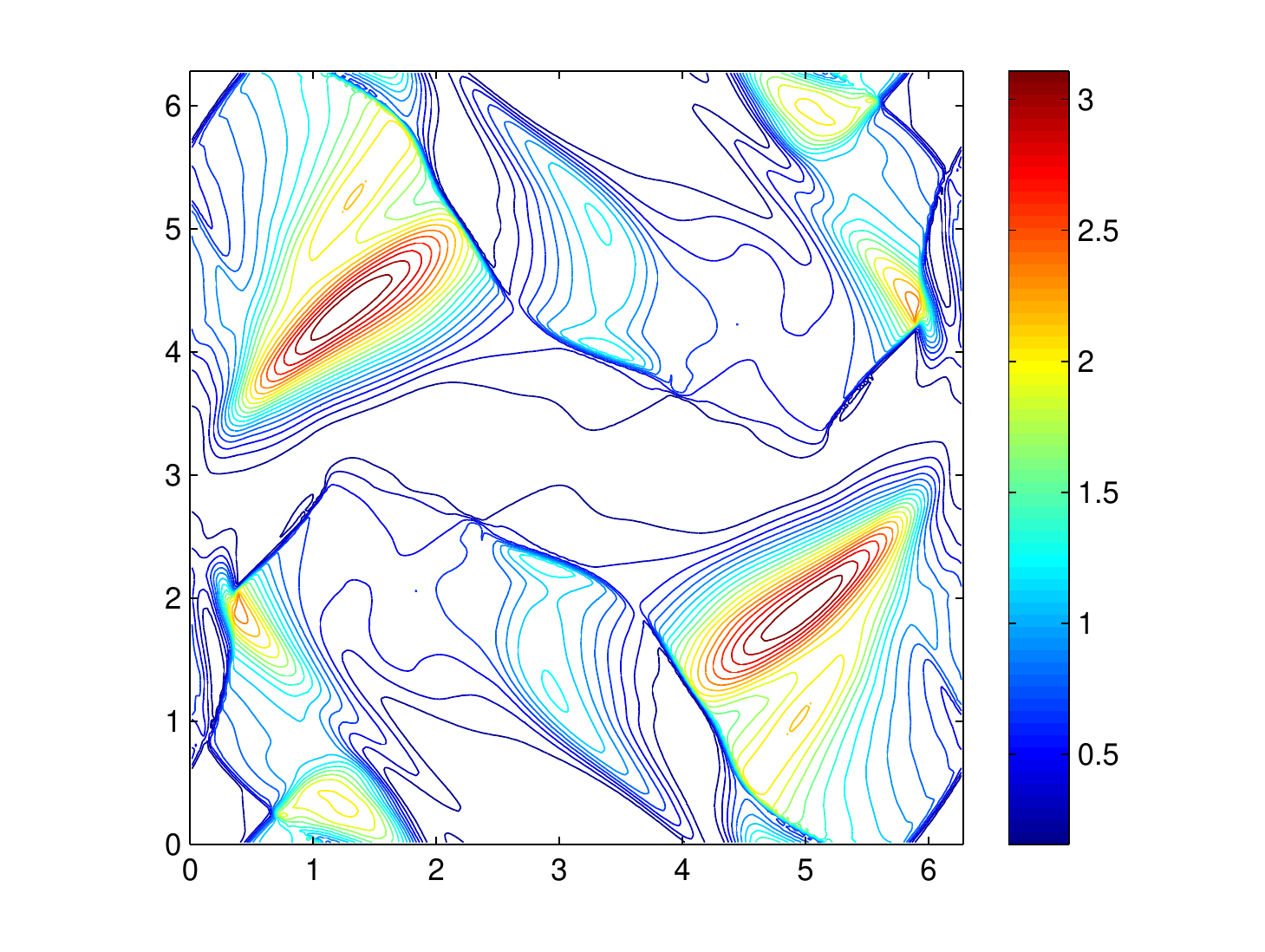}{d}
  \caption{The Orszag–-Tang MHD turbulence problem ($r_{L_i}=0$)
  at output time $t=2$. (a) density; (b) gas pressure; (c) magnetic pressure; (d) kinetic energy.}
  \label{orszag2}
\end{figure}

\begin{figure}
  \centering
  \includegraphics[width=0.45\textwidth]{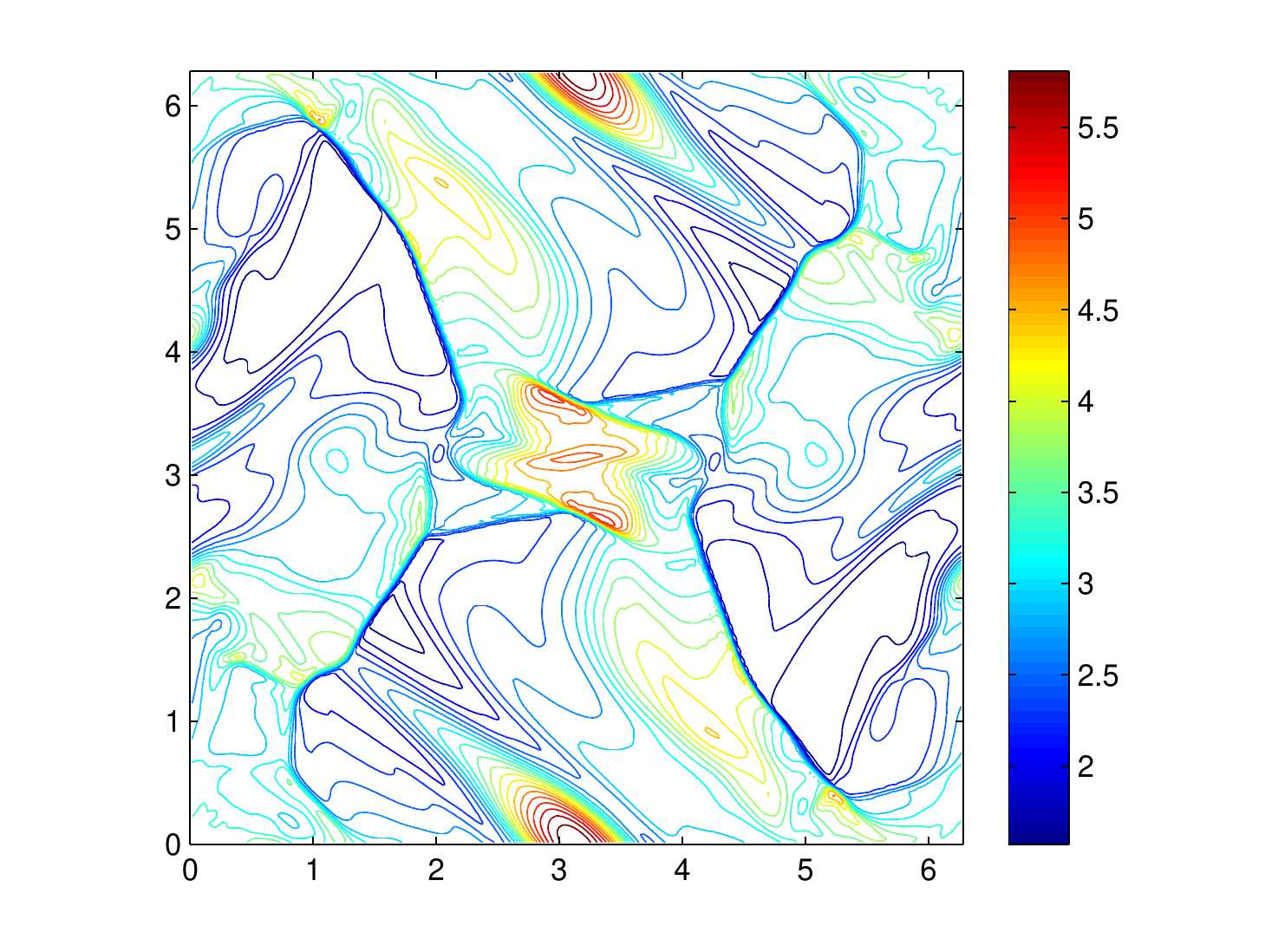}{a}
  \includegraphics[width=0.45\textwidth]{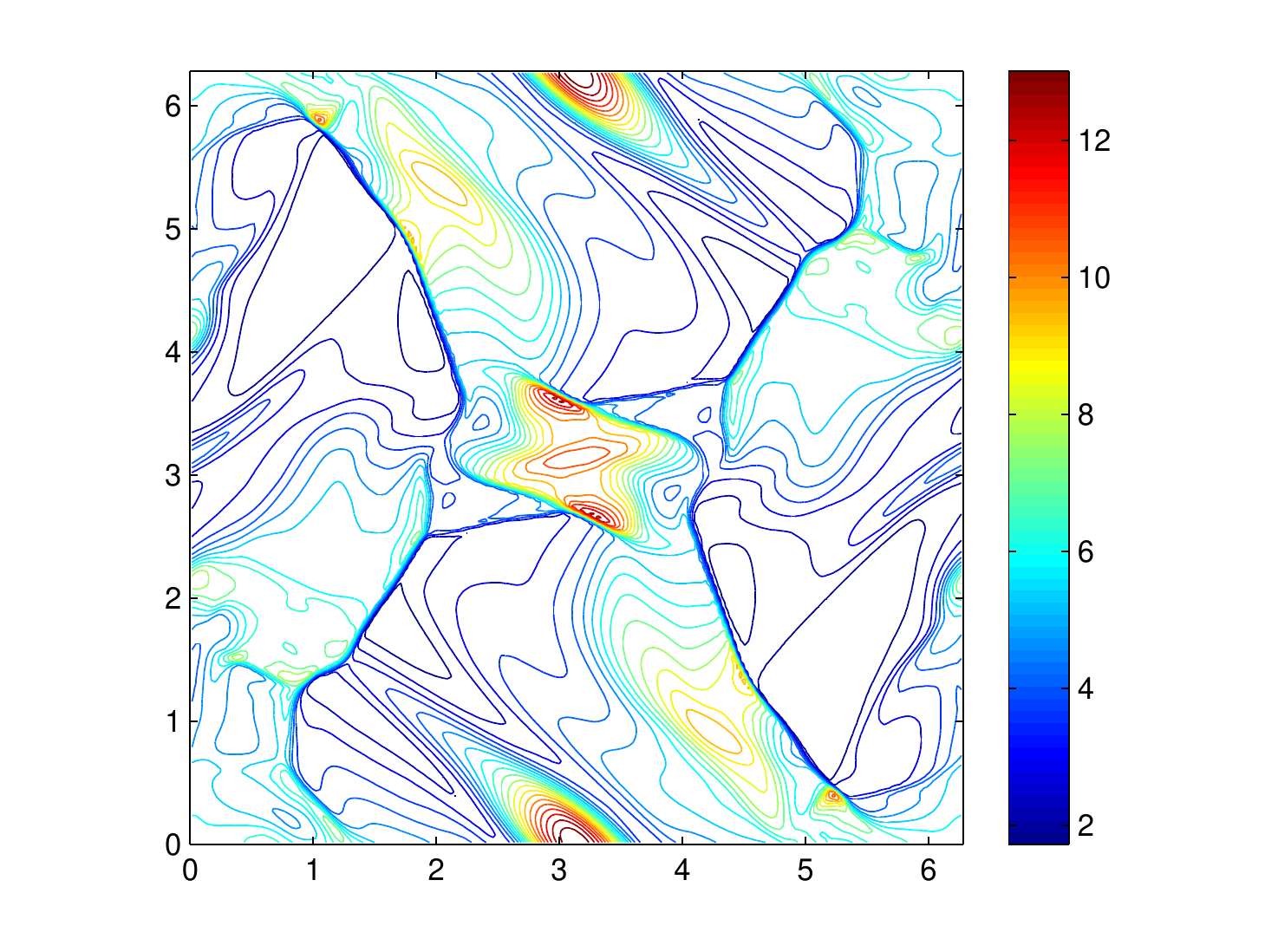}{b}
  \includegraphics[width=0.45\textwidth]{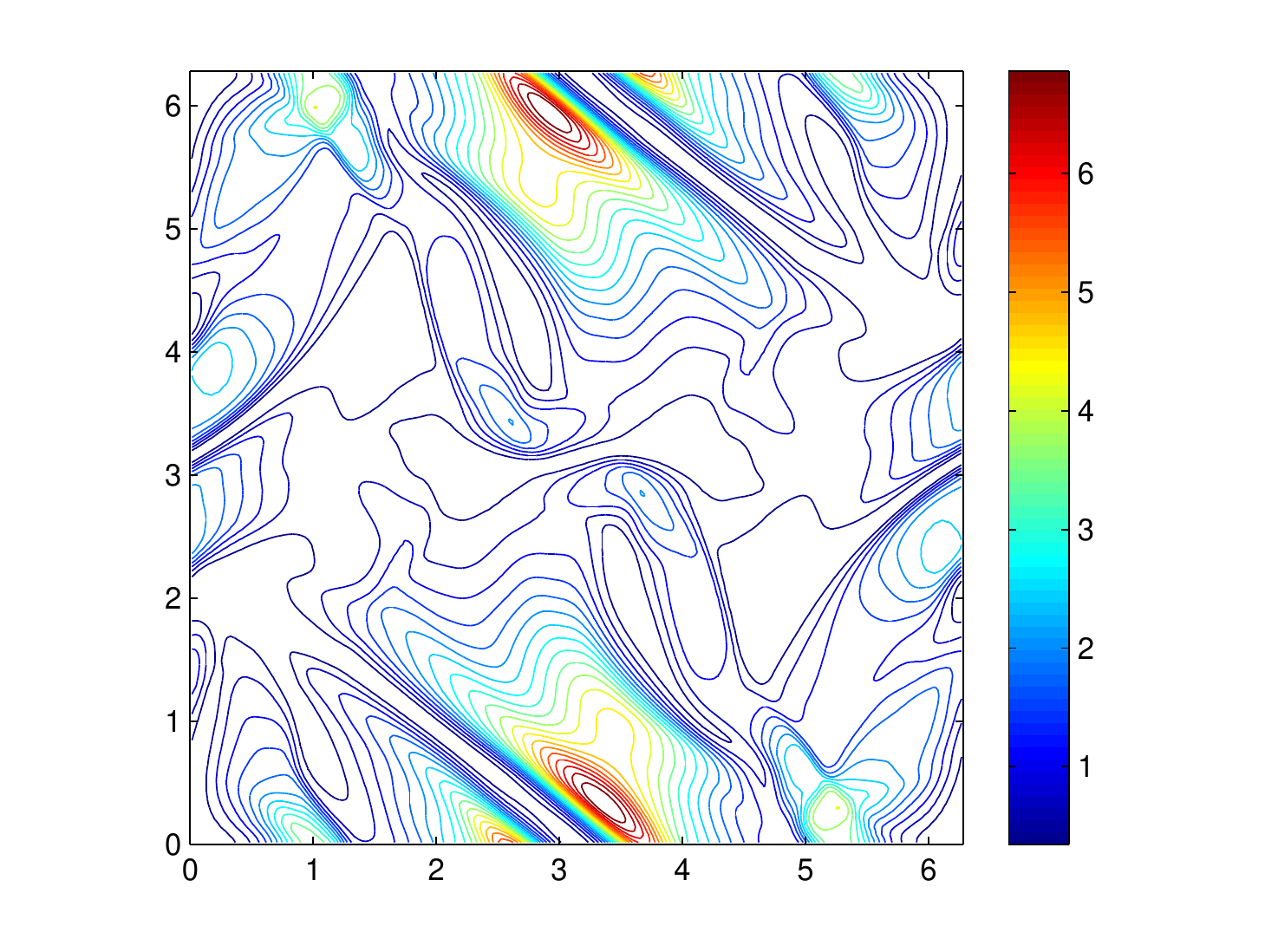}{c}
  \includegraphics[width=0.45\textwidth]{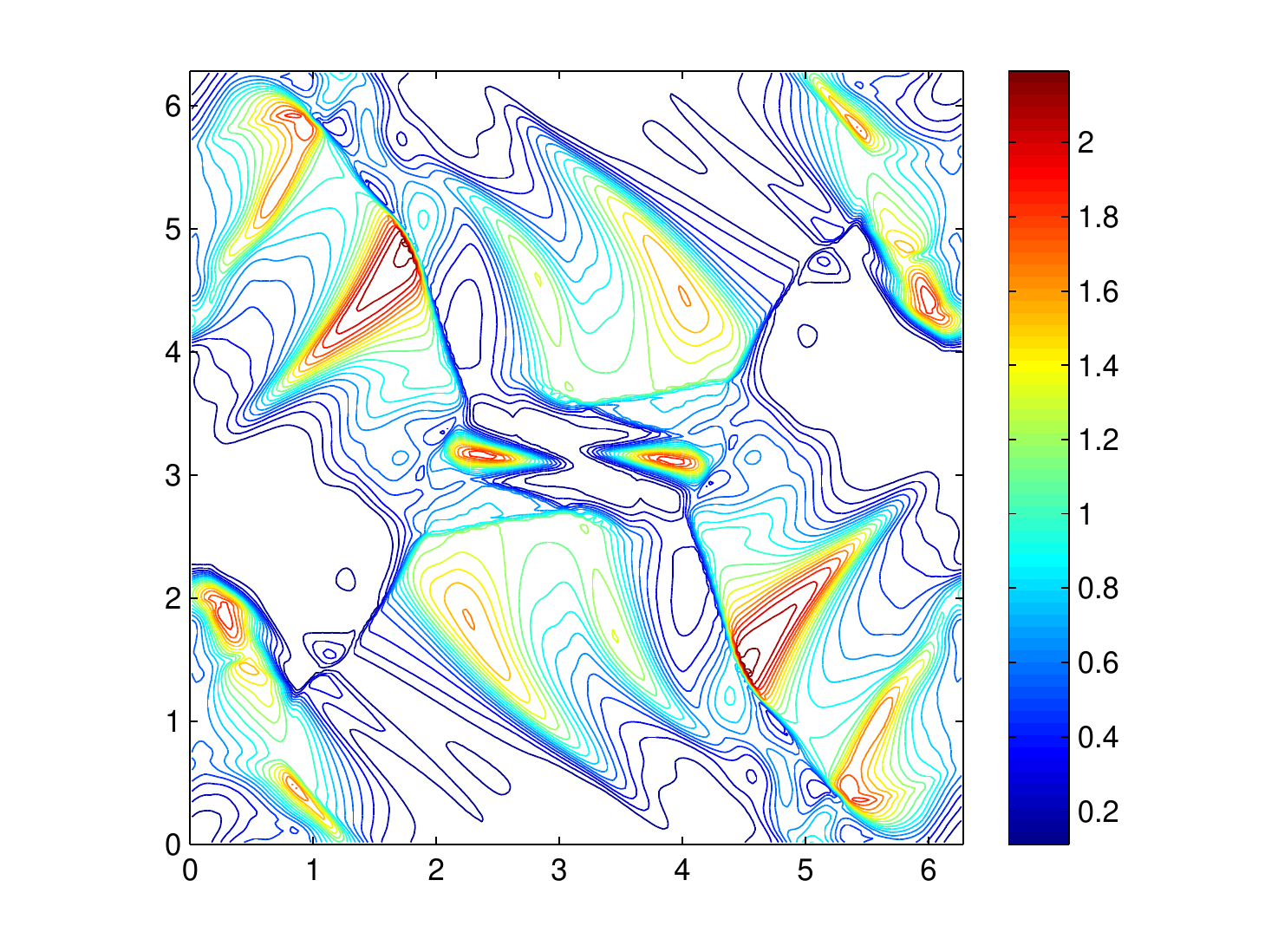}{d}
  \caption{The Orszag–-Tang MHD turbulence problem ($r_{L_i}=0$)
  at output time $t=3$. (a) density; (b) gas pressure; (c) magnetic pressure; (d) kinetic energy.}
  \label{orszag3}
\end{figure}

\begin{figure}
  \centering
  \includegraphics[width=0.45\textwidth]{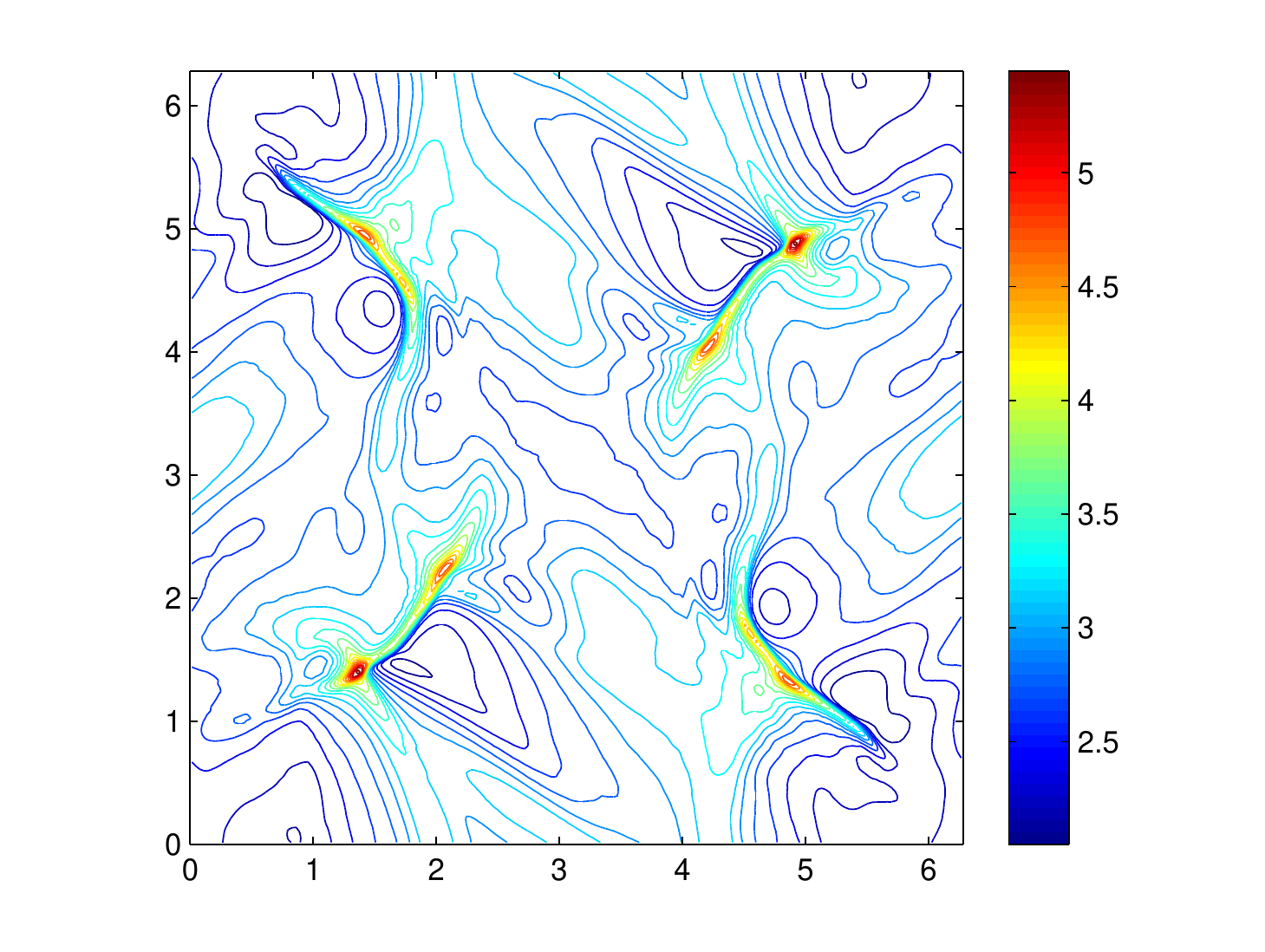}{a}
  \includegraphics[width=0.45\textwidth]{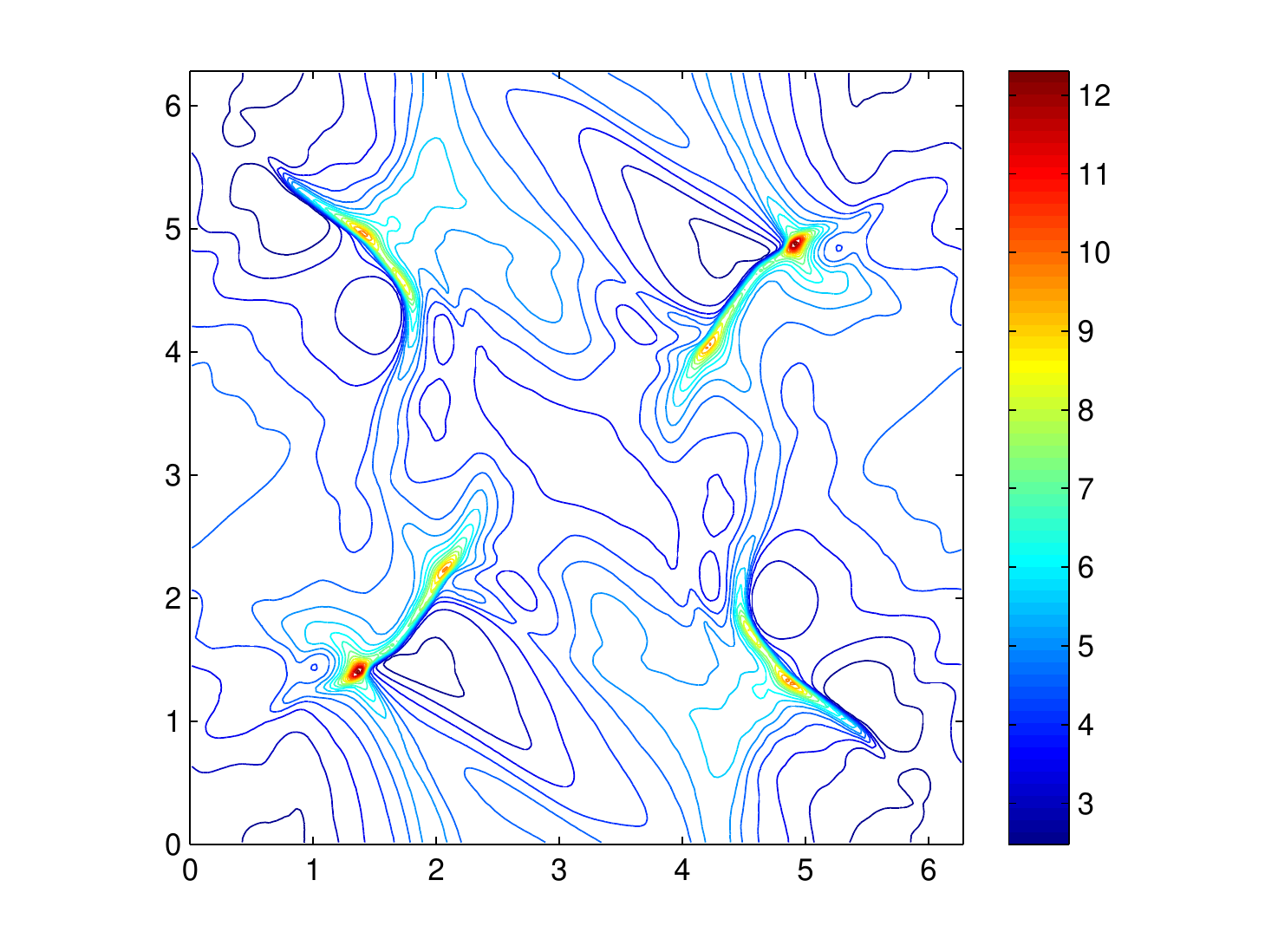}{b}
  \includegraphics[width=0.45\textwidth]{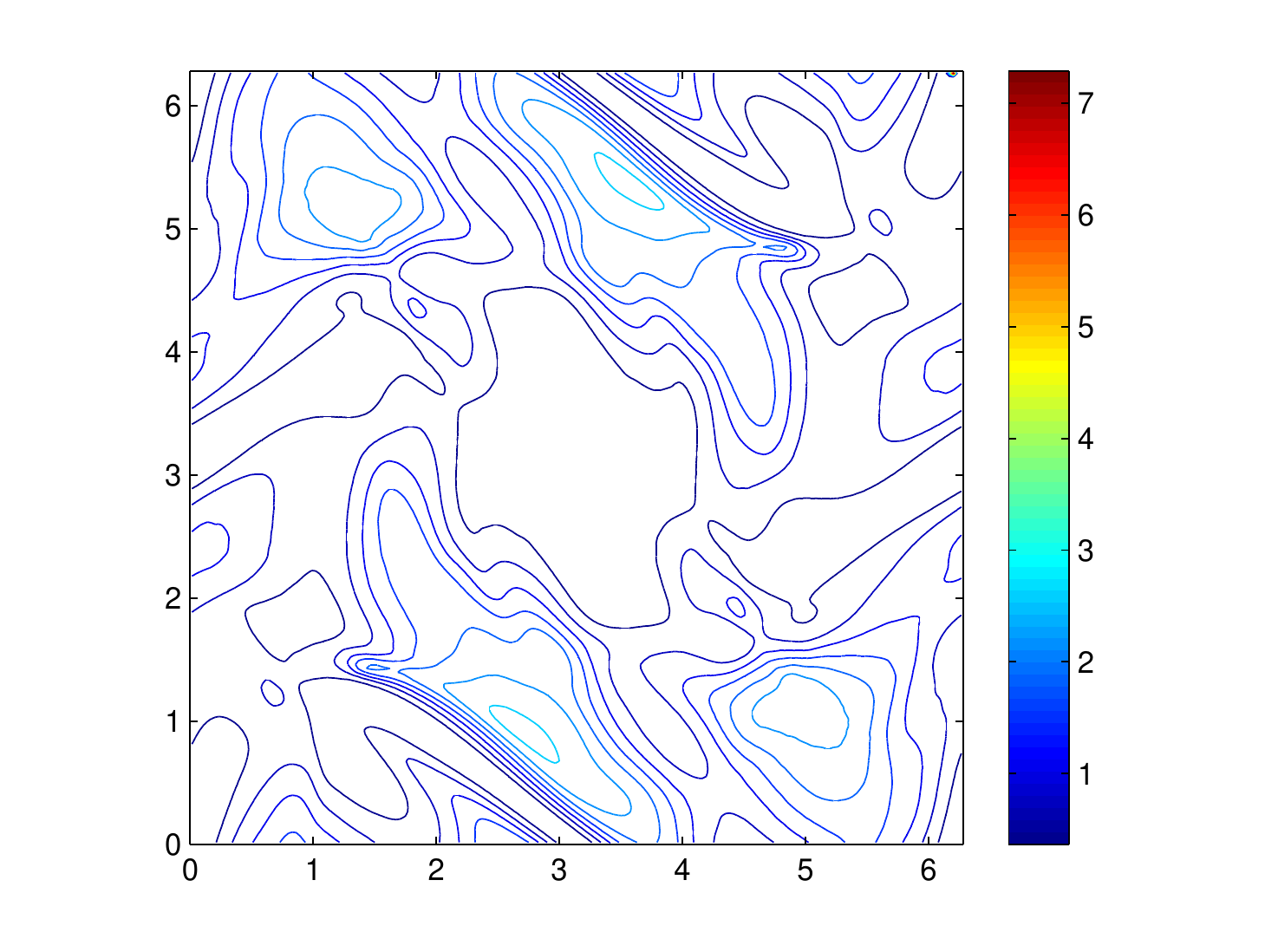}{c}
  \includegraphics[width=0.45\textwidth]{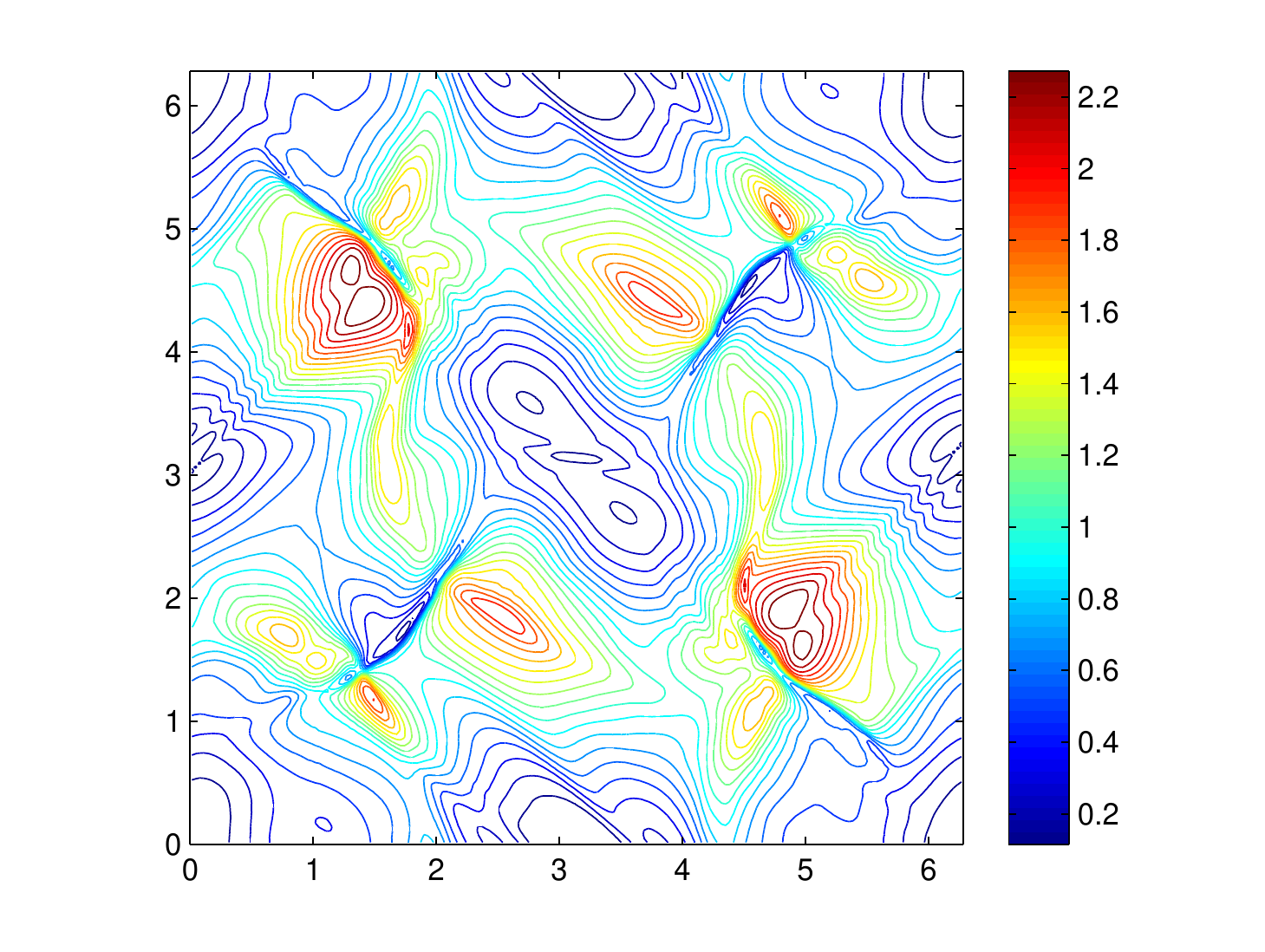}{d}
  \caption{The Orszag–-Tang MHD turbulence problem ($r_{L_i}=1.0$) at output time $t=3$.
  (a) density; (b) gas pressure; (c) magnetic pressure; (d) kinetic energy.}
  \label{orszag4}
\end{figure}

\begin{figure}
  \centering
  \includegraphics[width=0.45\textwidth]{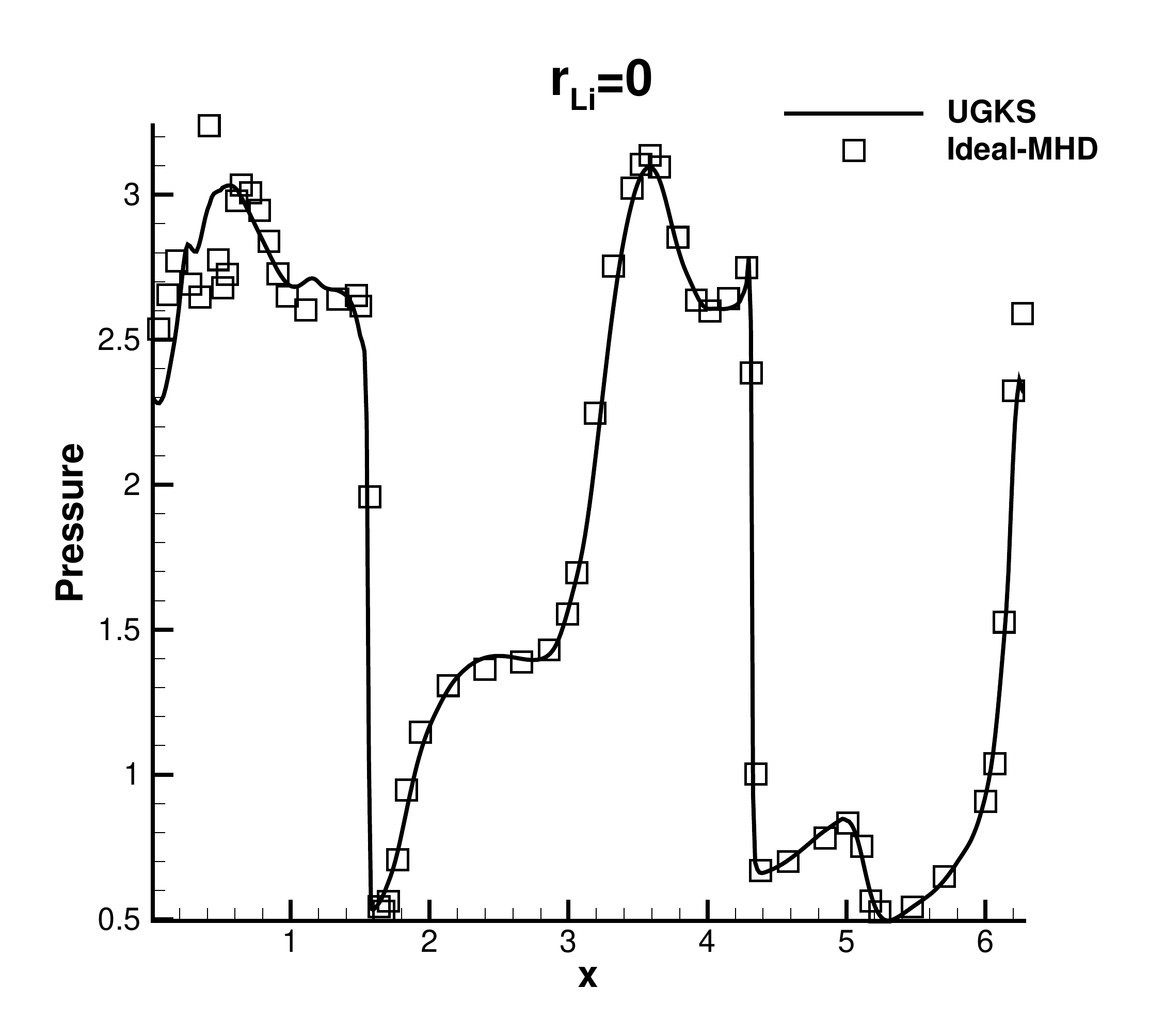}(a)
  \includegraphics[width=0.45\textwidth]{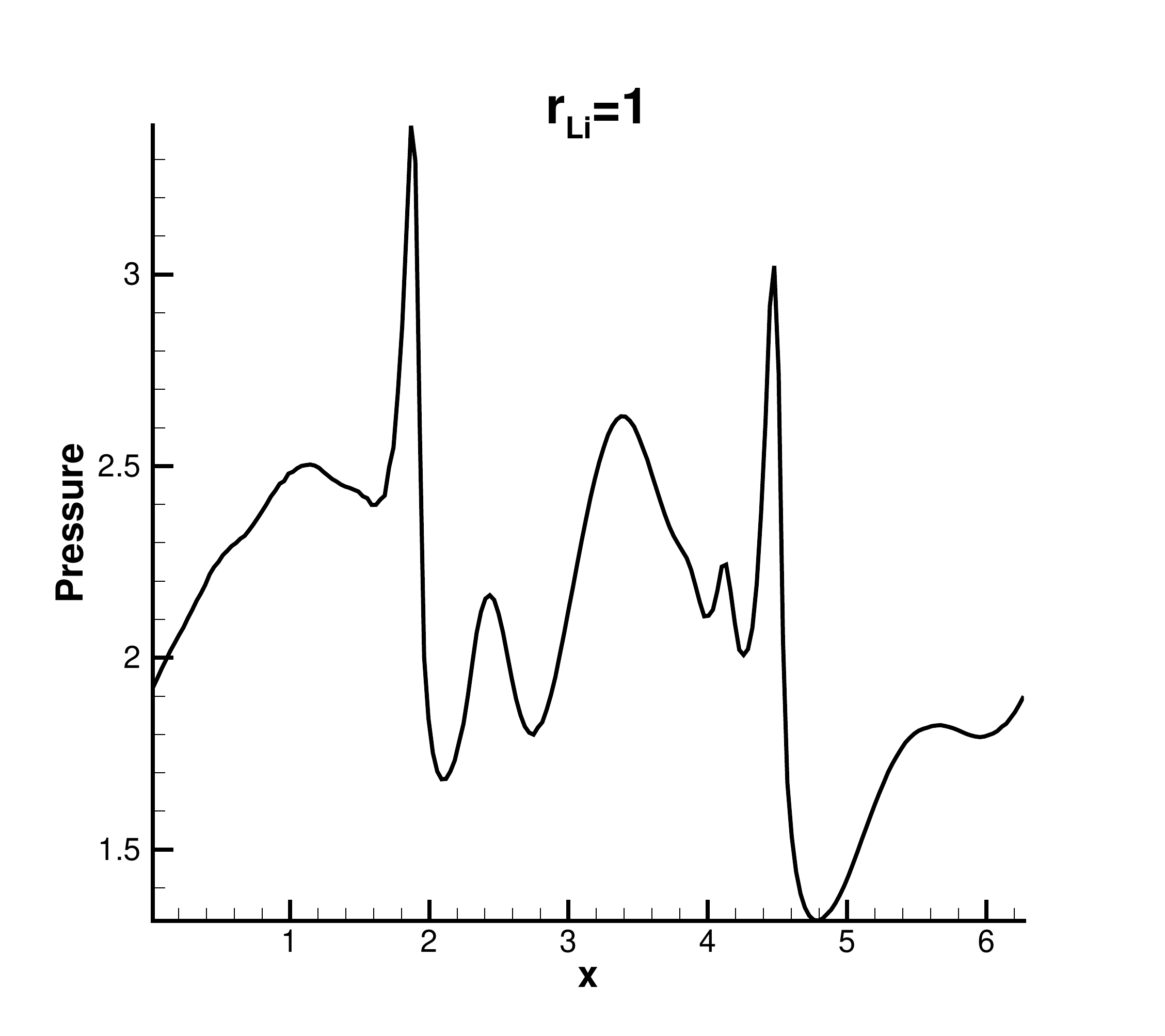}(b)
  \caption{Pressure distribution along the line $y=0.625\pi$: (a) $r_{L_i}=0$, UGKS solution and ideal-MHD solution \cite{tang2000high}. (b) $r_{L_i}=1.0$, UGKS solution only.}
\label{orszag5}
\end{figure}

This problem was introduced by Orszag and Tang as a simple model to study MHD turbulence \cite{orszag,tang2000high}. The mass ratio is set as $m_i/m_e=25$, the relaxation parameter $\tau=10^{-5}$ and the Larmor radius is set to be zero for ideal MHD solutions first.
The initial data for the current study is
\begin{equation}\nonumber
\begin{aligned}
&n_i=n_e=\gamma^2, P_i=P_e=\gamma, \ B_y=\sin(2x),\\ &u_{i,x}=u_{e,x}=-\sin(y), \ u_{i,y}=u_{e,y}=\sin(x),
\end{aligned}
\end{equation}
where $\gamma=5/3$.
The interspecies collision factor $\nu_{ie}$ is set to be zero.
The computation domain is $[0,2\pi]\times[0,2\pi]$ with a uniform mesh of $200\times200$ cells.
The velocity space for ion is $[-3, 3]$ and for electron is $[-5\sqrt{5}, 5\sqrt{5}]$ with $32\times32$ velocity grids.
Periodic boundary conditions are imposed in both x and y-directions.
Various values of Larmor radius $r_{L_i}=0, 0.5, 1.0, 2.0$ are used in our calculation.
The total density, total pressure, magnetic pressure, and total kinetic energy distributions
at output times $0.5, 2, 3$ for $r_{L_i}=0$
is shown in Fig. \ref{orszag1}-\ref{orszag3}.
Fig. \ref{orszag4} shows the results for $r_{L_i}=1.0$ at $t=3$. In Fig. \ref{orszag5}, we plot the pressure distribution along $y=0.625\pi$ for $r_{L_i}=1.0$ and $r_{L_i}=0$ cases,
and compare the results of $r_{L_i}=0$ with the ideal MHD solution.

\begin{figure}
  \centering
  \includegraphics[width=0.45\textwidth]{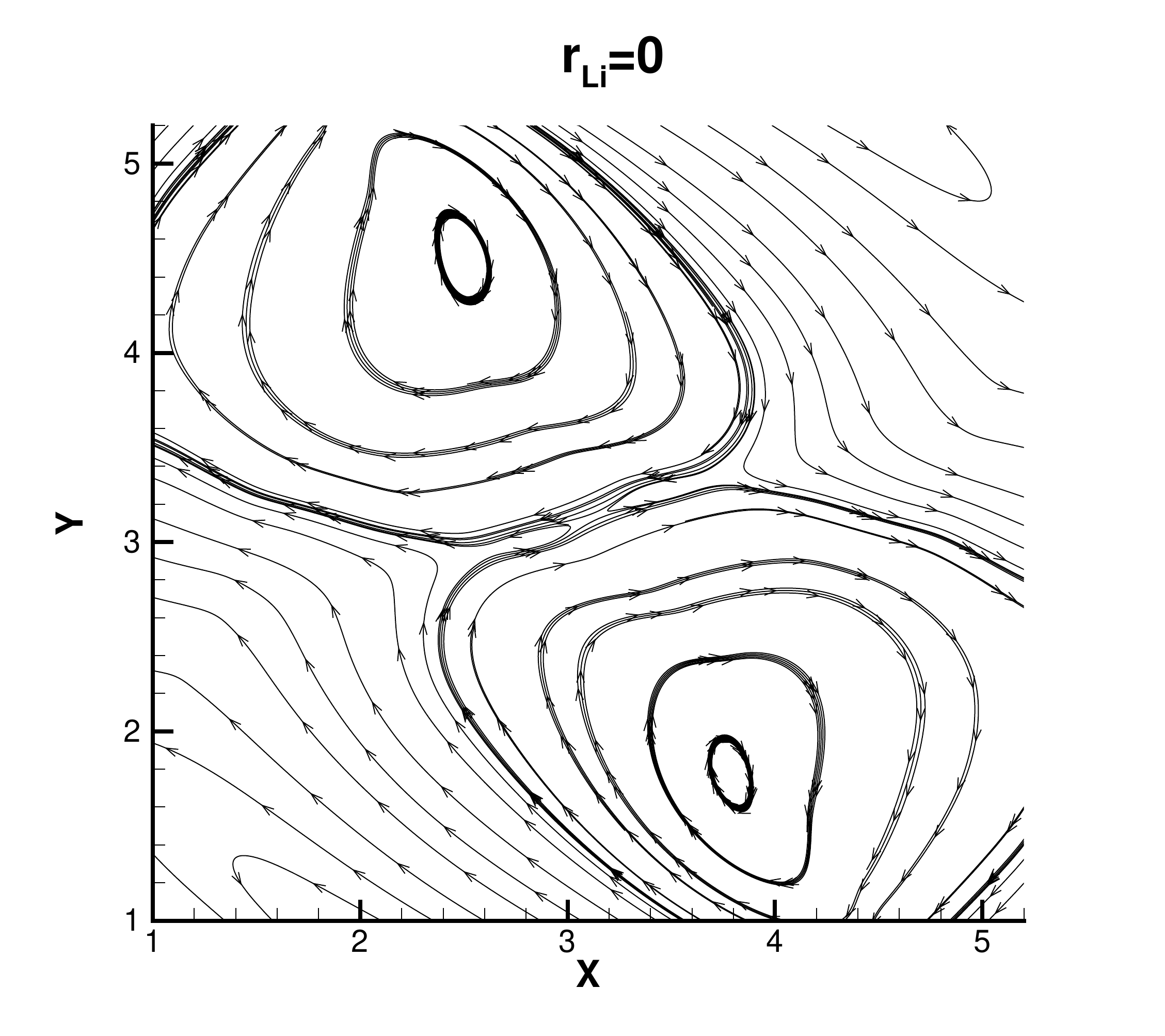}{a}
  \includegraphics[width=0.45\textwidth]{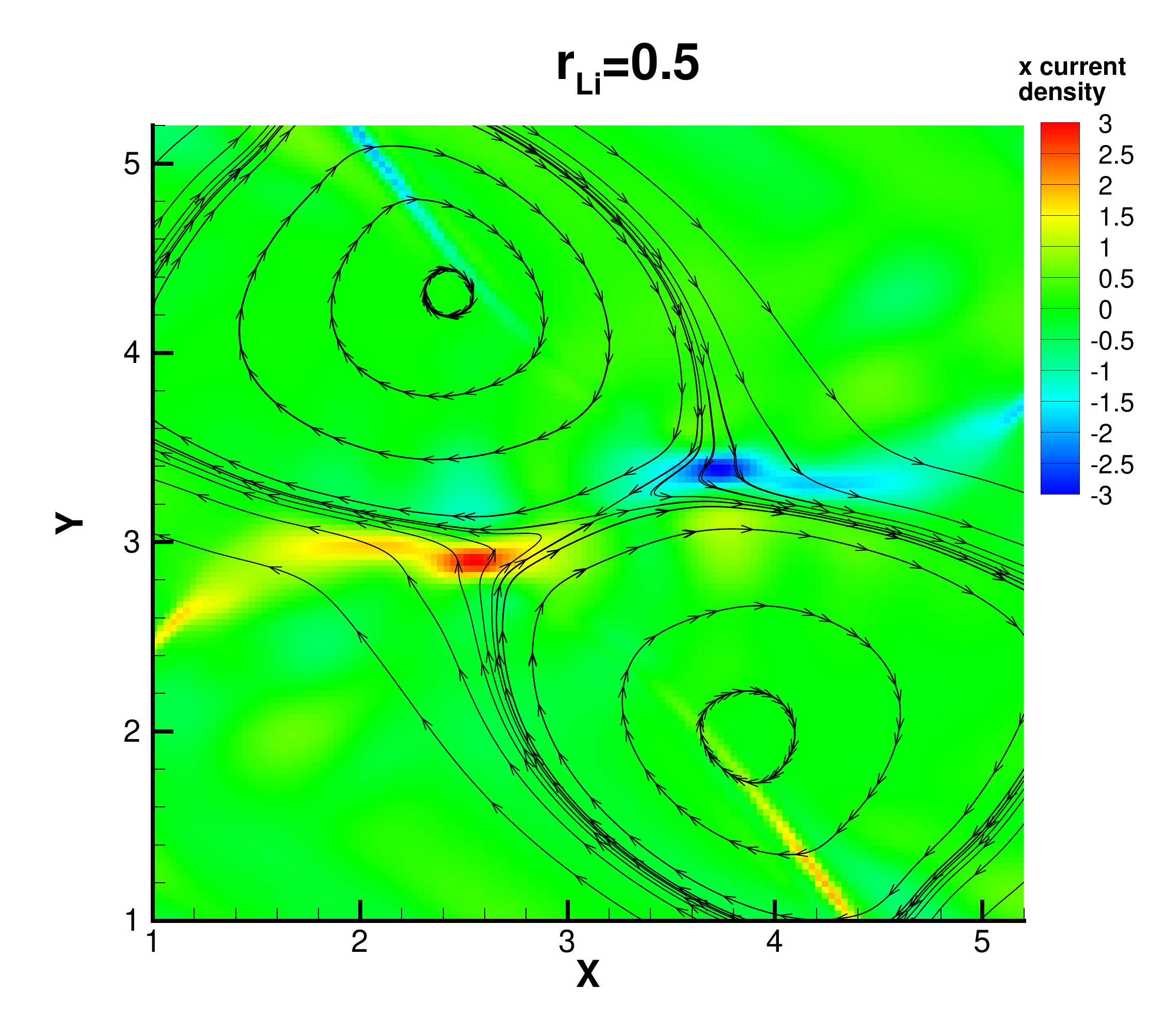}{b}
  \includegraphics[width=0.45\textwidth]{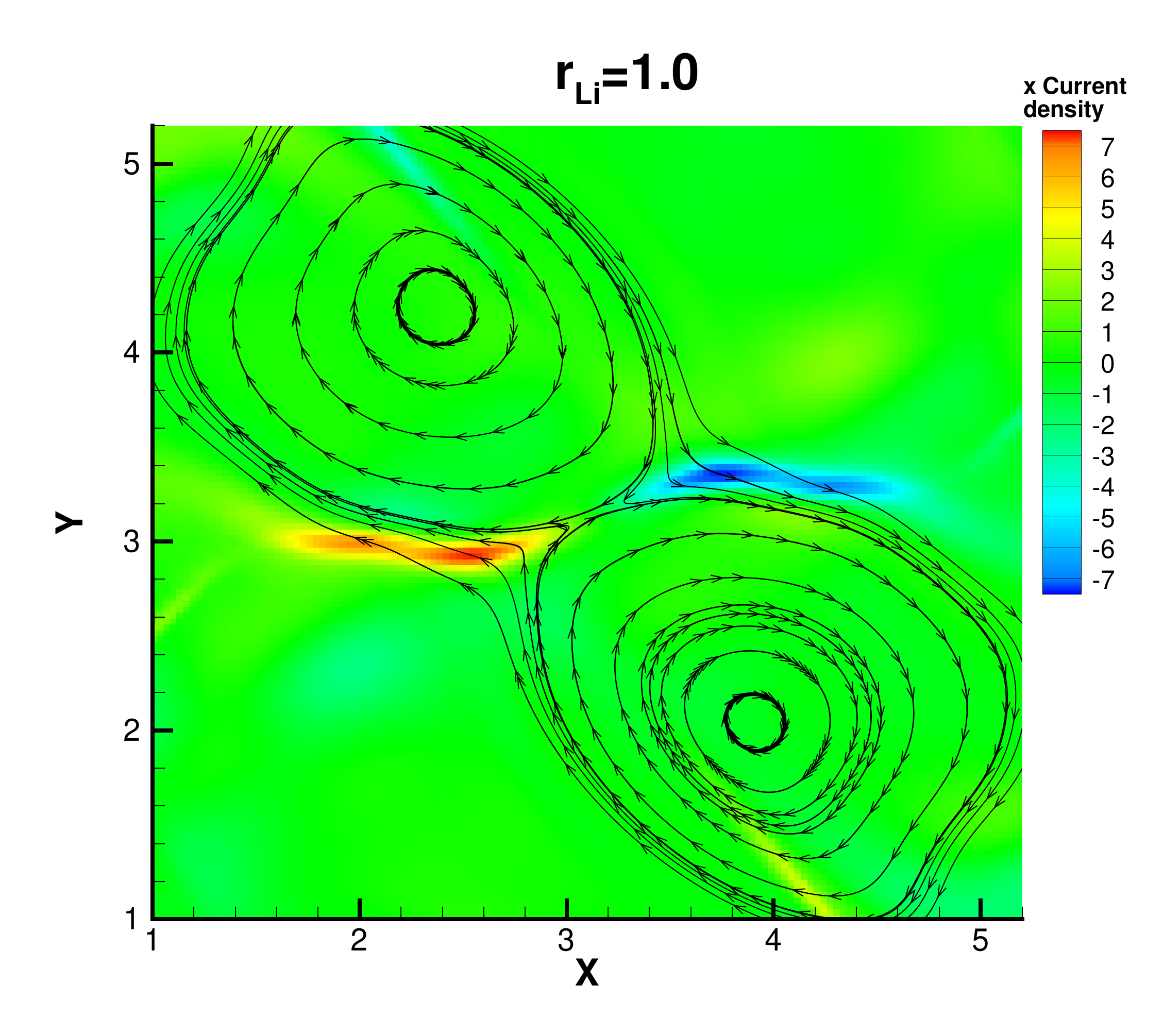}{c}
  \includegraphics[width=0.45\textwidth]{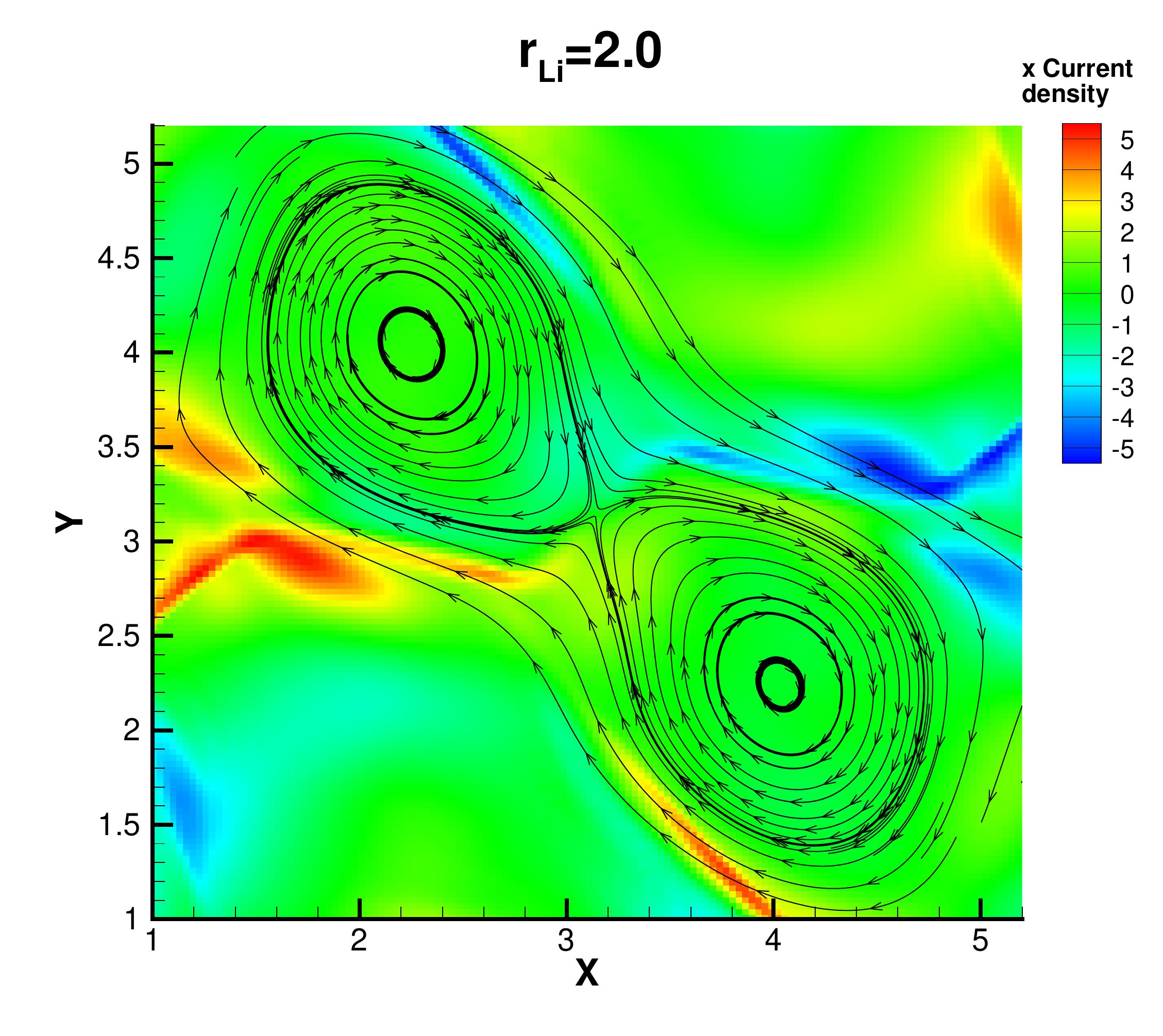}{d}
  \caption{The topology of magnetic lines near the center of computational domain
   at $t=2$ with (a) $r_{L_i}=0$,(b) $r_{L_i}=0.5$,(c) $r_{L_i}=1.0$
  and (d) $r_{L_i}=2.0$. The electron current sheets during the reconnection process are shown in (b), (c), and (d).}
  \label{orszag6}
\end{figure}

The magnetic reconnection happens near the center of the computational domain as shown in Fig. \ref{orszag6},
which merges two 'magnetic rings' into a single one with the time evolution.
The magnetic reconnection mechanism is different for different $r_{L_i}$, as shown in Fig. \ref{orszag6}.
For $r_{L_i}=0$ case, the aspect ratio of the reconnection layer is large where a double Y-point geometry is observed.
In this case, the reconnection is driven by the magnetic diffusion,
following the mechanism described by the Sweet-Parker model \cite{Parker1957}.
For $r_{L_i}=0.5, 1.0$ cases, the Hall effect shows up and the length of reconnection layer gets shorter.
Electron current sheet is observed along the reconnection layer.
For $r_{L_i}=2.0$ case, the Hall effect becomes dominant and an X-point geometry is observed.
The simulation shows that the reconnection rate is increased with a higher energy transfer efficiency as $r_{L_i}$ increases.

\subsection{Magnetic Reconnection}

Magnetic reconnection is a process in which the topology of the magnetic field lines changes \cite{geospace}.
In ideal MHD, the magnetic field lines cannot be changed as the field lines are 'frozen' into the fluid. Various models were used to describe this phenomenon, for example the electron MHD \cite{hesse}, MHD and Hall MHD \cite{birn2001geospace, ma2001hall}, full particle \cite{pritchett}, and hybrid model \cite{kuznetsova}. It was found that the reconnection initiates at a length scale on the order of the electron skin depth and the reconnection rate is governed by the ion dynamics. Our scheme is based on the Vlasov-BGK equation which can describe the physics at electron skin depth level. Hence it can be used to describe the reconnection process.

\begin{figure}[t!]
\centering
\includegraphics[width=0.6\textwidth]{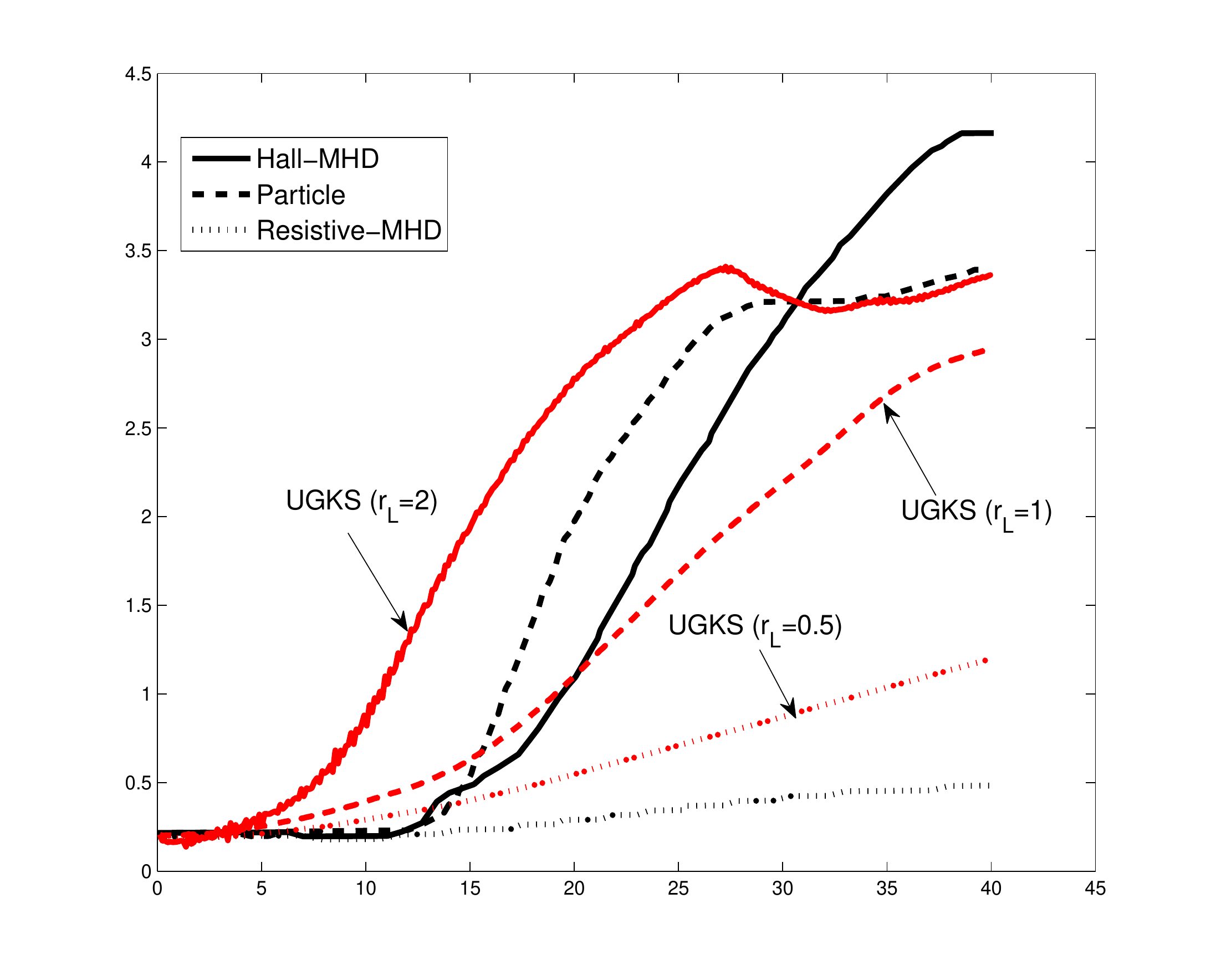}
\caption{The reconnected flux $\phi(t)$ from the UGKS and other GEM simulations.}
\label{reconnection}
\end{figure}

The simulation uses the same  initial conditions as the GEM challenge problem \cite{hakim2006high}.
The initial magnetic field is given by
\begin{equation}\nonumber
  \mathbf{B}(y)=B_0\tanh(y/\lambda)\mathbf{e_x},
\end{equation}
and a corresponding current sheet is carried by the electrons
\begin{equation}\nonumber
  \mathbf{J}_e=-\frac{B_0}{\lambda} \text{sech} ^2(y/\lambda)\mathbf{e_z}.
\end{equation}
The initial number densities of electron and ion are
\begin{equation}\nonumber
  n_e=n_i=1/5+\text{sech}^2(y/\lambda).
\end{equation}
The electron and ion pressures are set to be
\begin{equation}\nonumber
  P_i=5P_e=\frac{5B_0}{12}n(y),
\end{equation}
where $B_0=0.1$, $m_i=25m_e$ and $\lambda=0.5$.
The computational domain is $[-L_x/2, L_x/2]\times[-L_y/2,L_y/2]$ with $L_x=8\pi$, $L_y=4\pi$, which is divided into $200\times 100$ cells.
To initiate reconnection, the magnetic field is perturbed with $\delta \mathbf{B}=\mathbf{e_z}\times\nabla_\mathbf{x}\psi$, where
\begin{equation}\nonumber
  \psi(x,y)=0.1B_0\cos(2\pi x/L_x)\cos(\pi y/L_y).
\end{equation}
The velocity space for ion is $[-3, 3]$ and for electron is $[-25, 25]$ with $32\times32$ velocity grids.
The computational time for UGKS is about $1342$ mins on a 3.40GHz 4-core CPU.

\begin{figure}[t!]
\centering
\includegraphics[width=0.75\textwidth]{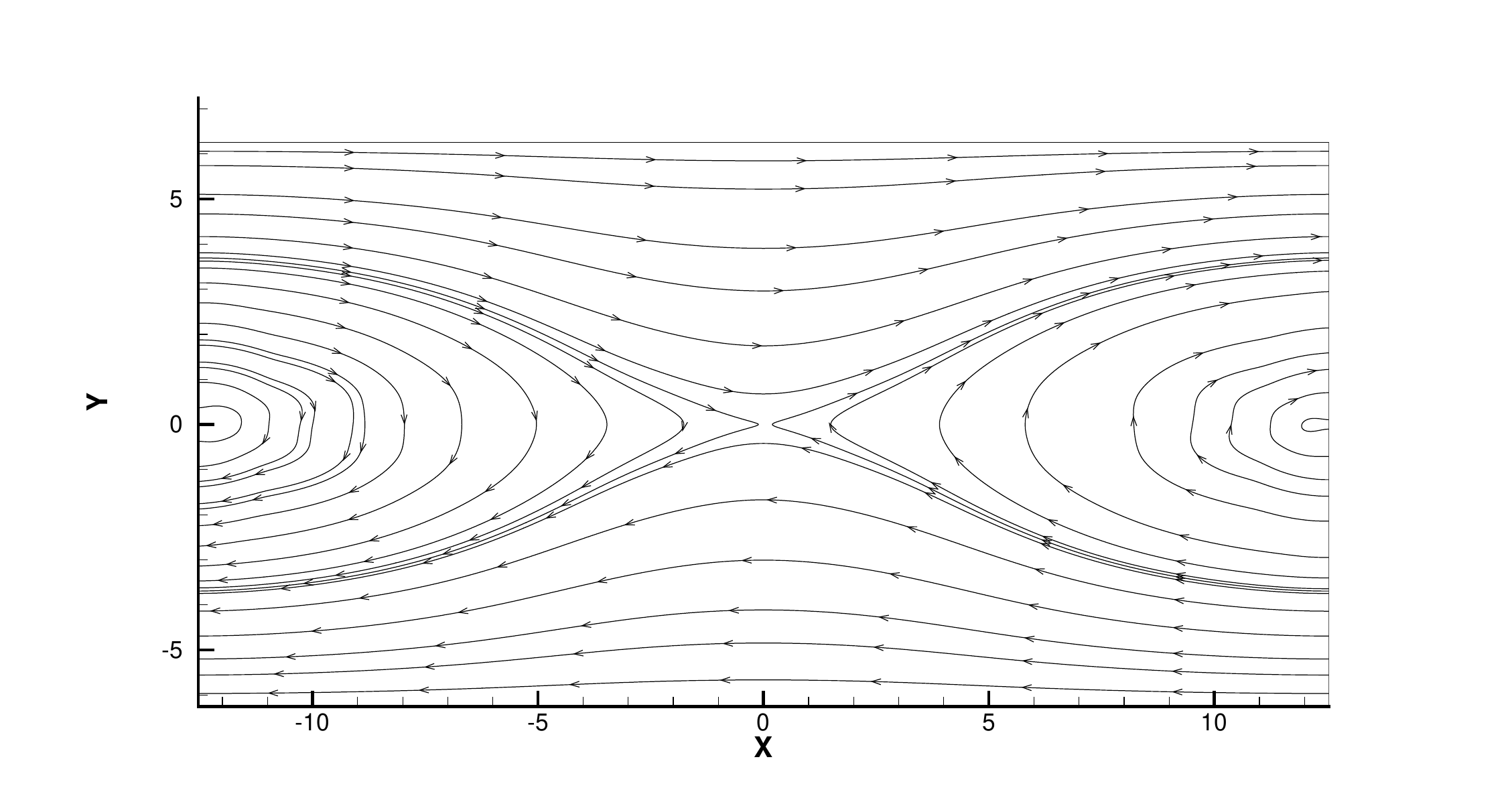}
\caption{Magnetic flux at $\omega t=40$ with $r_L=1$.}
\label{reconnection1}
\end{figure}

\begin{figure}
\centering
\includegraphics[width=0.5\textwidth]{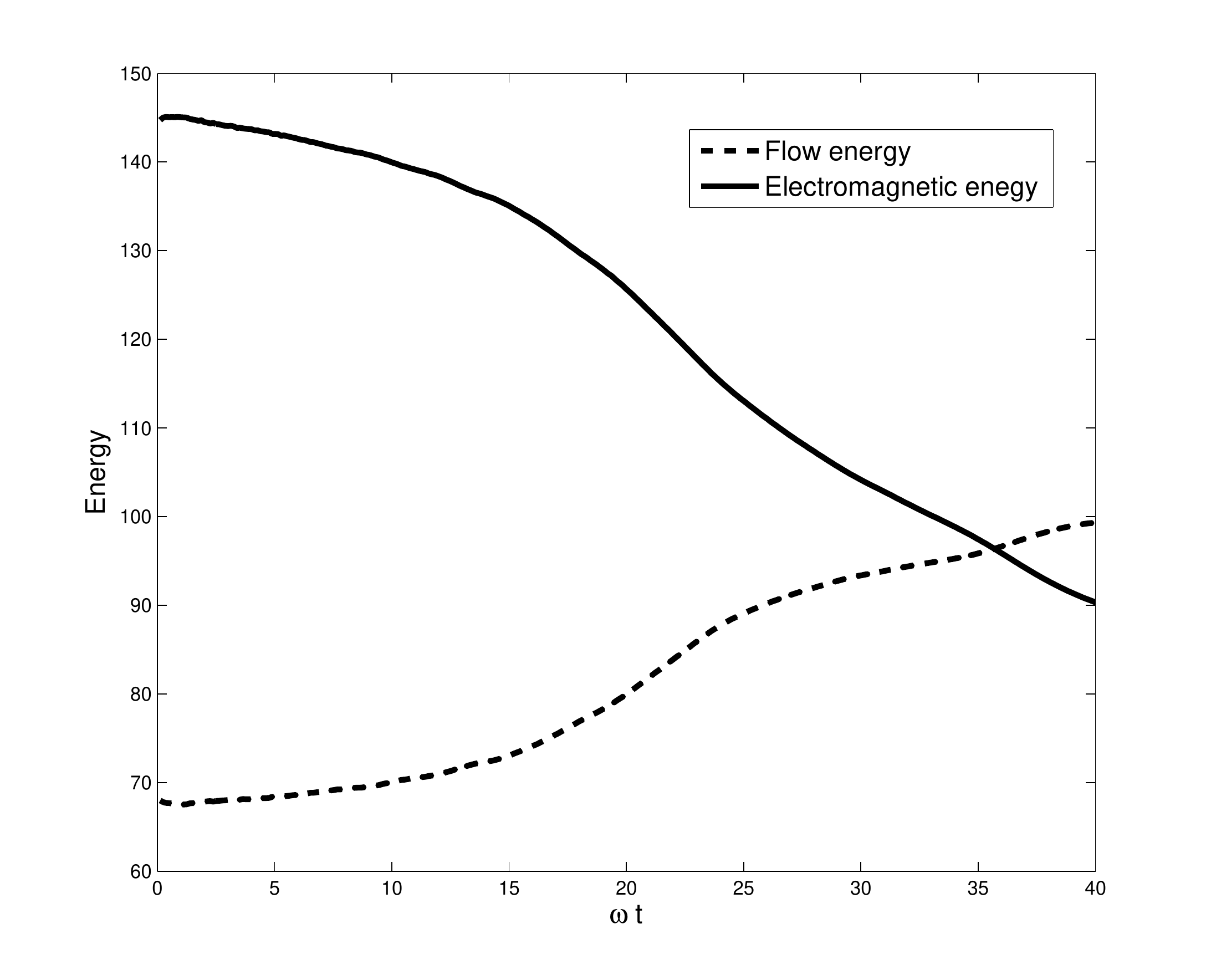}
\caption{Electromagnetic and flow energy evolution in the magnetic  reconnection process.}
\label{reconnection2}
\end{figure}

\begin{figure}
\centering
\includegraphics[width=0.45\textwidth]{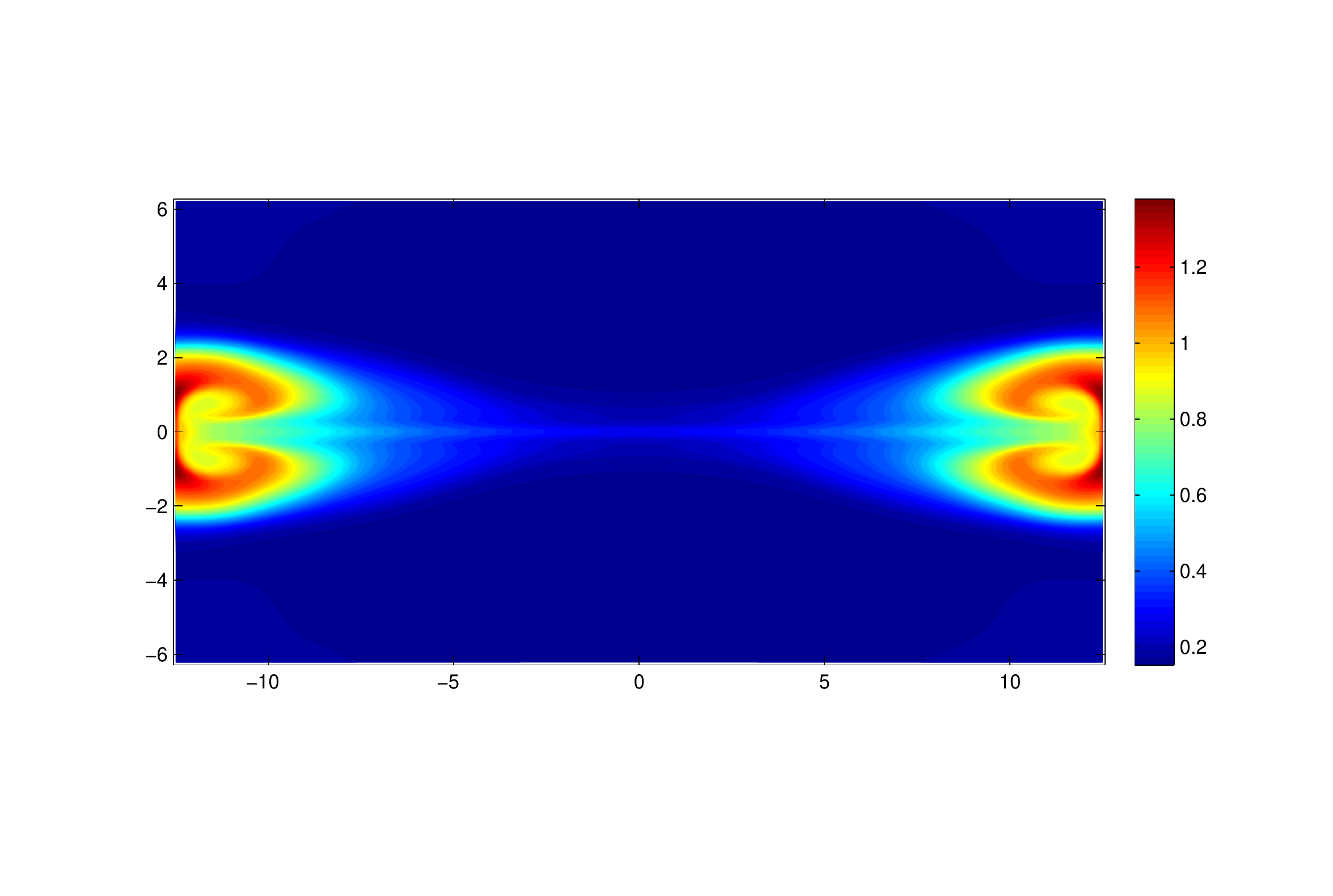}{a}
\includegraphics[width=0.45\textwidth]{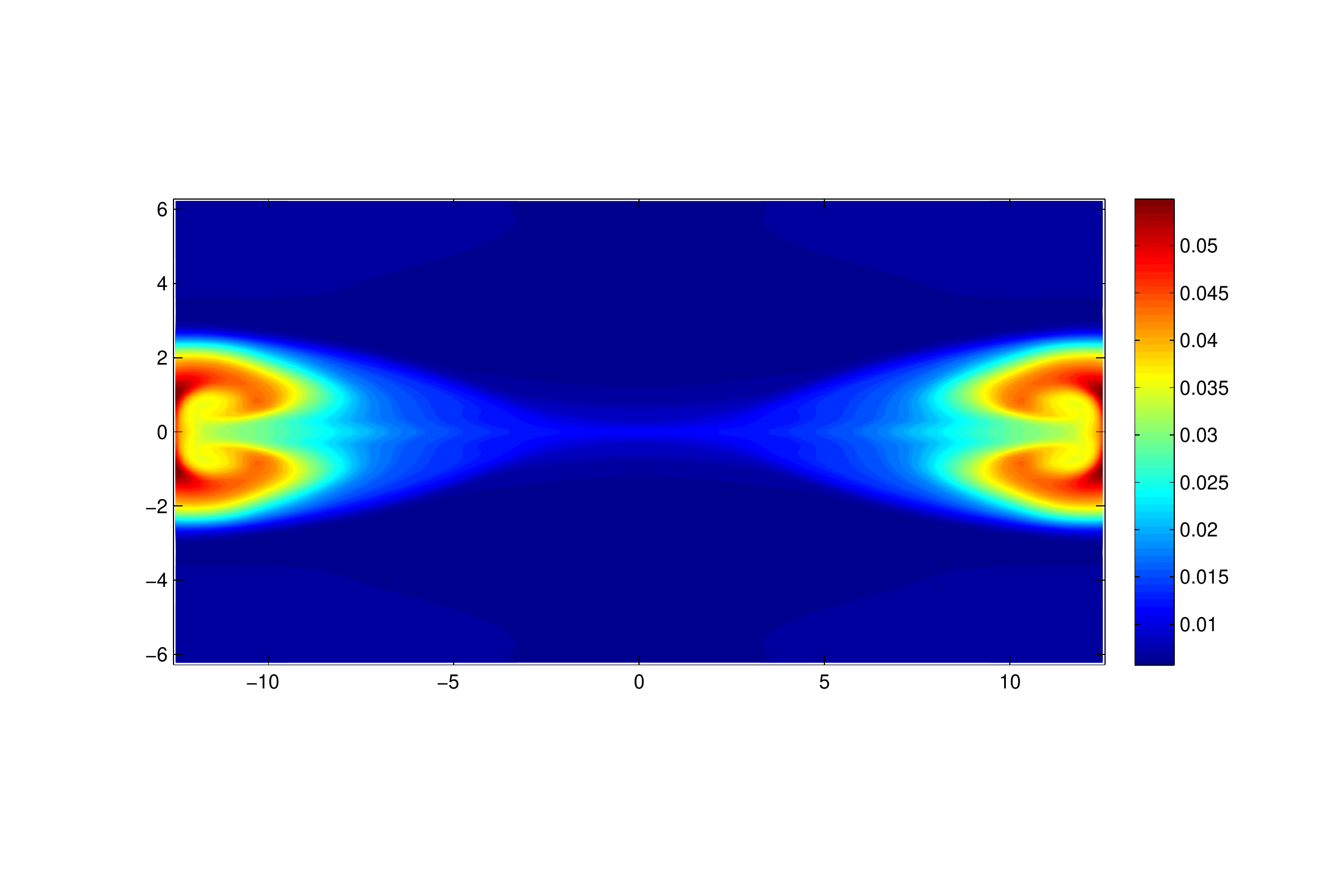}{b}\\
\includegraphics[width=0.45\textwidth]{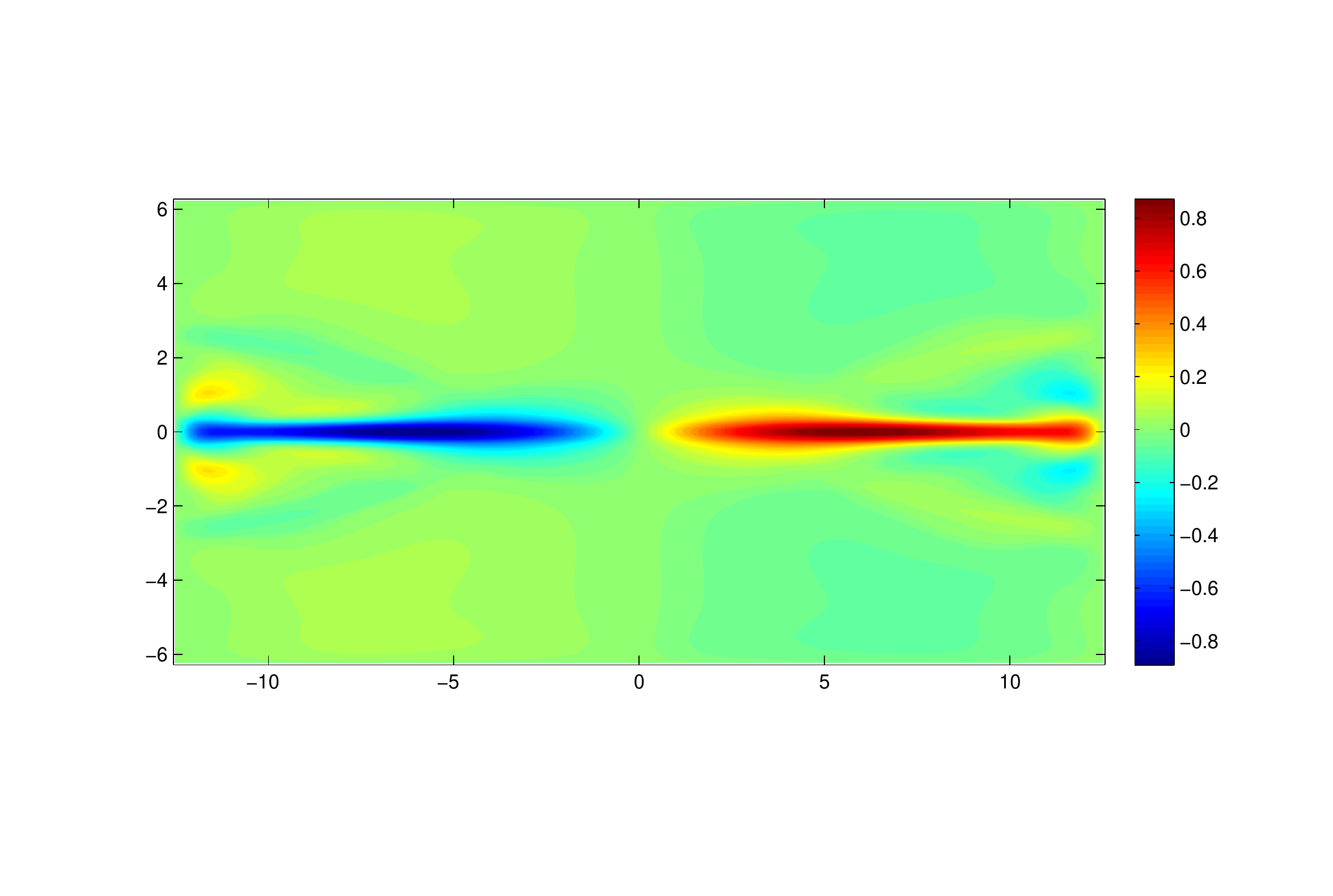}{c}
\includegraphics[width=0.45\textwidth]{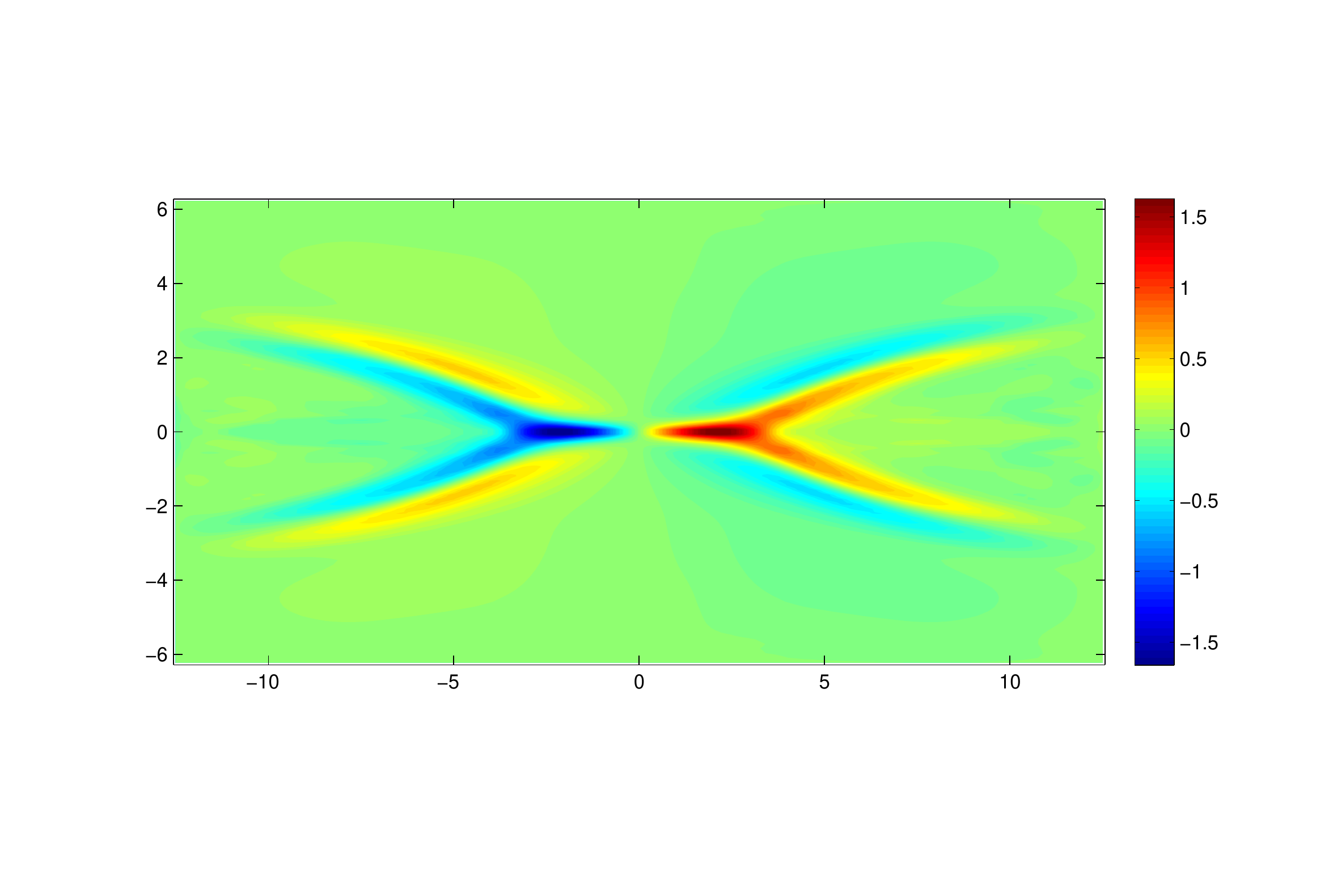}{d}\\
\includegraphics[width=0.45\textwidth]{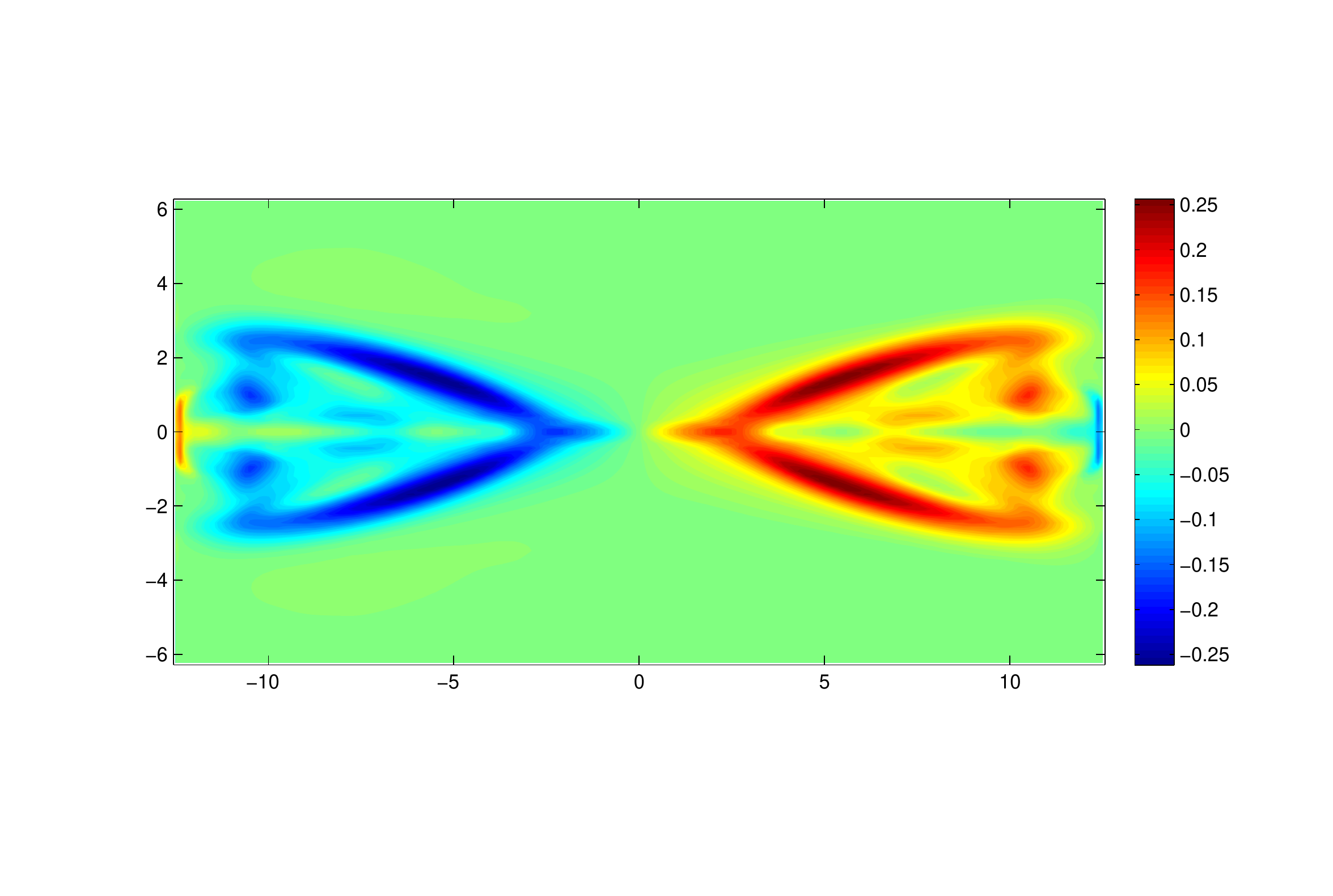}{e}
\includegraphics[width=0.45\textwidth]{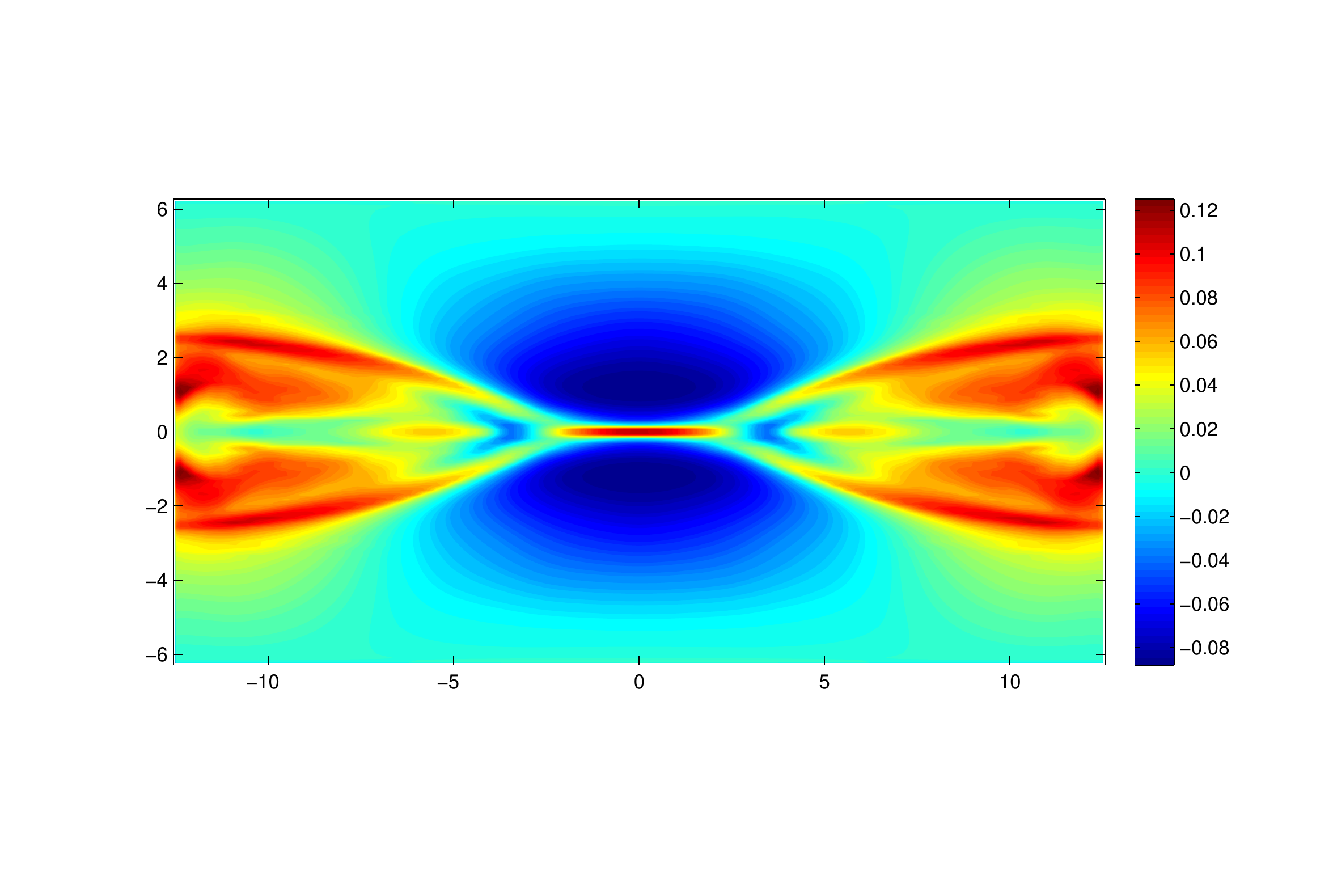}{f}
\caption{The reconnection process with $r_L = 1.0$
at $t=40\omega^{-1}$. (a) ion density; (b) electron density; (c) x momentum of ion; (d) x momentum of electron; (e) z-direction magnetic field; (e) z-direction electric field.}
\label{reconnection3}
\end{figure}

Fig. \ref{reconnection} shows the reconnected flux of UGKS defined by
\begin{equation}\nonumber
\phi(t)=\frac{1}{2L_x}\int_{-L_x/2}^{L_x/2} |B_y(x,0,t)| dx,
\end{equation}
 which are compared with other GEM results.
The fast reconnection rate can be predicted by UGKS, and the magnitude of reconnected flux
depends on the plasma conditions.
It can be observed from the results that the reconnected flux from UGKS behaves likes resistive-MHD result
when the normalized Larmor radius is small ($r_L=0.5$), and approaches to Hall-MHD solution when the normalized Larmor radius is large ($r_L=2$).
Fig. \ref{reconnection1} shows the magnetic flux at $\omega t=40$, with $r_L=1$.
Fig. \ref{reconnection2} shows the electromagnetic and flow  energy. The total energy of the system almost keeps a constant.
The electron and ion densities, and momentum distribution at $t=40\omega^{-1}$ are shown in Fig. \ref{reconnection3}, as well as the electromagnetic fields.

\section{Conclusion}

In this paper, a multi-scale numerical method for multi-species plasma simulation in the whole flow regimes
has been constructed.
The UGKS takes into account the collisions among electrons and ions, and their coupling through the full Maxwell equations.
The UGKS describes the plasma evolution on the mesh size and time step scales, which intrinsically provides the fundamental
multi-scale governing equations.
The flow physics covered by the current scheme is more general than those
from either the collisionless Vlasov equation or MHD equations
in the corresponding kinetic or hydrodynamic limit alone.
More importantly, the UGKS can give a reliable physical solution in the transitional regime as well,
which has not been fully explored before from the
particle-based and MHD-based numerical methods.

In the generalized Brio-Wu test case, the UGKS presents a smooth transition from neutral fluid results to the MHD solutions.
At the same time, with the time scale variation the results from the kinetic Vlasov equation to the hydrodynamic two-fluid system have been
obtained.
The study of the Orszag-Tang turbulence problem
 shows the ability of the UGKS in capturing all kinds of MHD solutions and recovering the magnetic reconnection mechanism
 under different conditions.
The UGKS is also able to capture the phenomena on the scale of Debye length, such as the fast magnetic reconnection in GEM case.
The direct modeling methodology makes it possible to construct UGKS for the study of multi-scale transport in the
rarefied gas dynamics, radiative transfer, and plasma physics.

\section*{Acknowledgement}
 The current work  is supported by Hong Kong Research Grant Council (620813, 16211014, 16207715), HKUST research fund
(PROVOST13SC01, IRS16SC42, SBI14SC11), and  National Science Foundation of China (91330203,91530319).

\section*{References}
\bibliographystyle{ieeetr}

\begin{thebibliography}{10}

\bibitem{chen1984plasma}
F.~F. Chen, {\em Introduction to plasma physics and controlled fusion}.
\newblock Plenum Press, New York and London, 2nd edition (1974).

\bibitem{vahedi1995monte}
V.~Vahedi and M.~Surendra, ``A {Monte Carlo} collision model for the
  particle-in-cell method: applications to argon and oxygen discharges,'' {\em
  Comput. Phys. Commun.}, vol.~87 (1995), no.~1, pp.~179--198.

\bibitem{filbet2003comparison}
F.~Filbet and E.~Sonnendr{\"u}cker, ``Comparison of {Eulerian {{Vlasov}}}
  solvers,'' {\em Comput. Phys. Commun.}, vol.~150 (2003), no.~3, pp.~247--266.

\bibitem{filbet2001conservative}
F.~Filbet, E.~Sonnendr{\"u}cker, and P.~Bertrand, ``Conservative numerical
  schemes for the {{{Vlasov}}} equation,'' {\em J. Comput. Phys.}, vol.~172 (2001),
  no.~1, pp.~166--187.

\bibitem{degon2010}
P.~Degon, F.~Deluzet, F.~Navoret, and A.B.~Sun, ``Asymptotic-preserving particle-in-cell method for the {Vlasov–Poisson} system near quasineutrality,'' {\em J. Comput. Phys.}, vol.~229 (2010),
  no.~1, pp.~5630--5652.

\bibitem{degon2017}
P.~Degon, F.~Deluzet, F.~Navoret, and A.B.~Sun, ``Asymptotic-preserving particle-in-cell method for the {Vlasov–Maxwell} system in the quasi-neutral limit,'' {\em J. Comput. Phys.}, vol.~330 (2017),
  no.~1, pp.~467-492.

\bibitem{degon2016}
P.~Degon, F.~Deluzet, ``Asymptotic-Preserving methods and multiscale models for plasma physics,'' {\em arXiv preprint}, (2016).

\bibitem{qiu2011conservative}
J.-M. Qiu and C.-W. Shu, ``Conservative semi-{{Lagrangian}} finite difference
  {WENO} formulations with applications to the {{{Vlasov}}} equation,'' {\em
  Commun. Comput. Phys.}, vol.~10 (2011), no.~4, p.~979.

\bibitem{guo2013hybrid}
W.~Guo and J.-M. Qiu, ``Hybrid {semi-{{Lagrangian}}} finite element-finite
  difference methods for the {{{Vlasov}}} equation,'' {\em J. Comput. Phys.},
  vol.~234 (2013), pp.~108--132.

\bibitem{xiong2014high}
T.~Xiong, J.-M. Qiu, Z.~Xu, and A.~Christlieb, ``High order maximum principle
  preserving semi-{{Lagrangian}} finite difference {WENO} schemes for the
  {{{Vlasov}}} equation,'' {\em J. Comput. Phys.}, vol.~273 (2014), pp.~618--639.

\bibitem{powell1999solution}
K.~G. Powell, P.~L. Roe, T.~J. Linde, T.~I. Gombosi, and D.~L. De~Zeeuw, ``A
  solution-adaptive upwind scheme for ideal magnetohydrodynamics,'' {\em J.
  Comput. Phys.}, vol.~154 (1999), no.~2, pp.~284--309.

\bibitem{brio1988upwind}
M.~Brio and C.~C. Wu, ``An upwind differencing scheme for the equations of
  ideal magnetohydrodynamics,'' {\em J. Comput. Phys.}, vol.~75 (1988), no.~2,
  pp.~400--422.

\bibitem{xu1999gas}
K.~Xu, ``Gas-kinetic theory-based flux splitting method for ideal
  magnetohydrodynamics,'' {\em J. Comput. Phys.}, vol.~153 (1999), no.~2,
  pp.~334--352.

\bibitem{araya2015magneto}
D.~B. Araya, F.~H. Ebersohn, S.~E. Anderson, and S.~S. Girimaji, ``Magneto-gas
  kinetic method for nonideal magnetohydrodynamics flows: verification protocol
  and plasma jet simulations,'' {\em J. Fluids Eng.}, vol.~137 (2015), no.~8,
  p.~081302.

\bibitem{shumlak2003approximate}
U.~Shumlak and J.~Loverich, ``Approximate {Riemann} solver for the two-fluid
  plasma model,'' {\em J. Comput. Phys.}, vol.~187 (2003), no.~2, pp.~620--638.

\bibitem{hakim2006high}
A.~Hakim, J.~Loverich, and U.~Shumlak, ``A high resolution wave propagation
  scheme for ideal two-fluid plasma equations,'' {\em J. Comput. Phys.},
  vol.~219 (2006), no.~1, pp.~418--442.

\bibitem{loverich2005discontinuous}
J.~Loverich and U.~Shumlak, ``A discontinuous {Galerkin} method for the full
  two-fluid plasma model,'' {\em J. Comput. Phys.}, vol.~169 (2005), no.~1,
  pp.~251--255.

\bibitem{loverich2011discontinuous}
J.~Loverich, A.~Hakim, and U.~Shumlak, ``A discontinuous {Galerkin} method for
  ideal two-fluid plasma equations,'' {\em J. Comput. Phys.}, vol.~9 (2011), no.~02,
  pp.~240--268.

\bibitem{srinivasan2011analytical}
B.~Srinivasan and U.~Shumlak, ``Analytical and
computational study of the ideal full two-fluid plasma model and asymptotic approximations for {Hall-magnetohydrodynamics},''
{\em Phys. Plasmas}, vol.~18 (2011), no.~9, p.~092113.

\bibitem{crestetto2012kinetic}
A.~Crestetto, N.~Crouseilles, and M.~Lemou, ``Kinetic/fluid micro-macro
  numerical schemes for {{{Vlasov}}-Poisson-{BGK}} equation using particles,''
  {\em Kin. Rel. Mod.}, vol.~5 (2012), no.~4, pp.~787--816.

\bibitem{dimarco2014asymptotic}
G.~Dimarco, L.~Mieussens, and V.~Rispoli, ``An asymptotic preserving automatic
  domain decomposition method for the {{{Vlasov}}-Poisson-{BGK}} system with
  applications to plasmas,'' {\em J. Comput. Phys.}, vol.~274 (2014), pp.~122--139.

\bibitem{jin2011class}
S.~Jin and B.~Yan, ``A class of asymptotic-preserving schemes for the
  {Fokker-Planck-Landau} equation,'' {\em J. Comput. Phys.}, vol.~230 (2011), no.~17,
  pp.~6420--6437.

\bibitem{dimarco2015numerical}
G.~Dimarco, Q.~Li, L.~Pareschi, and B.~Yan, ``Numerical methods for plasma
  physics in collisional regimes,'' {\em J. Plasma Phys.}, vol.~81 (2015), no.~1,
  305810106

\bibitem{degond}
P.~Degond, F.~Deluzet, L.~Navoret, A.-B. Sun, and M.-H. Vignal,
  ``{Asymptotic}-preserving particle-in-cell method for the {Vlasov--Poisson}
  system near quasineutrality,'' {\em J. Comput. Phys.}, vol.~229 (2010), no.~16,
  pp.~5630--5652.

\bibitem{xu2010}
K.~Xu and J.~Huang, ``A unified gas-kinetic scheme for continuum and rarefied
  flows,'' {\em J. Comput. Phys.}, vol.~229 (2010), no.~20, pp.~7747--7764.

\bibitem{Huangxuyu2}
J.~Huang, K.~Xu, and P.~Yu, ``A unified gas-kinetic scheme for continuum and
  rarefied flows \uppercase\expandafter{\romannumeral2}: Multi-dimensional
  cases,'' {\em Commun. Comput. Phys.}, vol.~12 (2012), no.~3, pp.~662--690.

\bibitem{Huangxuyu3}
J.~Huang, K.~Xu, and P.~Yu, ``A unified gas-kinetic scheme for continuum and
  rarefied flows \uppercase\expandafter{\romannumeral3}: Microflow
  simulations,'' {\em Commun. Comput. Phys.}, vol.~14 (2013), no.~5, pp.~1147--1173.

\bibitem{sun-wj}
W.~Sun, S.~Jiang, and K.~Xu, ``An asymptotic preserving unified gas kinetic
  scheme for gray radiative transfer equations,'' {\em J. Comput. Phys.},
  vol.~285 (2015), pp.~265--279.

\bibitem{sun-wj2}
W.~Sun, S.~Jiang, K.~Xu, and S.~Li, ``An asymptotic preserving unified gas
  kinetic scheme for frequency-dependent radiative transfer equations,'' {\em
  J. Comput. Phys.}, vol.~302 (2015), pp.~222--238.

\bibitem{guo-xu}
Z.~Guo and K.~Xu, ``{Discrete} unified gas kinetic scheme for multiscale heat transfer based on the phonon {Boltzmann} transport equation,'' {\em Int. J. Heat Mass},
  vol.~102 (2016),  pp.~944--958.

\bibitem{xu-book}
K.~Xu, ``Direct modeling for computational fluid dynamics: construction and
  application of unified gas-kinetic schemes,'' {\em World Scientific,
  Singapore} (2014).

\bibitem{AAP}
P.~Andries, K.~Aoki, and B.~Perthame, ``A consistent {{BGK}}-type model for gas
  mixtures,'' {\em J. Stat. Phys.}, vol.~106 (2002), no.~5-6, pp.~993--1018.

\bibitem{liu2016}
C.~Liu, K.~Xu, Q.~Sun, and Q.~Cai, ``A unified gas-kinetic scheme for continuum and rarefied flows IV: full {Boltzmann} and model equations,'' {\em
  J. Comput. Phys.}, vol.~314 (2016), pp.~305--340.

\bibitem{munz2000divergence}
C.-D. Munz, P.~Omnes, R.~Schneider, E.~Sonnendr{\"u}cker, and U.~Voss,
  ``Divergence correction techniques for {Maxwell} solvers based on a
  hyperbolic model,'' {\em J. Comput. Phys.}, vol.~161 (2000), no.~2, pp.~484--511.

\bibitem{LeVeque}
R.J.~LeVeque, `` Finite volume methods for hyperbolic problems,'' {\em Cambridge university press} (2002).

\bibitem{orszag}
S.~A. Orszag and C.-M. Tang, ``Small-scale structure of two-dimensional
  magnetohydrodynamic turbulence,'' {\em J. Fluid Mech.}, vol.~90 (1979), no.~01,
  pp.~129--143.

\bibitem{tang2000high}
H.-Z. Tang and K.~Xu, ``A high-order gas-kinetic method for multidimensional
  ideal magnetohydrodynamics,'' {\em J. Comput. Phys.}, vol.~165 (2000), no.~1,
  pp.~69--88.

\bibitem{Parker1957}
E.N.~Parker, ``{Sweet's} mechanism for merging magnetic fields in conducting fluids,'' {\em J. Geophys. Res.}, vol.~62 (1957), no.~4, pp.~509--520.

\bibitem{geospace}
J.~Birn, J.~Drake, M.~Shay, B.~Rogers, R.~Denton, M.~Hesse, M.~Kuznetsova,
  Z.~Ma, A.~Bhattacharjee, A.~Otto, {\em et~al.}, ``Geospace environmental
  modeling ({{GEM}}) magnetic reconnection challenge,'' {\em J. Geophys.
  Res.-Space}, vol.~106 (2001), no.~A3, pp.~3715--3719.

\bibitem{hesse}
M.~Hesse, J.~Birn, and M.~Kuznetsova, ``Collisionless magnetic reconnection:
  Electron processes and transport modeling,'' {\em J. Geophys. Res.-Space},
  vol.~106 (2001), no.~A3, pp.~3721--3735.

\bibitem{birn2001geospace}
J.~Birn and M.~Hesse, ``Geospace environment modeling ({{GEM}}) magnetic
  reconnection challenge: Resistive tearing, anisotropic pressure and hall
  effects,'' {\em J. Geophys. Res.-Space}, vol.~106 (2001), no.~A3, pp.~3737--3750.

\bibitem{ma2001hall}
Z.~Ma and A.~Bhattacharjee, ``Hall magnetohydrodynamic reconnection: The
  geospace environment modeling challenge,'' {\em J. Geophys. Res.-Space},
  vol.~106 (2001), no.~A3, pp.~3773--3782.

\bibitem{pritchett}
P.~Pritchett, ``Geospace environment modeling magnetic reconnection challenge:
  Simulations with a full particle electromagnetic code,'' {\em J. Geophys.
  Res.-Space}, vol.~106 (2001), no.~A3, pp.~3783--3798.

\bibitem{kuznetsova}
M.~M. Kuznetsova, M.~Hesse, and D.~Winske, ``Collisionless reconnection
  supported by nongyrotropic pressure effects in hybrid and particle
  simulations,'' {\em J. Geophys. Res.-Space}, vol.~106 (2001), no.~A3,
  pp.~3799--3810.

\end{thebibliography}

\clearpage

\section*{Appendix: eigen-system of perfectly hyperbolic Maxwell equations}

The PHM equations read
\begin{equation}
  \frac{\partial \mathbf{Q}}{\partial t}+\mathbf{A}_1\frac{\partial \mathbf{Q}}{\partial x}+\mathbf{A}_2\frac{\partial \mathbf{Q}}{\partial y}=s,
\end{equation}
where
$\mathbf{Q}=(E_1, E_2, E_3, B_1, B_2, B_3, \phi, \psi)^T$, $\mathbf{s}=(-J_1/\epsilon, -J_2/\epsilon, -J_3/\epsilon, 0, 0, 0, \chi\rho/\epsilon_0,0)^T$,
$$A_1=\left(
      \begin{array}{cccccccc}
        0 & 0 & 0 & 0 & 0 & 0 & c^2\chi & 0 \\
        0 & 0 & 0 & 0 & 0 & c^2 & 0 & 0 \\
        0 & 0 & 0 & 0 & -c^2 & 0 & 0 & 0 \\
        0 & 0 & 0 & 0 & 0 & 0 & 0 & \nu \\
        0 & 0 & -1 & 0 & 0 & 0 & 0 & 0 \\
        0 & 1 & 0 & 0 & 0 & 0 & 0 & 0 \\
        \chi & 0 & 0 & 0 & 0 & 0 & 0 & 0 \\
        0 & 0 & 0 & c^2\nu & 0 & 0 & 0 & 0 \\
      \end{array}
    \right),$$

    and

$$A_2=\left(
      \begin{array}{cccccccc}
        0 & 0 & 0 & 0 & 0 & -c^2 & 0 & 0 \\
        0 & 0 & 0 & 0 & 0 & 0 & c^2\chi & 0 \\
        0 & 0 & 0 & c^2 & 0 & 0 & 0 & 0 \\
        0 & 0 & 1 & 0 & 0 & 0 & 0 & 0 \\
        0 & 0 & 0 & 0 & 0 & 0 & 0 & \nu \\
        -1 & 0 & 0 & 0 & 0 & 0 & 0 & 0 \\
        0 & \chi & 0 & 0 & 0 & 0 & 0 & 0 \\
        0 & 0 & 0 & 0 & c^2\nu & 0 & 0 & 0 \\
      \end{array}
    \right).$$

    For $A_1$, the eigenvalues are $\{c,c,c\chi,c\nu,-c,-c,-c\chi,-c\nu\}.$ The right eigenvectors of $A_1$ are given by the columns of the matrix

    $$R_1=\left(
          \begin{array}{cccccccc}
            0 & 0 & c & 0 & 0 & 0 & -c & 0 \\
            0 & c & 0 & 0 & 0 & -c & 0 & 0 \\
            -c & 0 & 0 & 0 & c & 0 & 0 & 0 \\
            0 & 0 & 0 & 1/c & 0 & 0 & 0 & -1/c \\
            1 & 0 & 0 & 0 & 1 & 0 & 0 & 0 \\
            0 & 1 & 0 & 0 & 0 & 1 & 0 & 0 \\
            0 & 0 & 1 & 0 & 0 & 0 & 1 & 0 \\
            0 & 0 & 0 & 1 & 0 & 0 & 0 & 1 \\
          \end{array}
        \right).$$

    The left eigenvectors are the rows of the matrix

    $$L_1=\left(
           \begin{array}{cccccccc}
             0 & 0 & -\frac{1}{2c} & 0 & \frac12 & 0 & 0 & 0 \\
             0 & 0 & \frac{1}{2c} & 0& 0 & \frac12 & 0 & 0 \\
             \frac{1}{2c} & 0 & 0 & 0 & 0 &0 & \frac12 & 0 \\
             0 & 0 & 0 & 0 & \frac{c}{2} & 0 & 0 & \frac12 \\
             0 & 0 & \frac{1}{2c} & 0 & \frac12 & 0 & 0 & 0 \\
             0 & 0 & -\frac{1}{2c} & 0 & 0 & \frac12 & 0 & 0 \\
             -\frac{1}{2c} & 0 & 0 & 0 & 0 & 0 & \frac12 & 0 \\
             0 & 0 & 0 & 0 & -\frac{c}{2} & 0 & 0 & \frac12 \\
           \end{array}
         \right).$$

    For $A_2$, the eigenvalues are $\{c,c,c\chi,c\nu,-c,-c,-c\chi,-c\nu\}.$ The right eigenvectors of $A_2$ are given by the columns of the matrix

    $$R_2=\left(
          \begin{array}{cccccccc}
            0 & -c & 0 & 0 & 0 & c & 0 & 0 \\
            0 & 0 & c & 0 & 0 & 0 & -c & 0 \\
            c & 0 & 0 & 0 & -c & 0 & 0 & 0 \\
            1 & 0 & 0 & 0 & 1 & 0 & 0 & 0 \\
            0 & 0 & 0 & 1/c & 0 & 0 & 0 & -1/c \\
            0 & 1 & 0 & 0 & 0 & 1 & 0 & 0 \\
            0 & 0 & 1 & 0 & 0 & 0 & 1 & 0 \\
            0 & 0 & 0 & 1 & 0 & 0 & 0 & 1 \\
          \end{array}
        \right).$$

    The left eigenvectors are the rows of the matrix

    $$L_2=\left(
           \begin{array}{cccccccc}
             0 & 0 & \frac{1}{2c} & \frac12 & 0 & 0 & 0 & 0 \\
             -\frac{1}{2c} & 0 & 0 & 0& 0 & \frac12 & 0 & 0 \\
             0 & \frac{1}{2c} & 0 & 0 & 0 &0 & \frac12 & 0 \\
             0 & 0 & 0 & 0 & \frac{c}{2} & 0 & 0 & \frac12 \\
             0 & 0 & -\frac{1}{2c} & \frac12 & 0 & 0 & 0 &  0 \\
             \frac{1}{2c} &0 & 0 & 0 & 0 & \frac12 & 0 & 0 \\
             0 &-\frac{1}{2c} & 0 & 0 & 0 & 0 & \frac12 & 0 \\
             0 & 0 & 0 & 0 & -\frac{c}{2} & 0 & 0 & \frac12 \\
           \end{array}
         \right).$$
\end{document}